\documentstyle[psfig]{mn2e}
\input{psfig}
\def\bj{b_{\rm J}}

\title[]{The 2dF BL Lac Survey II}
\author[D.Londish et al.]
	{D. Londish$^{1,2,4}$, S.M. Croom$^{2}$, J.Heidt$^3$, B.J. Boyle$^{4,2}$,  E.M. Sadler$^{1}$, M. Whiting$^4$,\newauthor T. A. Rector$^5$, T. Pursimo$^6$, K. Chynoweth$^7$\\
${^1}$ University of Sydney, School of Physics, Sydney NSW 2006, Australia \\
${^2}$ Anglo-Australian Observatory, PO Box 296, Epping, NSW 1710, Australia\\
${^3}$ Landessternwarte Heidelberg, K\"onigstuhl, D$-$69117 Heidelberg, Germany\\
${^4}$ Australia Telescope National Facitity,PO Box 76, Epping NSW 1710, Australia\\
${^5}$ Dept of Physics and Astronomy, University of Alaska Anchorage, 3211 Providence Dr., Anchorage AK 99508 USA\\
${^6}$ Nordic Optical Telescope,E-38700 Santa Cruz de La Palma, Spain\\
${^7}$ NRAO, P.O. Box O, Socorro, NM 87801 USA\\}

\begin{document}

\pagerange{\pageref{firstpage}--\pageref{lastpage}} \pubyear{2006}

\maketitle


\begin{abstract}
We report on our further analysis of the expanded and revised sample of potential BL Lac objects (the 2BL) optically identified from two catalogues of blue-selected (UV excess) point sources, the 2dF and 6dF QSO Redshift Surveys (2QZ and 6QZ). The 2BL comprises 52 objects with no apparent proper motion, over the magnitude range $16.0 < \bj< 20.0$. Follow-up high signal-to-noise spectra of 36 2BL objects and NIR imaging of 18 objects, together with data for 19 2BL objects found in the Sloan Digital Sky survey (SDSS), show 17 objects to be stellar, while a further 16 objects have evidence of weak, broad emission features, although for at least one of these the continuum level has clearly varied. Classification of three objects remains uncertain,with  NIR results indicating a marked reduction in flux as compared to SDSS optical magnitudes. Seven objects have neither high signal-to-noise spectra nor NIR imaging. Deep radio observations of 26 2BL objects at the VLA resulted in only three further radio-detections, however none of the three is classed as a featureless continuum object. Seven 2BL objects with a radio detection are confirmed as candidate BL Lac objects while one extragalactic ($z=$0.494) continuum object is undetected at radio frequencies. One further radio-undetected object is also a potential BL Lac candidate. However it would appear that there is no significant population of radio-quiet BL Lac objects.
\end{abstract}

\section{Introduction}

Unification theories of active galactic nuclei AGN) (Antonucci 1993, Urry \& Padovani 1995) propose that BL Lac objects are the beamed counterparts of FRI/FRII\footnote{double-lobed radio sources classified by Fanaroff \& Riley (1974)} radio galaxies. At optical wavelengths BL Lac objects are dominated by Doppler boosted, synchrotron radiation, the resultant featureless  continuum making BL Lacs all but impossible to find in optical surveys.   Until recently, therefore, all complete, flux-limited samples of BL Lac objects have initially been identified from radio and/or X-ray surveys, unification theories seemingly ruling out the possibility of these objects being radio- and X-ray-quiet. Seven BL Lac objects are detected in the Palomar-Green Survey (PG, Green et al.\,1986). Three of these were originally classified as DC white dwarfs (a continuum white dwarf (WD) with no spectral absorption features), but were later found by Fleming et al.\,(1993) to be strong X-ray emitters;  these authors thus consider these sources to be X-ray selected BL Lac objects, while the remaining four PG BL Lacs are found in flux limited radio samples. 

Identification of the first optically selected sample of BL Lac objects from 
scrutiny of spectra in the 2dF QSO Redshift Survey (2QZ Croom et al.\,2001; 2004) was commenced in 2000 (Londish et al.\,2002, hereinafter Paper I).  A similar search of colour-selected point sources to optically identify BL Lac candidates has now also been conducted using the Sloan Digital Sky Survey (Collinge et al.\ 2005 ).  From the 2dF QSO Redshift Survey we initially selected 56 featureless continuum objects as candidate BL Lacs from 6772 spectra of objects with  $18.25<\bj\leq 20.00$ and signal-to-noise ratio (SNR) $\geq10.0$, observed up to August 2001. We have now expanded the survey to include scrutiny of a further 734 objects satisfying the same selection criteria, observed after August 2001 and up to completion of the 2QZ survey in April 2002.  In addition we have examined spectra of bright objects ($16.0 \leq \bj \leq 18.25$ and SNR $\geq 10$) from the 6dF QSO Redshift survey (6QZ, Croom et al.\,2004); only the southern strip of the 6QZ is complete. The survey area covered in the 2QZ is 646.8 sq deg and 332.9 sq deg for 6QZ. More accurate 1$\sigma$ proper motion errors in the updated SuperCosmos catalogue\footnote{SuperCosmos Sky Survey at www-wfau.roe.ac.uk/sss, maintained by the Institute for Astronomy, Royal Observatory, Edinburgh} led to a further revision of the original sample.

Full details of the selection criteria used (including the colour selection of 2QZ and 6QZ objects) are presented in paper I  and are outlined in part in Section 2 below. In Section 3 we report on results from the high signal-to-noise spectroscopy carried out using the Very Large Telescope\footnote{ VLT on Cerro Paranal (Chile) operated by the European  Southern Observatory, observing run 073.B-0082}, and in Section 4 on our investigation of the optical and near  infrared  colours of 2BL objects. Follow-up radio observations are presented in Section 5. A summary and conclusions follow in Section 6.  

\section{Selection of BL Lac candidates}
\begin{table*}
\caption{2BL sample of 49 objects from the 2QZ survey and 3 objects from the 6QZ survey }
\begin{tabular}{l l r c l c r r} 
\multicolumn{1}{c} { object} &
\multicolumn{1}{c}  { RA (2000)} & 
\multicolumn{1}{c} { dec (2000)} &
\multicolumn{1}{c} { $\bj$} &
\multicolumn{1}{c} {mJy radio flux$^a$} &
\multicolumn{1}{c} {Notes} &
\multicolumn{1}{c} {photometry$\dag$} &
\multicolumn{1}{c} {classification$\dag\dag$}\\
\hline
J002522.8$-$284034 & 00  25  22.80 & $-$28  40  34.9 & 18.95 & & & -- &star \\ 
J002746.6$-$293308 & 00  27  46.66 & $-$29  33  08.4 & 19.08 & & & -- &star \\ 
J002751.5$-$293506 & 00  27  51.56 & $-$29  35  06.9 & 18.83 &  $S_{8.4} <$0.15& & -- &star\\
J003058.2$-$275629 & 00  30  58.27 & $-$27  56  29.8 & 19.03 &  $S_{1.4} < $0.15 & $z=$1.606& -- & AGN\,I$^b$  \\  
J004106.8$-$291114 & 00  41  06.83 & $-$29  11  14.7 & 18.97 &   $S_{8.4} < $0.21& &  --  & star \\  
J004950.6$-$284907 & 00  49  50.69 & $-$28  49  07.9 & 19.70 &   $S_{8.4} < $0.11 && -- & star \\  
$^*$J014149.8$-$310738 & 01 41 49.79 & $-$31 07 38.5 & 19.33 & $S_{\rm 843MHz} =10.5$ &  & -- & --\\
J014310.1$-$320056 & 01  43  10.10 & $-$32  00  56.7 & 19.36 &  $S_{1.4}=76$&  $z=$0.375, BLL$^c$ & -- & AGN\,I\\  
$^*$J015407.1$-$320104 & 01 54 07.18 & $-$32 01 03.1 & 19.24 & && -- & --\\
$^*$J020837.0$-$294826 & 02 08 36.99 & $-$29 48 26.1 & 19.37 & & &-- & --\\
$^{**}$J022044.7$-$303713 & 02 20 44.75 & $-$30 37 13.3 & 17.41  &  & $z=$1.439& --& AGN\,I \\
$^*$J022120.5$-$303722 & 02 21 20.58 & $-$30 37 22.2 & 18.98 & & &-- & --\\
J023405.5$-$301519 & 02  34  05.58 & $-$30  15  19.5 & 18.54 &   $S_{8.4} < $0.11& & -- & --\\  
J023536.7$-$293843 & 02  35  36.70 & $-$29  38  43.6 & 18.64 & $S_{1.4}=5$ & BLL$^c$  & -- & BLLac \\  
J030416.3$-$283217 & 03  04  16.33 & $-$28  32  17.9 & 19.54  &$S_{1.4}=8$ & BLL$^c$ & -- & BLLac \\	
J031056.9$-$305901 & 03  10  56.91 & $-$30  59  01.3 & 18.70   &  $S_{8.4} = \,$0.129? &$z=$1.911 & -- & AGN\,I\\  
J103607.4+015658   & 10  36  07.48 &  +01  56  58.4 & 19.23   & $S_{8.4} < $0.11 & $z=$1.875  & nonthermal & AGN\,I\\	
J105355.1$-$005538 & 10  53  55.18 & $-$00  55  38.5 & 19.53  & & & thermal  &star\\ 
J105534.3$-$012617 & 10 55  34.36 & $-$01 26 17.3 &  18.96   & $S_{1.4}=11$& BLL$^c$ & -- & BLLac \\  
J110644.5+000717 &  11  06  44.52 &  +00  07  17.4 & 19.86 & && thermal & star \\
$^*$J113039.1$-$004023 & 11 30 39.10  & $-$00 40 23.8 &  19.01   & & $z=$1.7652  &  thermal & AGN\,I\\
J113413.4+001041   & 11  34  13.47 &  +00  10  41.3 & 18.80  & $S_{8.4} < $0.11&   $z=$1.4856 & ?? & AGN\,I   \\  
J113900.5$-$020140 & 11  39  00.54 & $-$02  01  40.9 & 19.68 &   $S_{8.4} < $0.11 & & variable? & --    \\ 
J114137.1$-$002730 & 11  41  37.10 & $-$00  27  30.8 & 19.97 &  $S_{8.4} = \,$3.58 & & variable? & --   \\  
J114221.4$-$014812 & 11  42  21.42 & $-$01  48  12.2 & 19.30  & $S_{8.4} < $0.11 & & thermal& star  \\  
J114327.3$-$005050 & 11 43  27.30 & $-$00 50 50.6 & 19.97  & $S_{8.4} < $0.11&   & nonthermal? & --     \\  
J114554.8+001023  & 11  45  54.85 &   +00 10 23.6 & 19.59 && & thermal  & star \\
J115909.6$-$024534 & 11  59  09.61 & $-$02  45  34.9 & 19.24 &   $S_{8.4} =$0.296 &  & ?? & AGN\,I  \\  
J120015.3+000552   & 12  00  15.35 &  +00  05  52.6 & 19.81  & $S_{8.4} < $0.11 &  $z=$1.650  & nonthermal? &AGN\,I \\
J121834.8$-$011955 & 12  18  34.88 & $-$01  19  55.9 & 19.70 &   $S_{1.4} = $244 & & nonthermal  & BLLac \\  
J123437.6$-$012953 & 12  34  37.64 & $-$01  29  53.1 & 19.44  & $S_{8.4} < $0.11 & $z=$1.716 & nonthermal? & AGN\,I \\ 
J125435.7$-$011822 & 12  54  35.76 & $-$01  18  22.5 & 19.44 &  $S_{8.4} < $0.11 & & thermal   & star   \\  
$^*$J125501.2+015513 & 12 55 01.32 & +01 55 13.3 & 19.66  && thermal& -- &\\
J140207.7$-$013033 & 14  02  07.70 & $-$01  30  33.3 & 19.71 & $S_{8.4} = \,$0.101 & & thermal  & star \\ 
J141040.2$-$023020 & 14  10  40.28 & $-$02  30  20.7 & 19.43 &  $S_{8.4} < $0.09 & & thermal  &  star \\  
J142526.2$-$011826 & 14  25  26.20 & $-$01  18  26.3 & 19.91 &    $S_{1.4} = \,$10 & & nonthermal& BLLac\\
J215454.3$-$305654 & 21  54  54.35 & $-$30  56  54.3 & 19.55 &   $S_{1.4} < $0.21 & $z=$0.494  & --  & BLLac  \\  
$^{**}$J220222.7$-$285306 & 22 02 22.70 & $-$28 53 05.5  & 18.12 & &$z=$1.231& -- & AGN\,I\\ 
J220850.0$-$302817 & 22  08  50.02 & $-$30  28  17.6 & 18.40  && $z=$1.711& -- & AGN\,I\\
J221105.2$-$284933 & 22 11 05.25 & $-$28 49 33.0 & 18.57  & $S_{1.4} = \,$81&$z=$1.840&  -- & AGN\,I\\
J221450.1$-$293225 & 22  14  50.11 & $-$29  32  25.2 & 19.21  &	&$z=$1.626 & -- & AGN\,I\\
J223233.5$-$272859 & 22  32  33.57 & $-$27  28  59.9 & 19.33  & $S_{8.4} < $0.11 & & -- & star\\  
$^*$J223245.4$-$284731 & 22 32 45.41 & $-$28 47 31.2 & 19.47 & && -- & --\\
J224559.1$-$312223 & 22  45  59.10 & $-$31  22  23.3 & 19.38 &   $S_{8.4} < $0.11 &$z=1.324$ or 0\,? & -- & ?? \\  
J225453.2$-$272509 & 22  54  53.20 & $-$27  25  09.4 & 18.83 &    $S_{1.4} = \,$52& $z=$0.333 & -- & BLLac\\  
J230306.0$-$312737 & 23  03  06.04 & $-$31  27  37.4 & 18.76 &  $S_{8.4} < $0.11 &$z=$2.459 & -- & AGN\,I\\
J230443.6$-$311107 & 23  04  43.60 & $-$31  11  07.5 & 19.49 &  $S_{8.4} < $0.11 & & -- & star\\  
$^*$J231713.6$-$314942 & 23  17  13.62 & $-$31  49  42.2 & 19.81 &  $S_{1.4} = \,$17.4 && -- & --\\
J231749.0$-$285350 & 23  17  49.00 & $-$28  53  50.2 & 19.51 &  $S_{8.4} < $0.11 & & --    & star\\ 
J232531.3$-$313136 & 23  25  31.36 & $-$31  31  36.0 & 19.54 &  & & -- & star   \\  
$^{**}$J233508.1$-$283035 & 23 35 08.19 & $-$28 30 35.3 & 16.41  & $S_{1.4} = \,$6 & $z=$1.386& -- & AGN\,I\\
J234414.7$-$312304 & 23  44  14.70 & $-$31  23  04.3 & 19.65 &     $S_{1.4} = \,$5.4 & & -- & BLLac  \\ 
\hline
\end{tabular}
\begin{flushleft}  {$^*$ These objects have been added to the original sample presented in Paper I;$^{**}$ signifies an object from the 6QZ survey.  \\
$^a$ 1.4 GHz fluxes are from NVSS (NRAO/VLA Sky Survey, Condon et al.\,1998) and/or FIRST (Faint Images of the Radio Sky at 20cm, White et al.\,1997) and upper limits from Australia Telescope Compact Array measurements (see \S5); 8.4 GHz fluxes and upper limits are from our own VLA observations (\S5); 843 MHz flux from the SUMSS Catalogue (Bock et al.\,1999).\\
$^b$ AGN\,I: broad/weak line AGN\\
$^c$ BLL signifies that the object is classified in the literature as a BL Lac (V\`eron-Cetty \& V\`eron 2001; Voges et al.\,1999)\\
$\dag$Optical photometry from SDSS, NIR photometry from Kitt Peak, Nordic Optical Telescope and Calar Alto\\
$\dag\dag$ Overall classification from our VLT observations (\S3) and/or SDSS (Sloan Digital Sky Survey, Stoughton et al.\,2002) spectra (Fig A1), and NIR imaging (\S4).\\
Redshifts are as quoted in SDSS and/or measured from VLT spectra - see Figs 3--5}
\end{flushleft}
\end{table*}
\begin{table*}
\caption{The 15 featureless continuum objects removed from the 2BL in June 2003 on account of their $> 2.5\sigma$ proper motion. Objects marked $\dag$ have SDSS photometry (see \S4) that indicates the object is indeed most likely stellar, whereas for the two objects marked $\dag \dag$ no blackbody curve could be fit to the SDSS optical magnitudes}
\begin{tabular}{l l r c  c} 
\multicolumn{1}{c} { object} &
\multicolumn{1}{c}  { RA (2000)} & 
\multicolumn{1}{c} { dec (2000)} &
\multicolumn{1}{c} { $\bj$} &
\multicolumn{1}{c}  {tot pm/$\sigma_{pm}$} \\
\hline
J024659.5$-$294822 & 02 46 59.58 & $-$29 48 22.2 & 19.94 & 3.95\\
J100253.3$-$001727$\dag$     &  10 02 53.31   & $-$00 17 27.9    &  19.16 & 11.58\\
J102615.3$-$000629$\dag$     &  10 26 15.35   & $-$00 06 30.0    &  19.28 & 5.70\\
J104519.7+002615$\dag$     &  10 45 19.74     & +00 26 15.3 &  18.67 & 7.29\\
J114010.5$-$002935$\dag$     &  11 40 10.54   & $-$00 29 35.5    &  19.85 & 3.34\\
J114521.6$-$024757$\dag$     &  11 45 21.64   & $-$02 47 57.4    &  18.45 & 3.70\\
J120558.1$-$004215$\dag$     &  12 05 58.18   & $-$00 42 15.8    &  19.23 & 12.40\\
J120801.8$-$004219$\dag$  & 12  08 01.85 & $-$00 42  19.5 & 18.96 & 2.72 \\
J122338.0$-$015619$\dag$  & 12  23 38.05 & $-$01 56  19.1 & 19.22 &  2.56\\
J130009.9$-$022600$\dag$  &  13 00 09.95 & $-$02 26 00.5    &  19.22 & 10.21 \\
J132811.5+000227$\dag$   & 13  28 11.54 &  +00  02  27.8 & 19.79 &  2.51 \\
J131635.1$-$002810$\dag \dag$ & 13 16 35.15   & $-$00 28 10.6  &  19.92& 9.07\\
J140021.0+001956$\dag \dag$     &  14 00 21.06   & +00 19 56.7    &  19.90&  3.87\\
J140916.3$-$000011$\dag$  &  14 09 16.36 & $-$00 00 11.1    &  18.89&  3.92 \\
J220515.8$-$311536     &  22 05 15.80   & $-$31 15 36.3    &  19.77 & 3.63 \\
\hline
\end{tabular}
\end{table*}
Spectroscopic observations for the 2QZ and 6QZ surveys were made using the 2-degree field (2dF) instrument at the Anglo-Australian Telescope and the 6-degree field (6dF) instrument at the UK Schmidt Telescope respectively. Data reduction and redshift estimation was carried out using the 2dF/6dF data reduction pipeline (Bailey et al.\,2003, Bailey \& Glazebrook 1999).  All spectra with  signal-to-noise ratio (SNR) $\geq10.0$ and $16.0 \leq \bj \leq 20.00$ were  visually 
re-examined with the express purpose of identifying objects with a featureless spectrum. For consistency we adopted the standard operational
definition of a BL Lac as an extragalactic object with no emission
lines greater than $W_{\lambda}=$5\AA\ and CaII H\&K break
contrast of less than 0.25 (Stocke et al.\ 1990). Objects from 2QZ fields with high levels of scattered moonlight were excluded from the sample.

Proper motion studies of the 2BL sample were repeated in June 2003 using the revised SuperCosmos catalogue (Hambly et al.\,2001);  for the majority of 2dF fields a proper motion error of 0.011 arcsec is quoted in each co-ordinate, i.e total $1\sigma$ error of $\sim$0.015 arcsec. A cutoff of 2.5$\sigma$ was adopted for inclusion in the 2BL sample, thus we are sensitive to transverse proper motions of $ >$45 mas\,yr$^{-1}$ for the typical base lines of 11-16 years in the Super Cosmos Sky Survey.  The greater accuracy of the June 2003 SuperCosmos catalogue led to  the rejection of 15 candidate BL Lac objects (see Table 2) included in Paper 1.  Large 1$\sigma$ proper motion errors in the original SuperCosmos catalogue in certain fields had resulted in a calculation of $<2.5\sigma$ proper motion for the 15 objects. One of these 15 objects, J140021.0+001956, was detected at 8.4 GHz (0.336 mJy, see \S4) and appears to have a nonthermal SED from SDSS photometry (Sloan Digital Sky Survey, Stoughton et al.\,2002) as well as our own follow-up NIR observations (\S4.2). Nevertheless this object has been removed from the 2BL sample. In contrast J113039.1-004023 which was originally found to have $>2.5\sigma$ proper motion has been re-instated in the sample.\

The revised 2BL sample of 52 candidate BL Lac objects is listed in Table 1. This updated 2BL sample now includes 41 objects from the original sample plus the one re-instated object, three objects from the 6QZ catalogue and seven objects observed with 2dF after August 2001.  

Cross-correlation of this revised sample with the NVSS and FIRST radio surveys revealed radio detections for only 25 per cent of 2BL objects. We therefore conducted our own radio observations of the sample, as well as further high signal-to-noise spectroscopy and infra-red imaging studies in an attempt to rule out contamination of the sample by weak-lined AGN and cool, featureless DC white dwarf stars. \

\section{High signal-to-noise spectroscopy}
\subsection{Observations}
\begin{figure}
\psfig{file=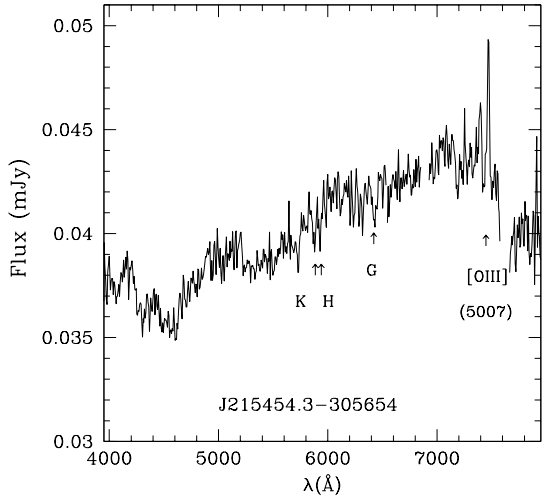,height=3.0in}
\caption{VLT spectrum of the radio-quiet featureless continuum object J215454.3-305654 taken in August 2003. }
\label{2154}
\end{figure}
It was necessary to obtain high signal-to-noise spectra of all 2BL objects in order to search for possible weak features not distinguishable in the original 2QZ spectra. A spectroscopic observing campaign was thus carried out using the FORS1 instrument at the Very Large Telescope (VLT) with time being allocated in service observing mode (program ID 073.B-0082) during 2004. 
 To date high SNR spectra of 35 2BL objects have been obtained at the VLT. J215454.3$-$305654 was first observed as a test case in August 2003 (during observing run 71.A-0174) and 
 confirmed as an extragalactic featureless continuum object at $z=0.494$ from weak absorption features (Ca\,H\&K and G-band) and [OIII] emission (Figure \ref{2154}; see Londish et al.\,2004 for full details). Spectroscopy of the remaining 34 objects was obtained during 2004 (April-August), using the long-slit spectrograph (1\,arsec slit) and 300V grism with central wavelength 5900 \AA.  Exposure times varied from 2\,$\times 10$ to 2\,$\times 15$  mins, depending on the brightness of the source. The observations were made in generally poor seeing conditions (0.7''--2.2'') and six objects were observed during bright time. Flux standards were not observed on all nights. Flux calibration was nevertheless performed using standards taken on another night in order to correct for instrument response. This, together with slit losses, has resulted in the flux calibration of our 2BL objects being only approximate, with spectra showing a drop off in flux below $\sim$4200 \AA; contamination from second-order overlap redward of 7500 \AA\ is also evident in the spectra of blue objects.
 
Data reduction was carried out using standard IRAF\footnote{Image Reduction and Analysis Facility supported by the National Optical Astronomy Observatories (NOAO) in Tucson, Arizona. The NOAO is operated by the Association of Universities for Research in Astronomy (AURA), Inc. under cooperative agreement with the National Science Foundation.} routines. These high signal-to-noise spectra (Figs.\,\ref{bb1}$-$\ref{bb3}) show features not clearly defined in the original 2QZ spectra.    \
\begin{figure*}
\centerline{\psfig{file=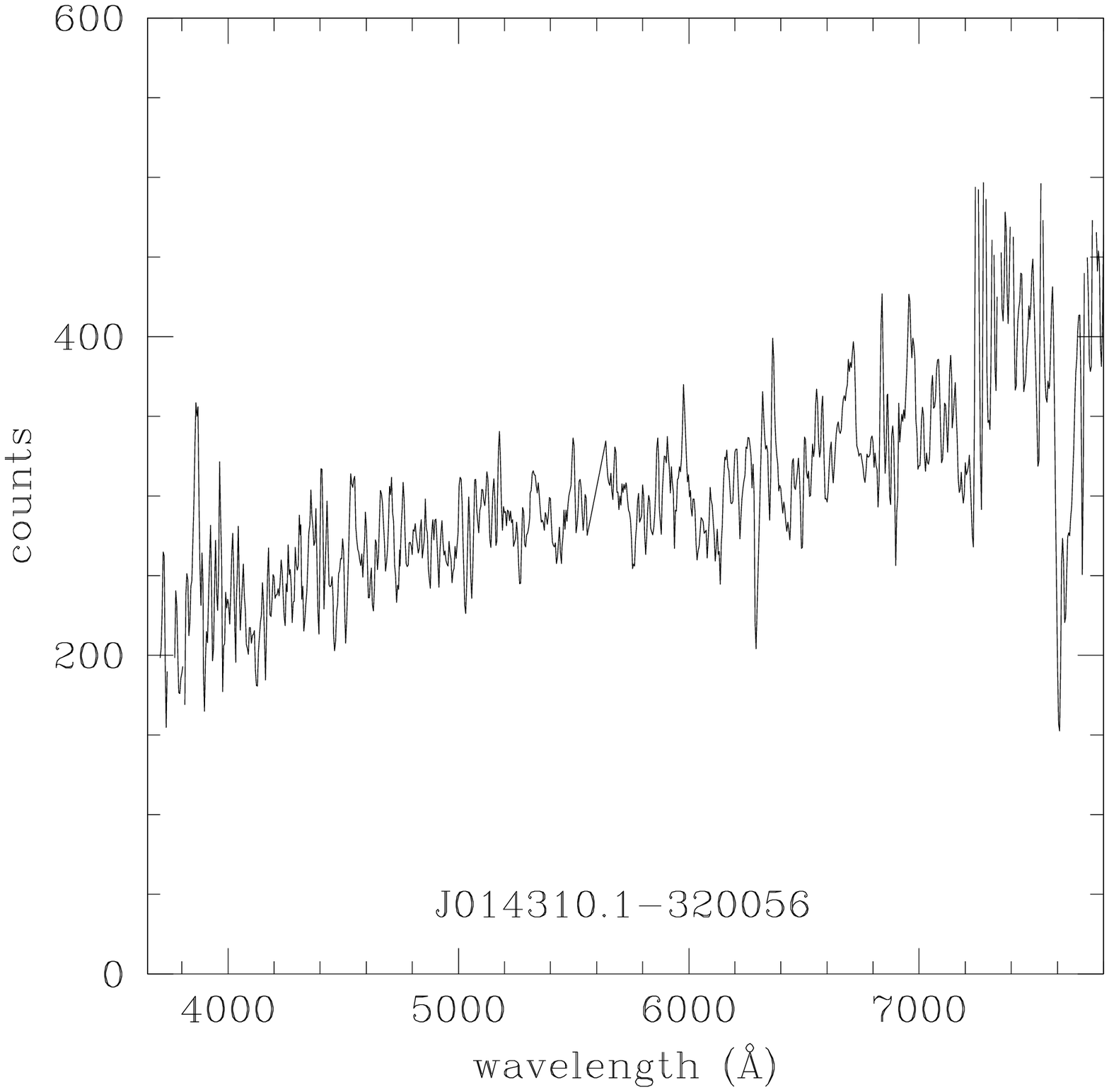,height=3.2in}\psfig{file=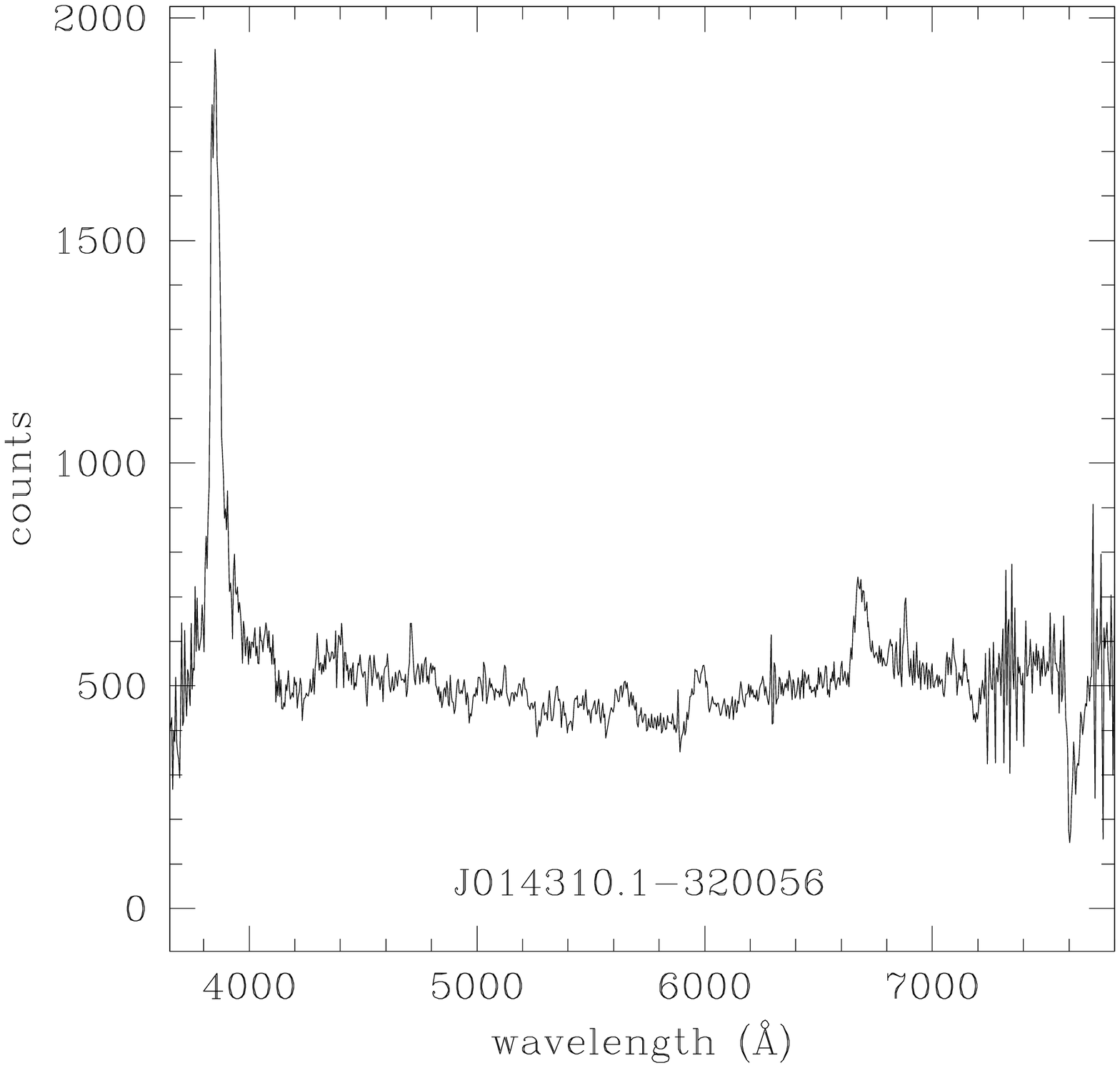,height=3.2in}}
\caption{2QZ spectra of J014310.1$-$320056 taken on 12/01/00 and 17/09/01. Emission features in the spectrum on the right put the object at $z=0.3751$}
\label{0143}
\end{figure*}
\subsection{Curve fitting of VLT spectra}
In order to classify 2BL objects with VLT spectra as either thermal or nonthermal a curve fitting procedure was performed using a least-squares fit to the data. A list of results is shown in Table \ref{bb_pl}.  For each spectrum, two models are fit, each with two free parameters: a blackbody function
(temperature and normalisation) and a power law function (slope and
normalisation). Fitting was performed over the wavelength range $\sim$4000-8000\AA\,such that instrumental effects do not contribute to the fit. In addition those parts of the spectrum strongly
affected by features such as atmospheric lines or emission and
absorption features are blanked out interactively.  In this way the models are fit to the estimated continuum via a least-squares (or $\chi^2$-minimisation)
technique. A fixed error on each data point of
$\sigma_i=2\mu$Jy was assumed.  The model with the lower value of $\chi^2 = \sum_i (f_i - m_i)^2/\sigma_i^2$ (where $f_i$ are the measured fluxes, and
$m_i$ are the model fluxes) is chosen as the best fitting model. Note
that the value of $\chi^2$ is used only to ascertain a relative goodness of fit and to discriminate between the
two models, so the exact value of the error $\sigma_i$ is not
crucial. However, if the reduced-$\chi^2$ value (defined as
$\chi^2_\nu \equiv \chi^2/\nu$ where $\nu$ is the number of degrees of
freedom, in this case two less than the number of data points) is
greater than 5, the fits are rejected. In four cases where emission and absorption features dominate the spectrum neither a power law nor a blackbody function could be fit.  In all other cases the fit clearly indicated either a blackbody curve or positive powerlaw slope (i.e increasing flux for increasing  wavelength).The actual value of T or $\alpha$ returned by the fitting routine will differ depending  on the choice of regions excised from the fitting procedure.  For thermal sources the estimated 1$\sigma$ error on the best fit blackbody temperature is 100-250 K. \
\subsection{Results}
Spectra of 14 objects appear to be well fit by a thermal blackbody curve, 
although one of these objects -- J224559.1-312223 -- has a weak emission feature that could be MgII at $z = 1.324$. Of the remaining 21 objects seven appear to be featureless continuum objects (only one of which, J215454.3$-$305654, is radio-quiet), and 
13 show weak, often very broad emission features as well as absorption systems in their spectra. One of these 13 objects, J014310.1$-$320056,  is classified in the literature as a BL Lac, and does indeed appear to have a featureless continuum in the 2QZ spectrum initially used to select the object as a candidate BL Lac (Fig \ref{0143},\,left). However in a second 2QZ spectrum taken 9 months later (Fig \ref{0143},\,right), and in our VLT spectrum, the object is clearly a broad line AGN ($z=0.375$). The spectrum of the remaining object (J113413.4+001041) is too noisy to distinguish the presence or absence of lines; this object is classed as a QSO at $z= 1.4856$ from its SDSS spectrum (Fig A1).\

In addition to J113413.4+001041  another eight objects now also have spectra available in the Sloan Digital Sky Survey Data Release 4 (SDSS, Stoughton et al.\,2002). Five of these flux calibrated SDSS spectra (Fig \ref{sdss4}) confirm the identity of two objects as candidate BL Lacs, two as QSOs and one as a WD (white dwarf) star; SDSS spectra of a further three 2BL objects without VLT spectra identify the objects as QSOs.  Classification of all 2BL objects is listed in Table 1.\
\begin{figure*}
\centerline{\psfig{file=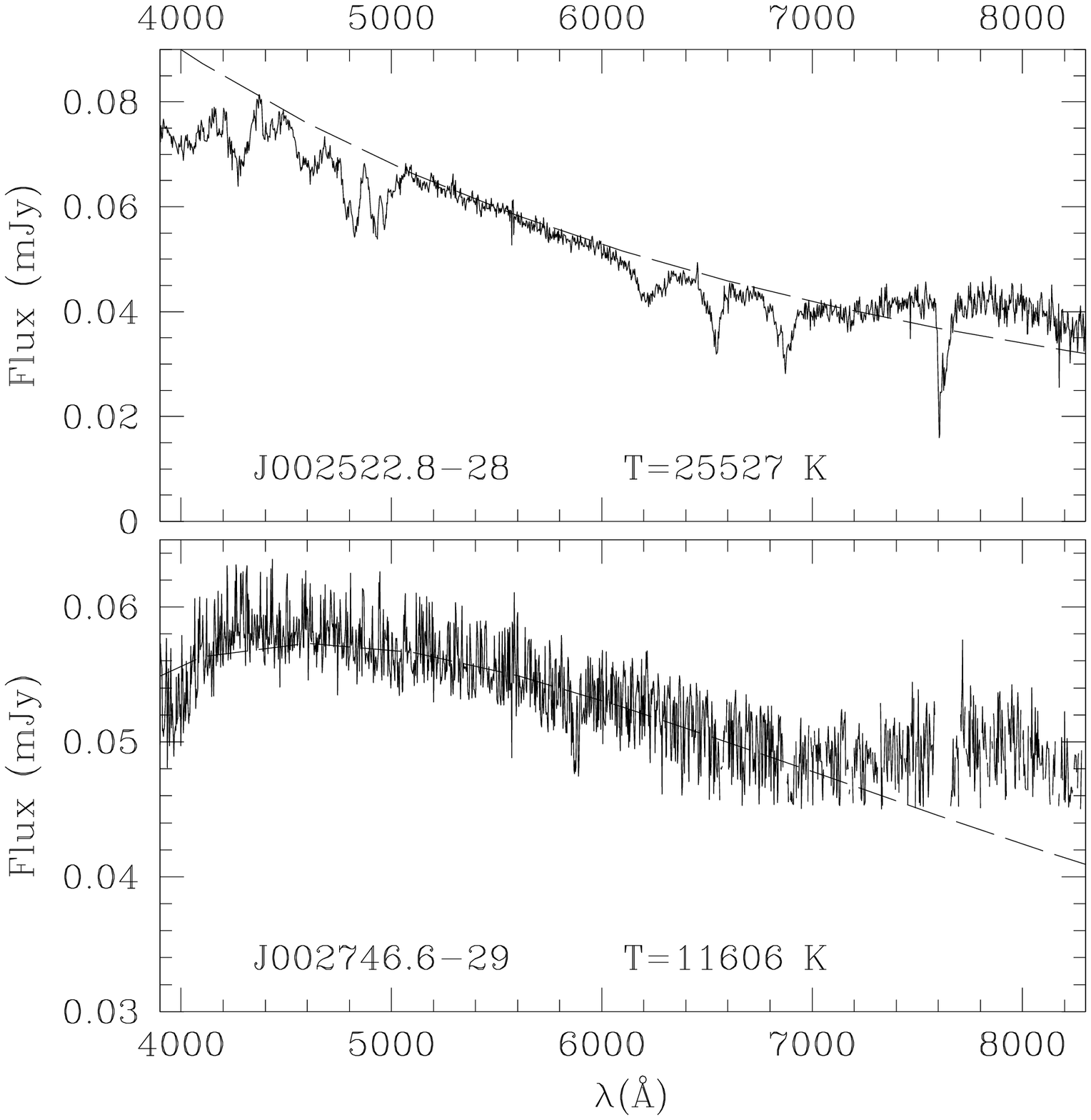,height=2.9in}\psfig{file=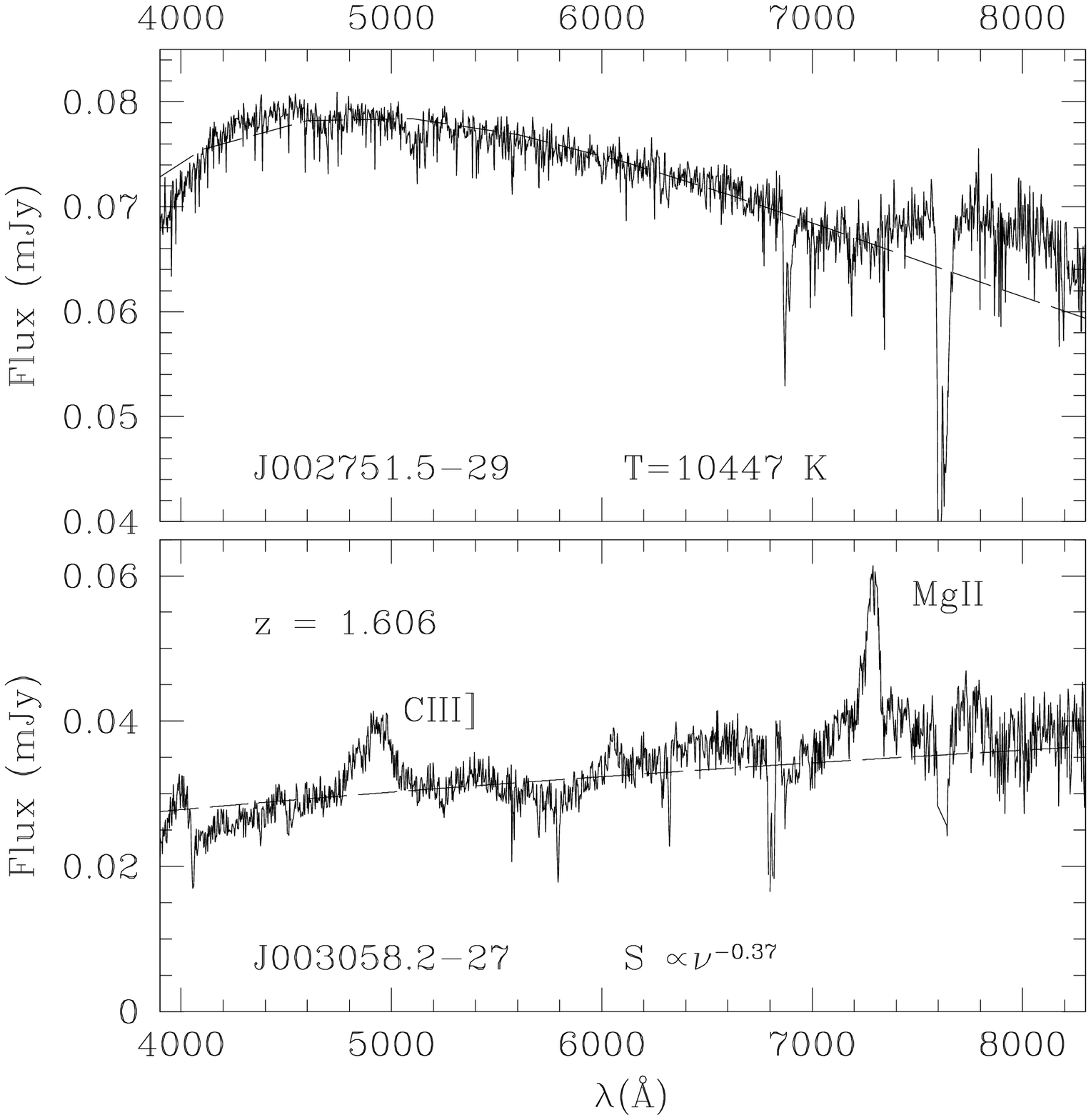,height=2.9in}} 
\centerline{\psfig{file=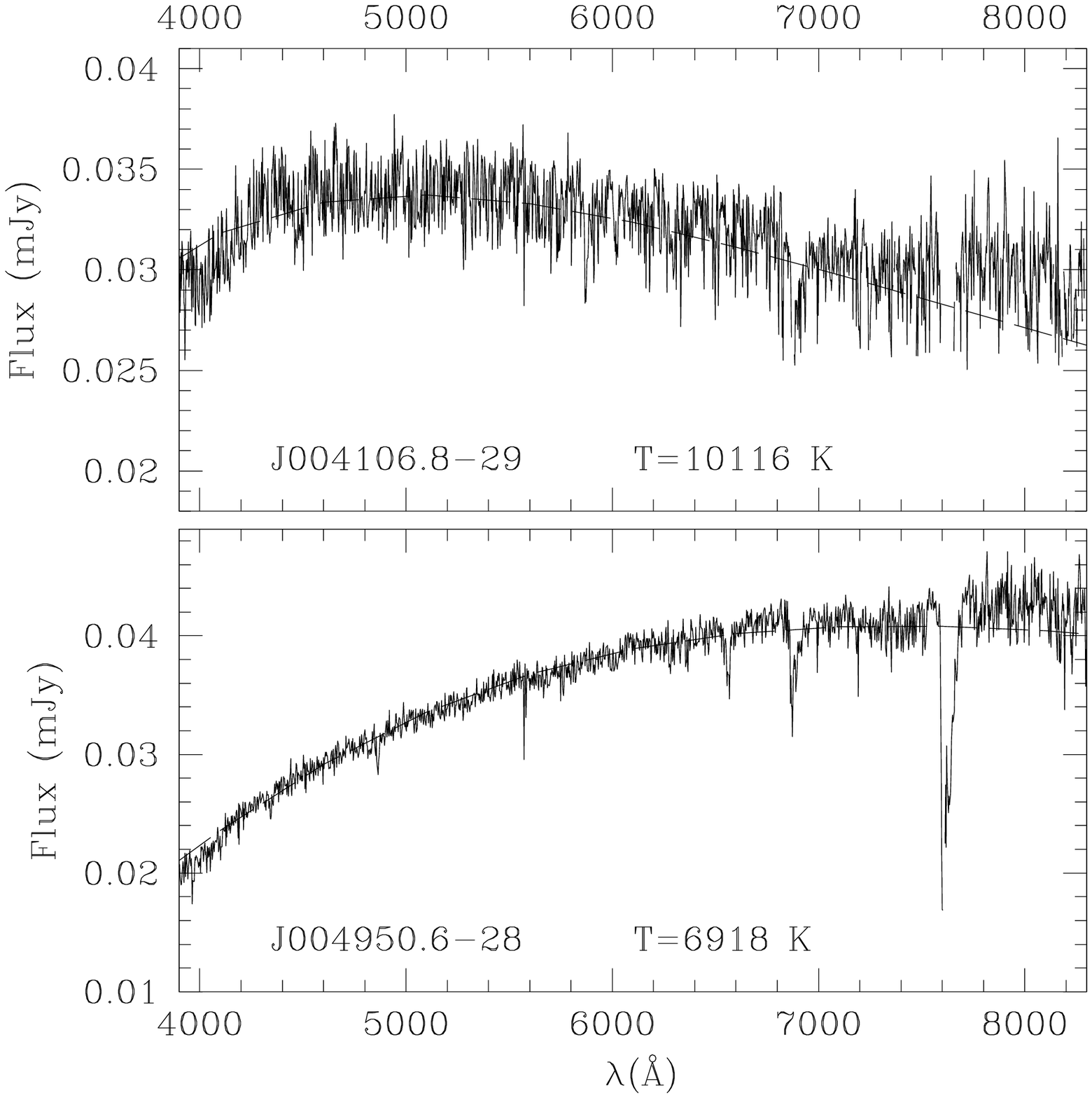,height=2.9in}\psfig{file=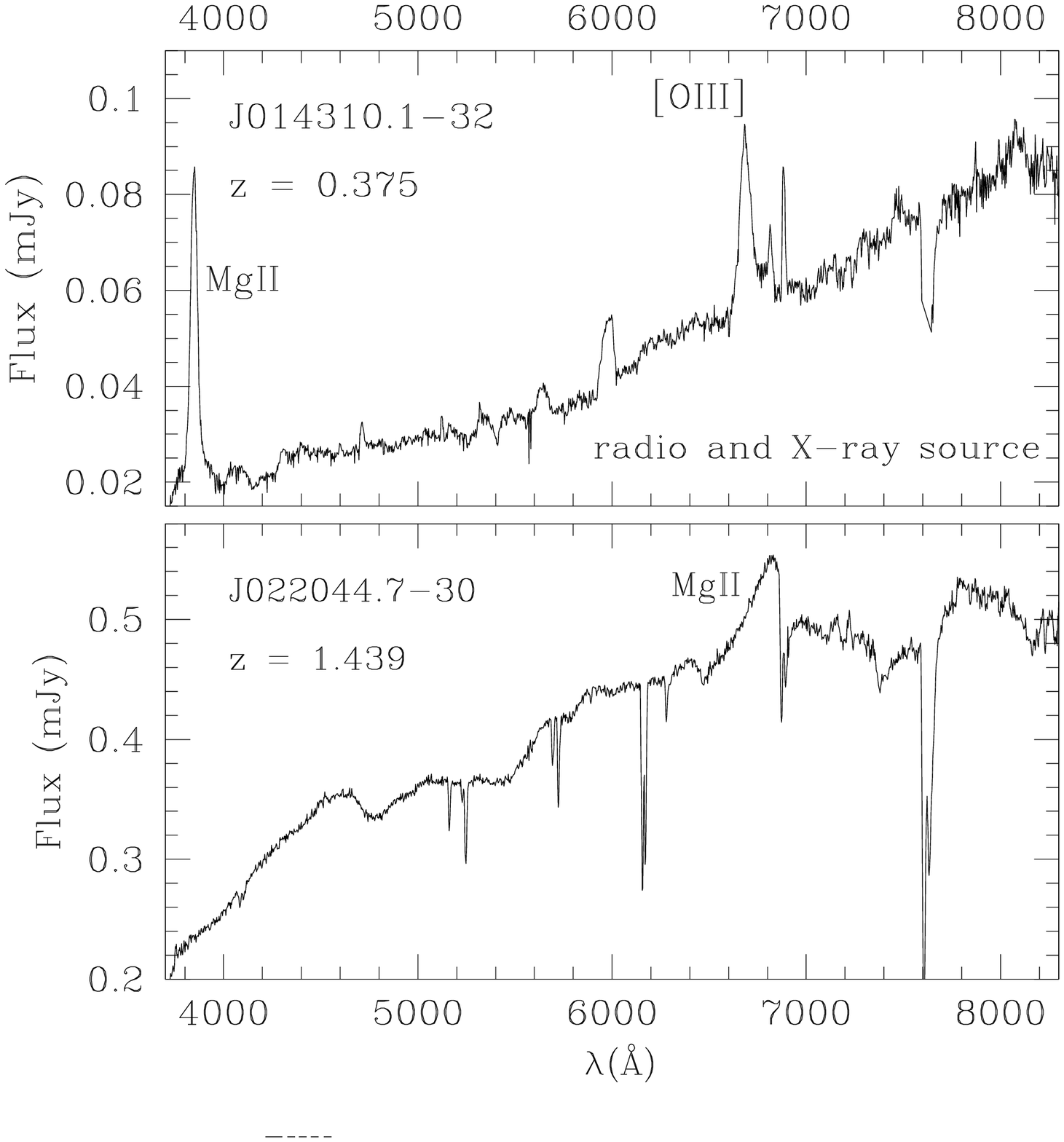,height=2.9in}}
\centerline{\psfig{file=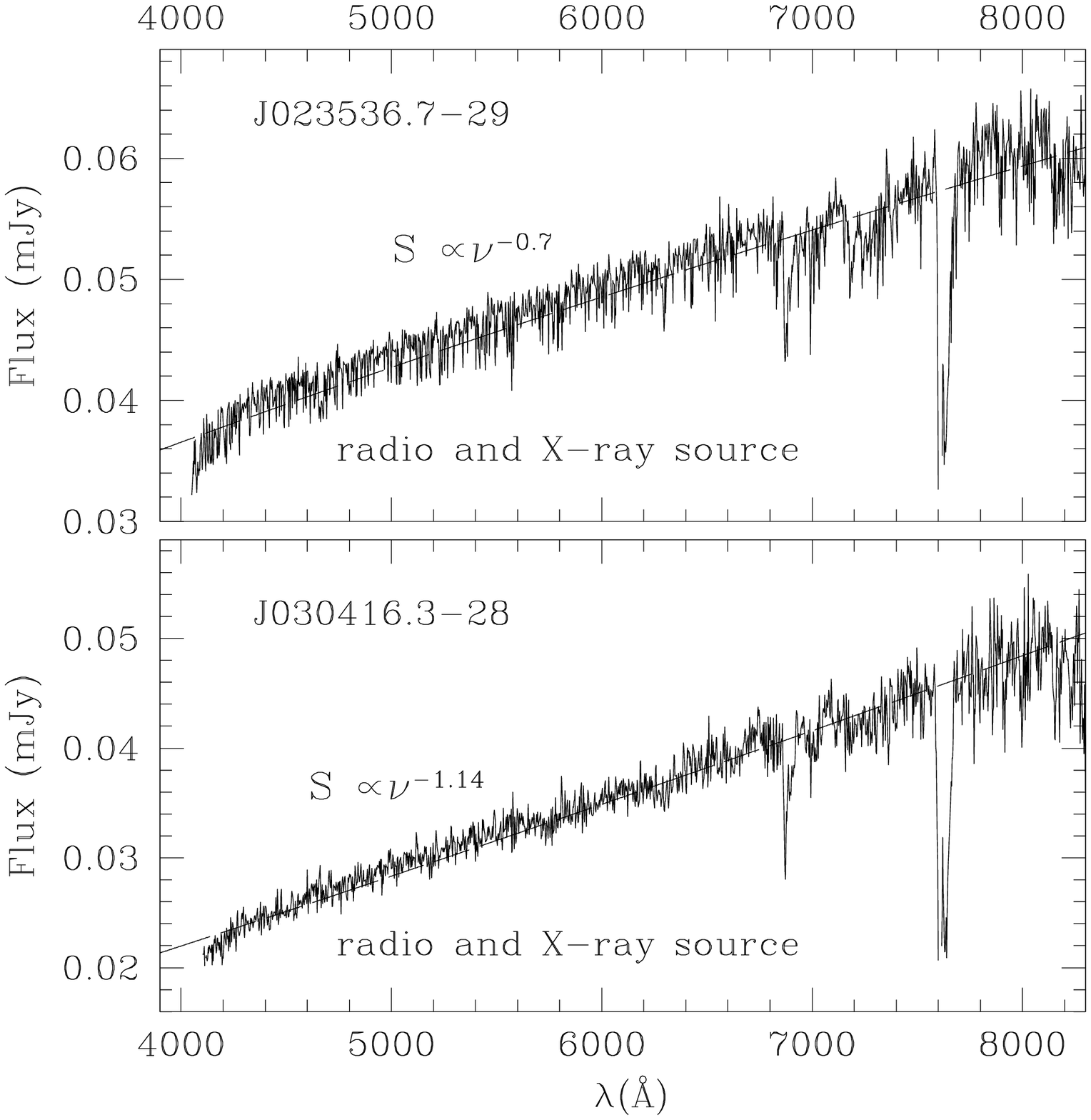,height=2.9in}\psfig{file=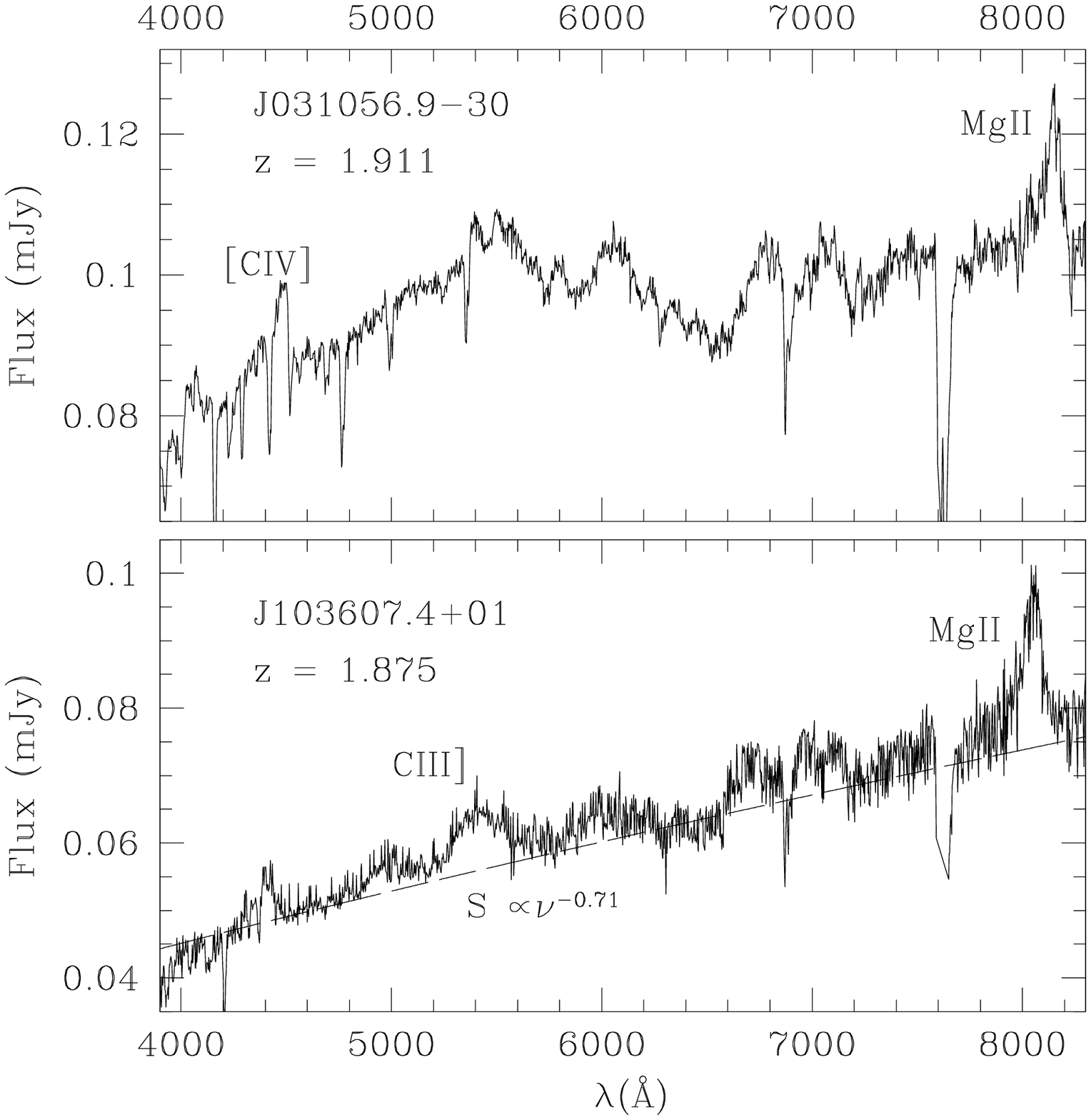,height=2.9in}}
\caption{VLT spectra of 2BL objects, fitted with either a blackbody curve or power-law slope. In three cases  no slope could be fitted.}
\label{bb1}
\end{figure*}
\begin{figure*}
\centerline{\psfig{file=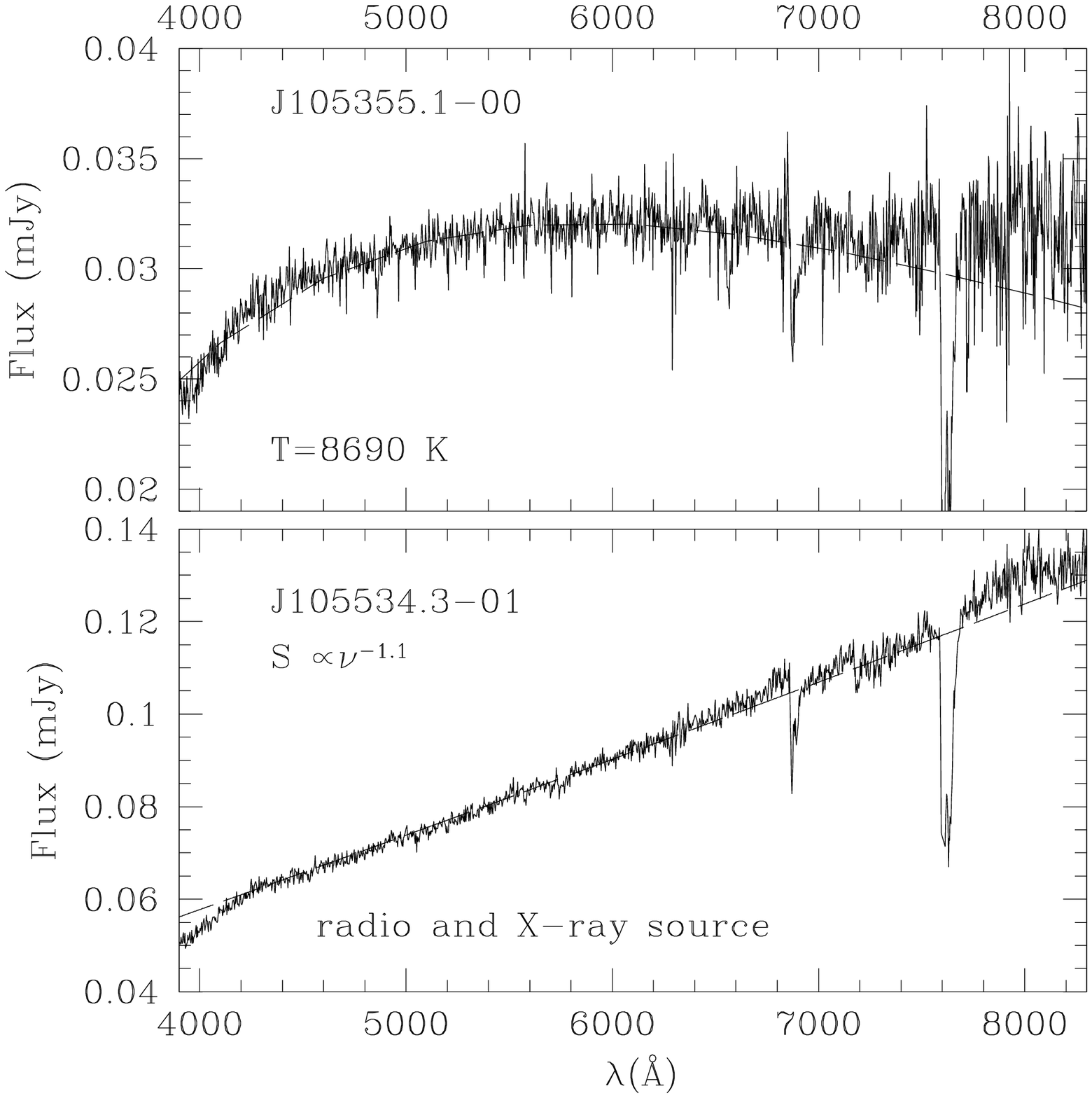,height=2.9in}\psfig{file=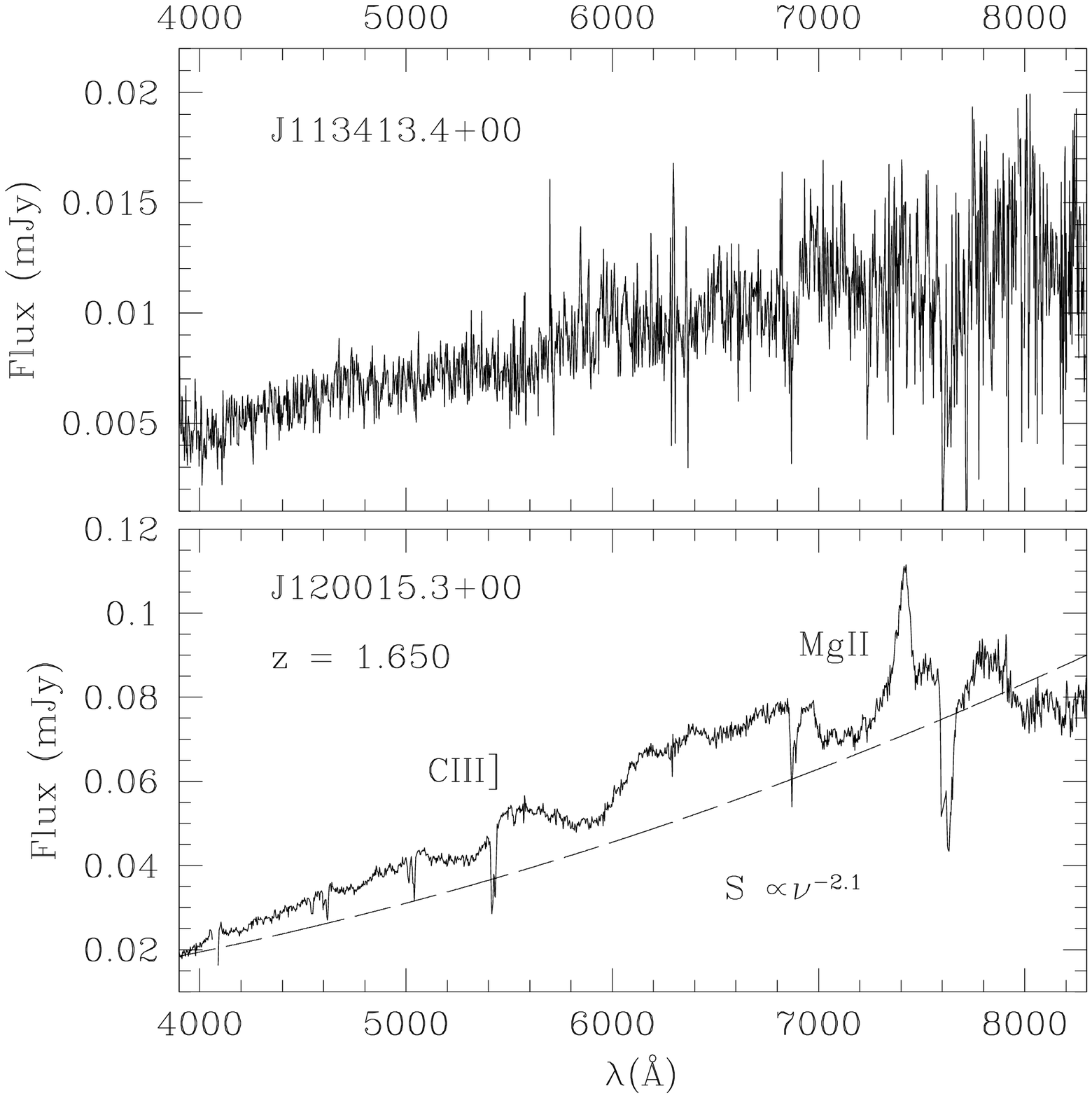,height=2.9in}} 
\centerline{\psfig{file=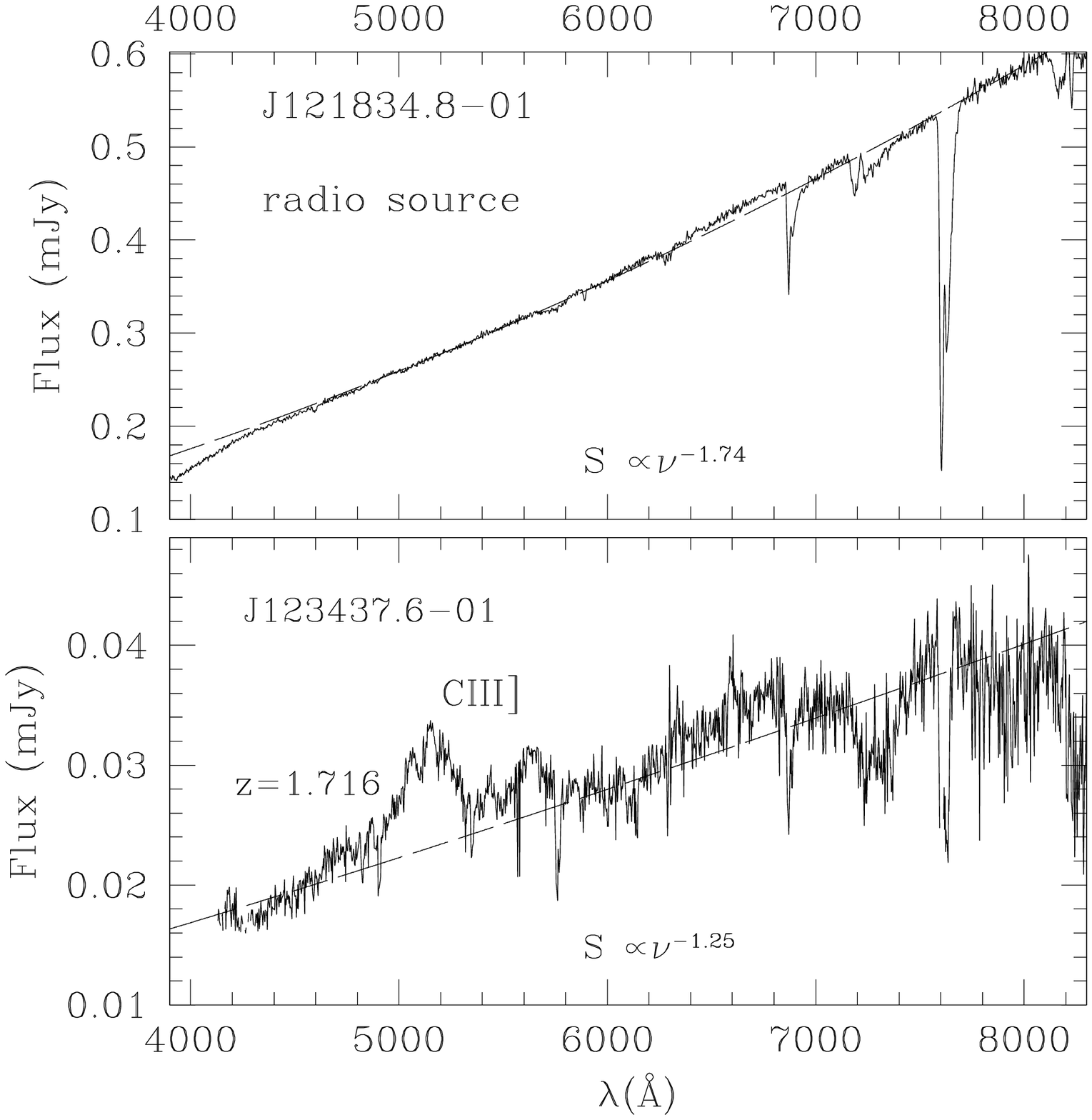,height=2.9in}\psfig{file=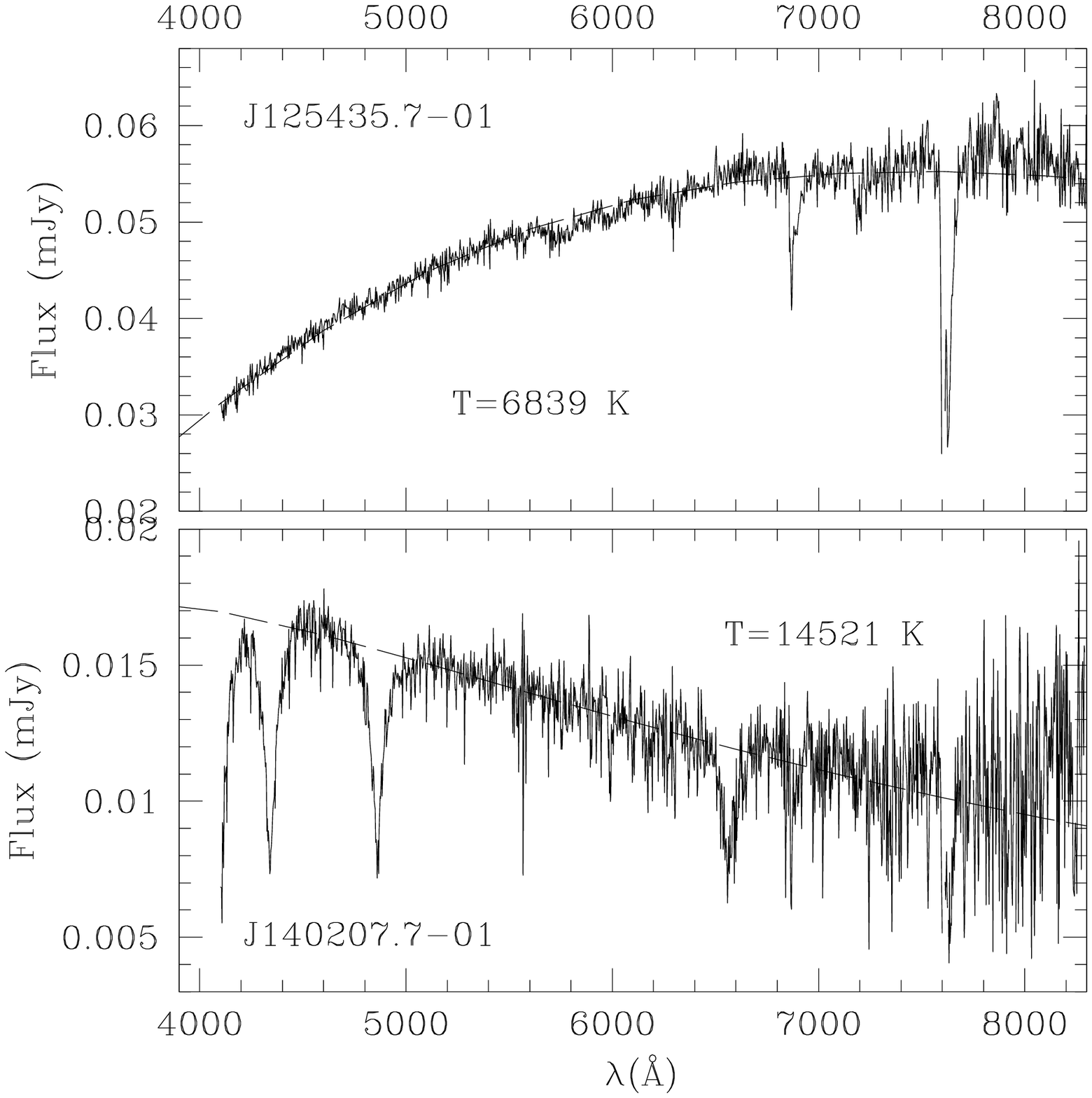,height=2.9in}}
\centerline{\psfig{file=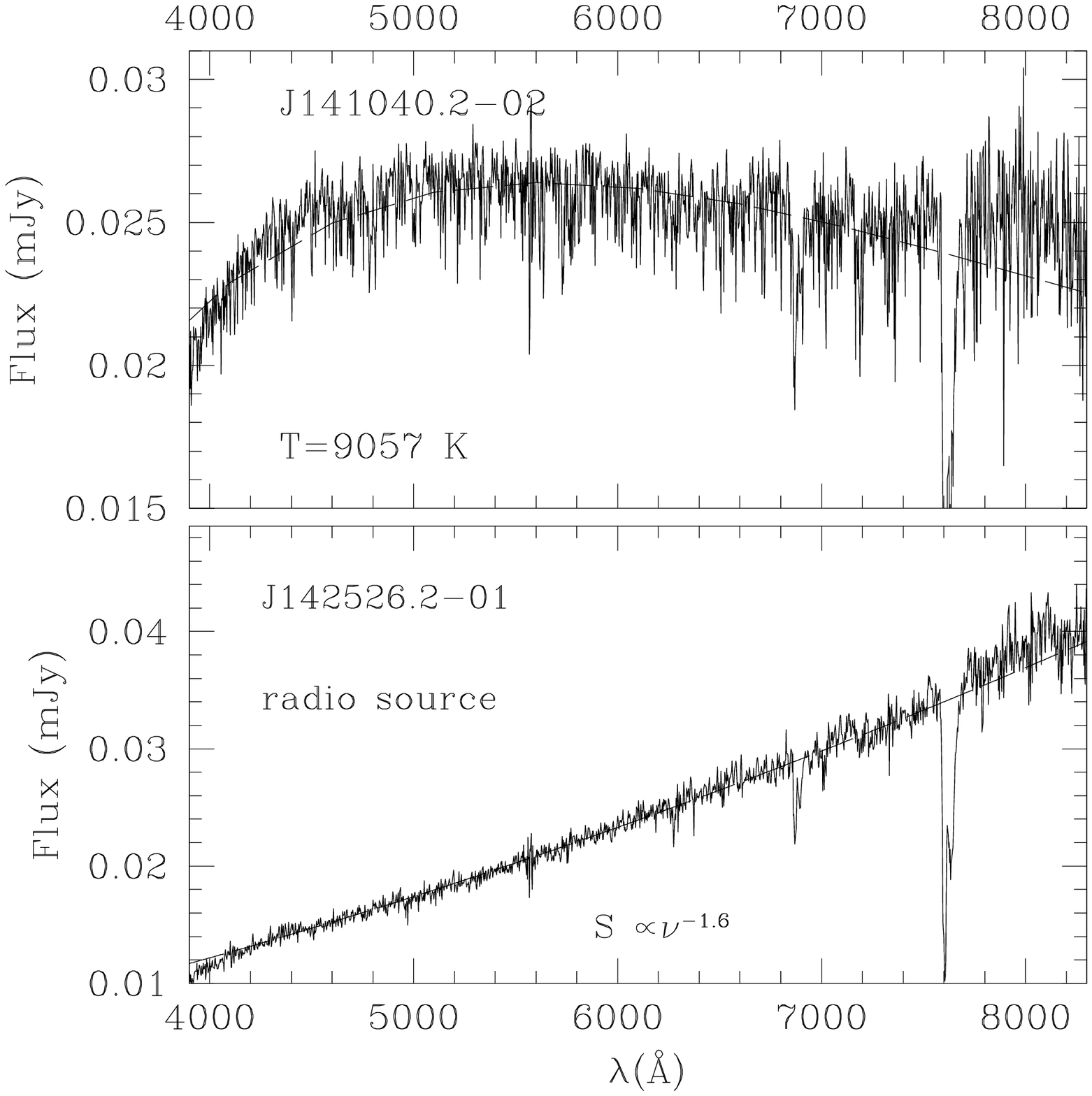,height=2.9in}\psfig{file=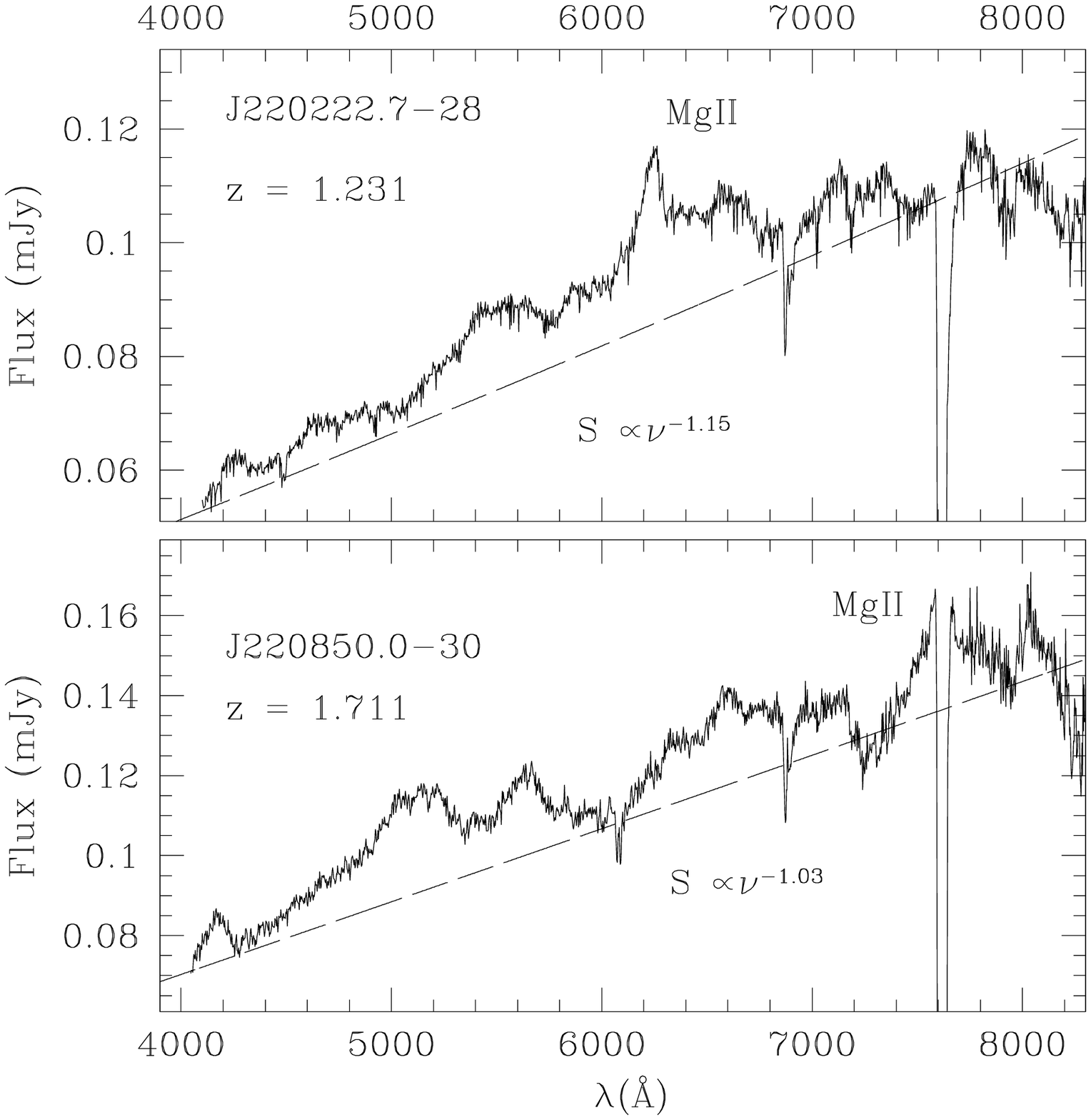,height=2.9in}}
\caption{VLT spectra of 2BL objects, fitted with either a blackbody curve or power-law slope.}
\label{bb2}
\end{figure*}
\begin{figure*}
\centerline{\psfig{file=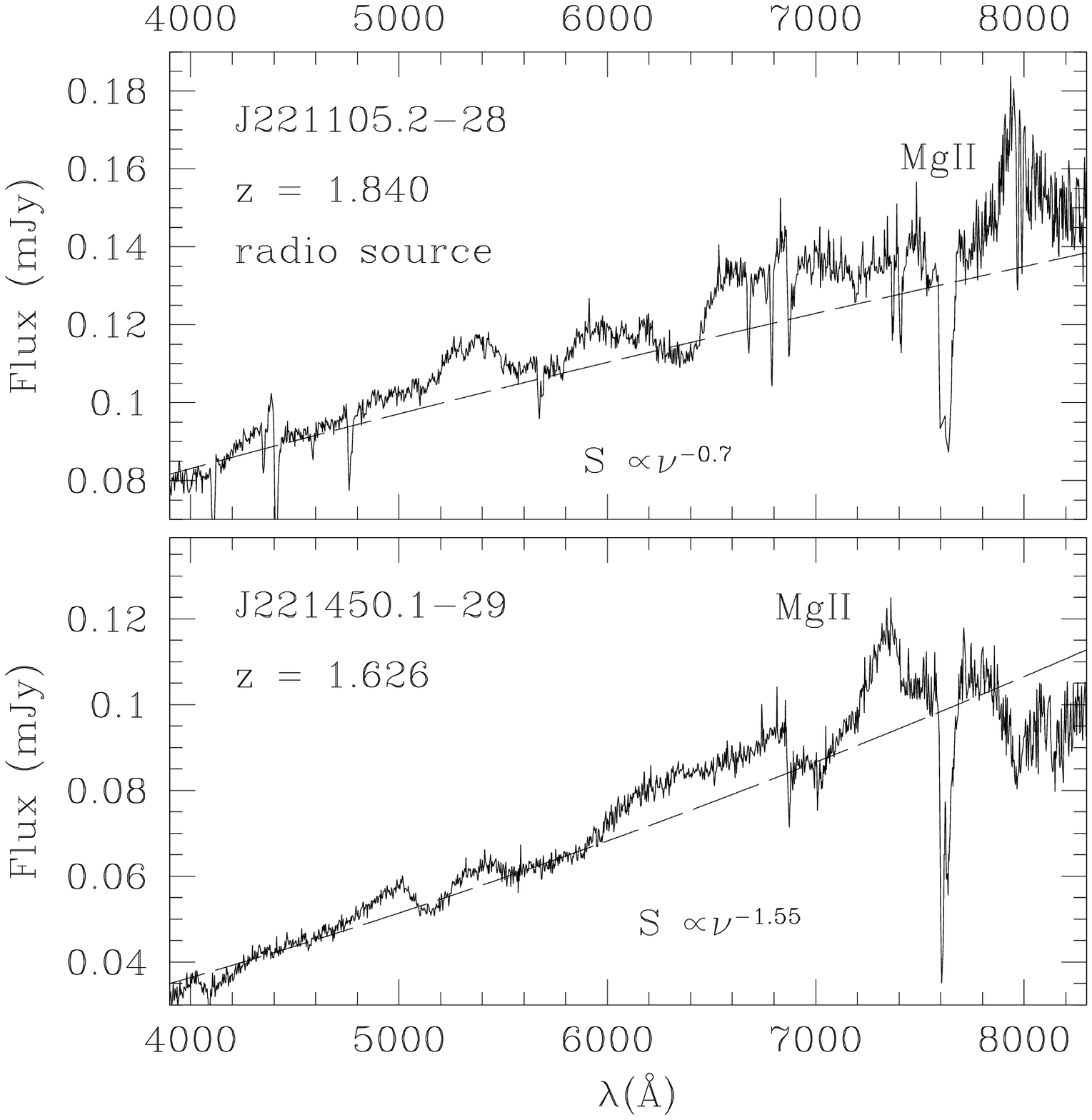,height=2.9in}\psfig{file=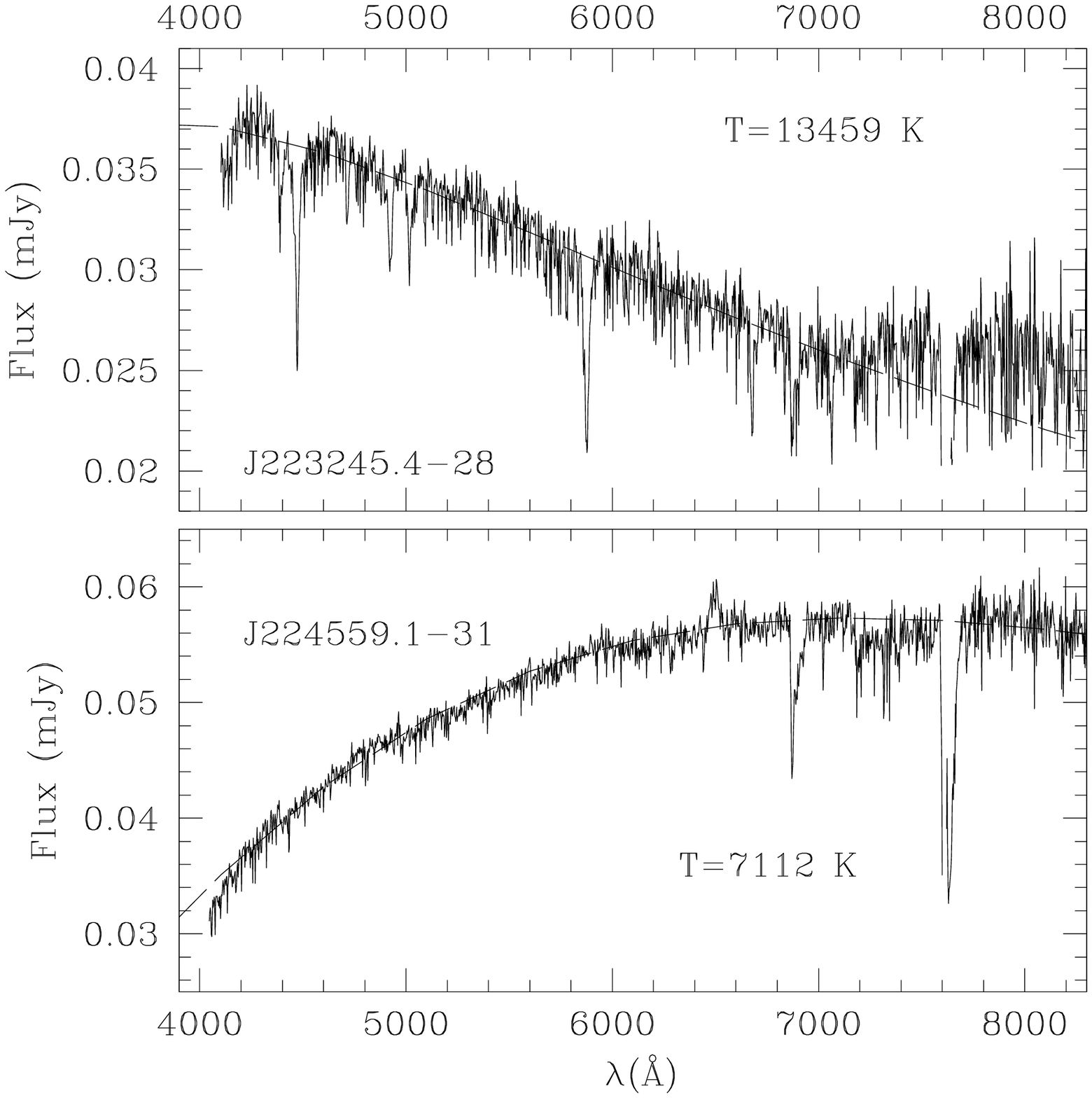,height=2.9in}} 
\centerline{\psfig{file=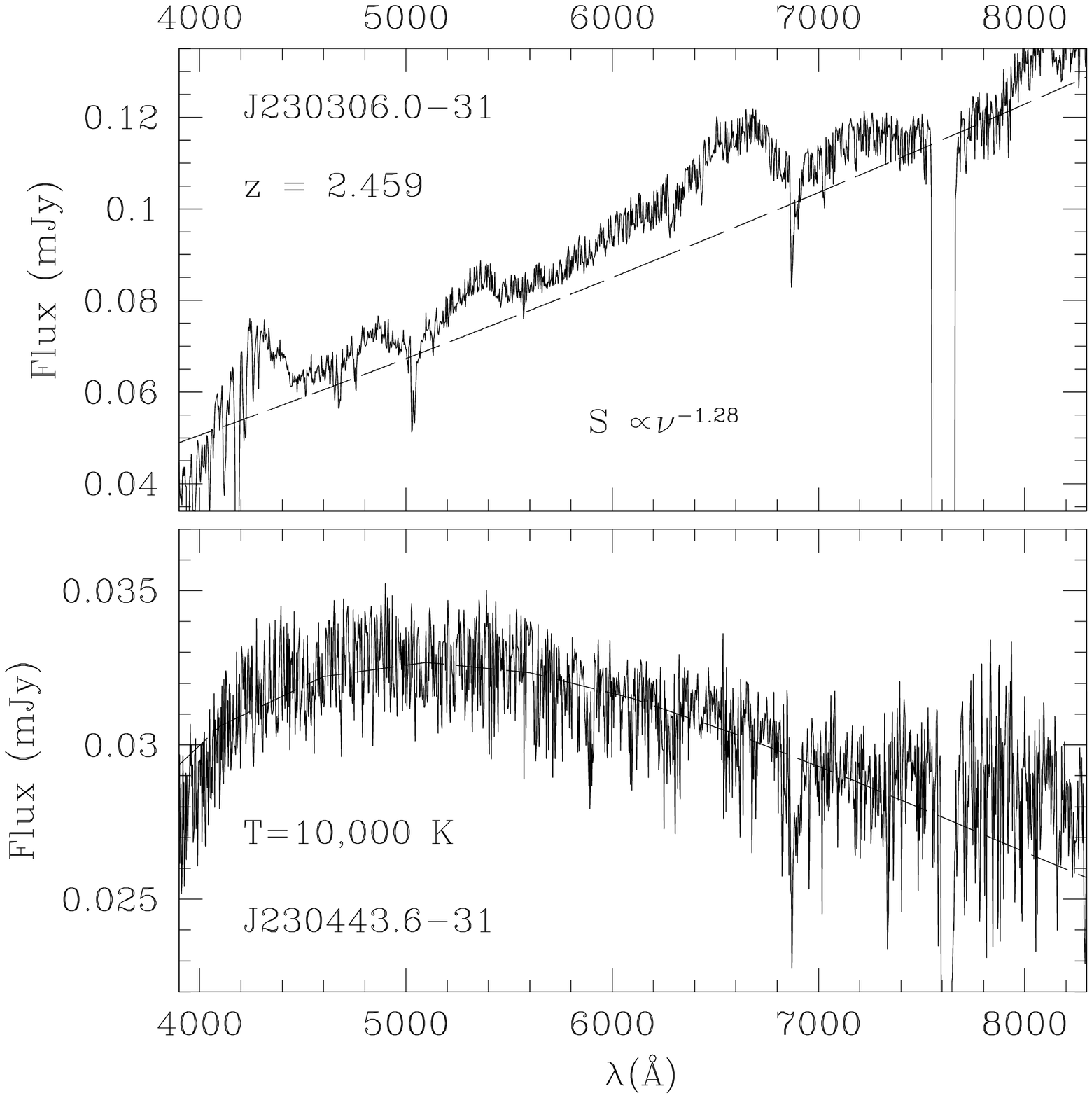,height=2.9in}\psfig{file=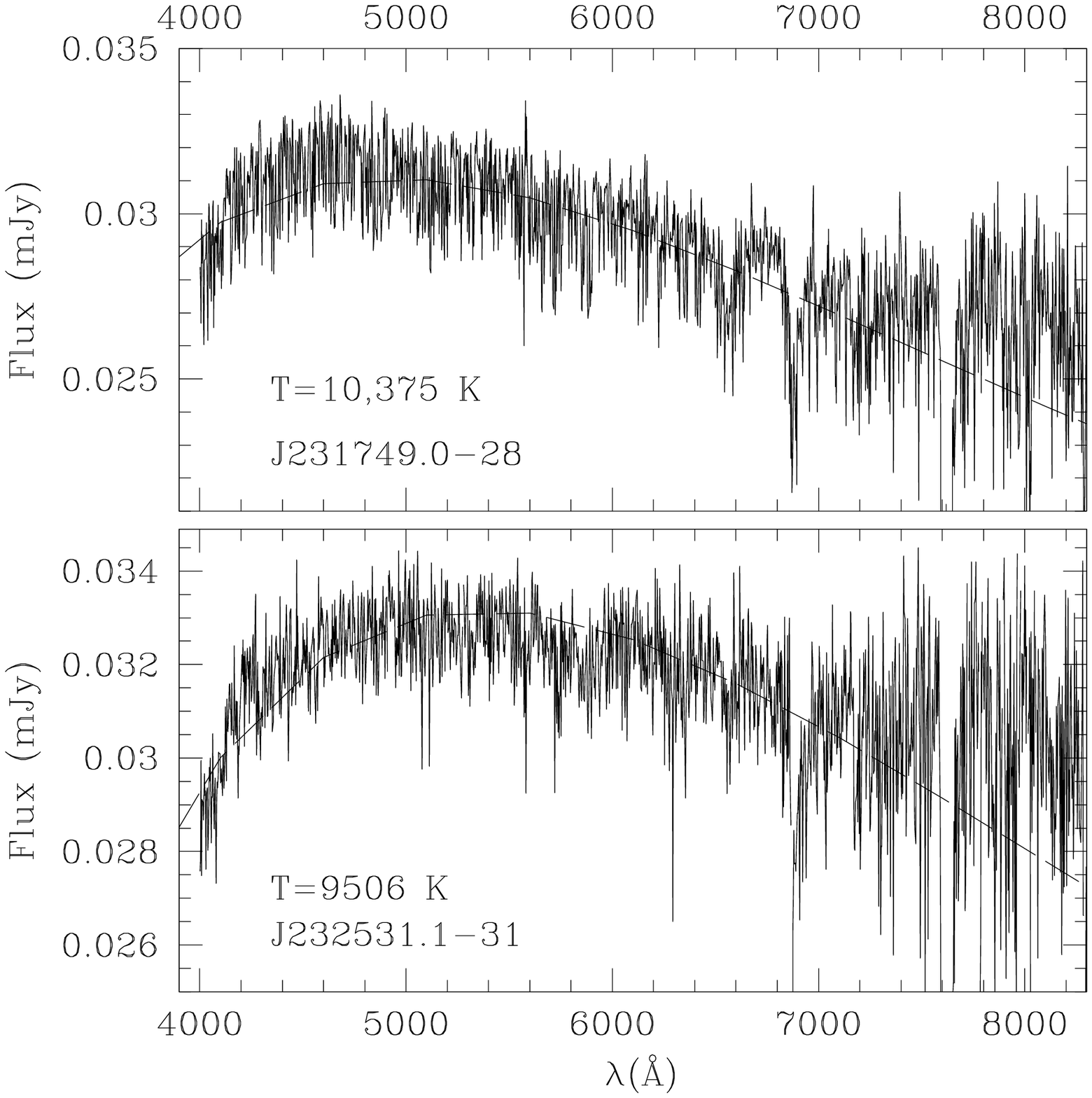,height=2.9in}}
\centerline{\psfig{file=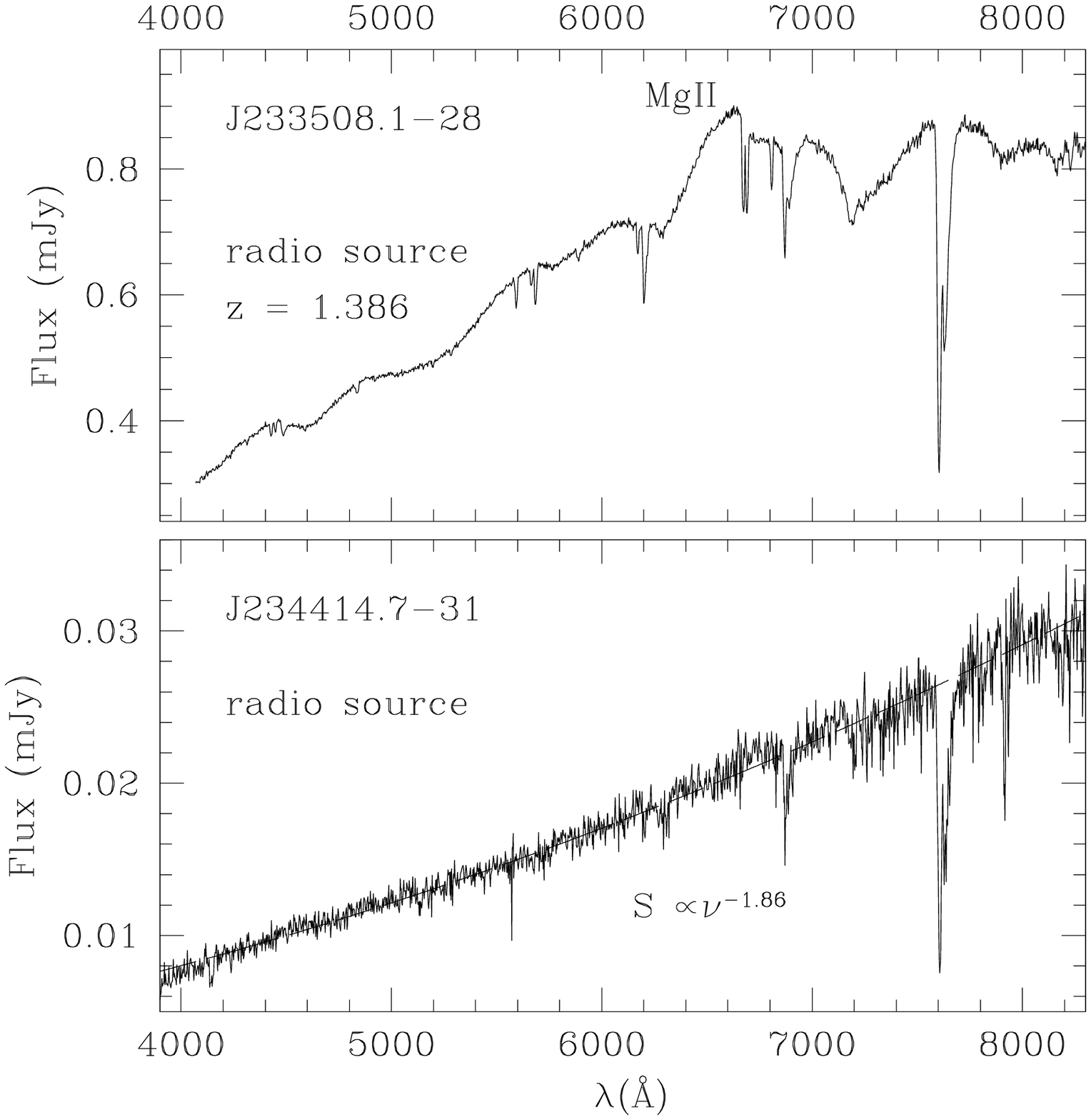,height=2.9in}}
\caption{VLT spectra of 2BL objects, fitted with either a blackbody curve or power-law slope.In one case where absorption and emission features dominate the continuum no slope has been fitted }
\label{bb3}
\end{figure*}

VLT spectroscopy indicates $\sim$40 per cent contamination of the sample by stellar objects with a similar percentage of contamination by AGN with weak or broad features. In the case of the latter it is not possible to determine from the original 2QZ/6QZ spectra (which are not flux calibrated) whether objects were  misclassified or whether continuum levels have subsequently changed.  A comparison of original 2QZ/6QZ spectra (corrected for 2dF/6dF instrument response) with the corresponding VLT spectrum is shown in Appendix C. 

\begin{figure}
\psfig{file=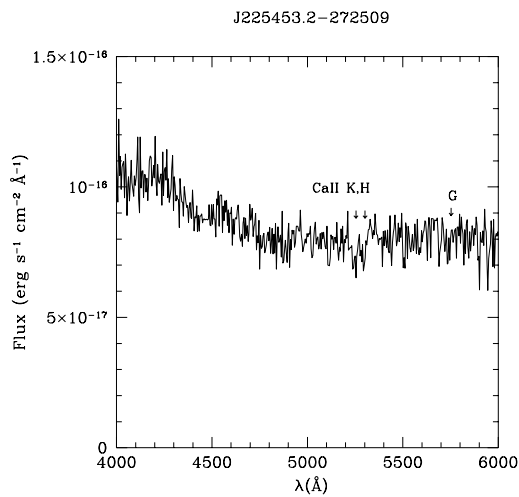,height=2.8in}
\caption{J225453.2$-$272509, a 52mJy continuum object at $z=$0.333}
\label{mssso}
\end{figure}

Note that one radio-loud 2BL source without a VLT spectrum (J225453.2$-$272509) was observed in 2001 at the Siding Spring Observatory, Australia (Fig \ref{mssso} -- see Paper 1 for details); it is classified as a continuum source at a redshift of $z=$0.333 identified from weak Ca\,II H\,\&\,K absorption features and G-band in the spectrum. Paper 1 also includes high signal to noise Keck spectra of the two radio-loud objects J105534.3$-$012617 and J121834.8$-$011955 (observed by Brotherton et al.\,1998), showing these objects to be featureless continuum sources.
\begin{table*}
\caption{Curve fitting results for 2BL objects with VLT spectra and SDSS optical photometry }
\label{bb_pl}
\begin{tabular}{l r l  r r} 
\multicolumn{1}{c} { object} &
\multicolumn{1}{c} { $z$} &
\multicolumn{1}{c} {notes} &
\multicolumn{1}{c}  {temp/$\alpha$ } &
\multicolumn{1}{c}  {temp/$\alpha$} \\
\multicolumn{3}{c} { } &
\multicolumn{1}{c}  {from VLT spectra}&
\multicolumn{1}{c}  {from SDSS photometry}\\
\hline
J002522.8$-$284034 & 0.00 & Odd WD with split lines & 25,500 K &\\
J002746.6$-$293308 & 0.00 & WD with He and H$\alpha$ &  11,600 K & \\  
J002751.5$-$293506 & 0.00 & no obvious features& 10,400 K& \\
J003058.2$-$275629 & 1.606 & Broad MgII, CIII] & 0.37   & \\
J004106.8$-$291114 &0.00 & DB with HeI and H$\alpha$ & 10,100 K  & \\  
J004950.6$-$284907 &0.00 & Weak Balmer lines  &  6,900 K & \\ 
J014310.1$-$320056 &0.375 & Strong Balmer lines, MgII ($S_{1.4}=$76mJy) & 1.1/1.85 & \\
J022044.7$-$303713 &1.439 & Weak, broad MgII, CIII], abs at z=1.202 &-- & \\
J023536.7$-$293843 &-- & BL Lac obj  & 0.70   & \\ 
J030416.3$-$283217 &-- & BL Lac obj & 1.14 & \\	
J031056.9$-$305901 &1.911 & Broad MgII, weak CIV, CIII]  & --  & \\ 
J103607.4+015658   &  1.875& Broad MgII, CIV, CIII] & 0.71 &0.43\\
J105355.1$-$005538 &0.00 & Weak Balmer lines &  8,700 K& 9,500 K\\
J105534.3$-$012617 & -- & BL Lac obj & 1.1 & \\
J110644.5+000717 & -- & -- & -- & 27,000 K\\
J113413.4+001041   &  & Noisy spectrum & -- & \\
J113900.5$-$020140 & -- & -- & -- & 1.57\\
J114137.1$-$002730 & -- & -- & -- & 0.67\\
J114221.4$-$014812 & -- & -- & -- & 10,800 K\\
J114554.8+001023 & -- & -- & -- & 6,800 K\\
J120015.3+000552   & 1.650 &  Broad MgII, weak CIV, CIII] & 2.1& \\
J121834.8$-$011955 & -- & strong continuum object ($S_{1.4}=$244mJy)& 1.74 & 1.44\\
J123437.6$-$012953 &1.716 & Broad CIV, CIII], MgII & 1.25& \\
J125435.7$-$011822 &0.00 & no obvious features &  6,800 K    & 7,200 K\\
J125501.2+015513& -- & -- & -- & 10,400 K\\
J140207.7$-$013033 &0.00 &DA  &  14,500 K & 13,400 K \\ 
J141040.2$-$023020 &0.00 & no obvious features & 9,100 K  &10,500 K \\ 
J142526.2$-$011826 &-- & continuum object ($S_{1.4}=$10 mJy) & 1.6& 1.12 \\
J215454.3-305654 & 0.494 & weak [OIII], Ca H \& K, G-band &0.36 & \\
 J220222.7$-$285306 & 1.231 &  Weak broad MgII and CIII] plus FeII & 1.15& \\ 
J220850.0$-$302817 &1.711 &  Weak broad MgII, CIII] and CIV plus FeII& 1.03& \\
J221105.2$-$284933 & 1.840 &  Weak broad MgII, CIII] and CIV ($S_{1.4}=$81mJy)&  0.7& \\
J221450.1$-$293225 &1.626 &  Weak broad MgII, CIII] and CIV plus FeII & 1.55& \\J223233.5$-$272859 &0.00 &DB  & 13,500 K  & \\ 
J224559.1$-$312223 &?? & star or MgII at z=1.324  &  7,100 K & \\
J230306.0$-$312737 & 2.459 &  Weak broad CIII], CIV and Ly-alpha & 1.28 & \\
J230443.6$-$311107 & 0.00 &WD, DO, Weak HeII at 4686 \AA & 10,000 K& \\ 
J231749.0$-$285350 &0.00 &WD, H$\alpha$ + HeI lines  & 10,400 K& \\
J232531.3$-$313136 &0.00 & WD, weak H$\alpha$, H$\beta$ &  9,500 K & \\  
J233508.1$-$283035 & 1.386 &  Very weak broad MgII and CIII] ($S_{1.4}=$6mJy)& --& \\
J234414.7$-$312304 &-- & continuum object ($S_{1.4}=$5.4mJy) & 1.86     & \\ 
\hline		     
\end{tabular}
\begin{flushleft}  {WD classifications: DA denotes a WD with a hydrogen-rich atmosphere or outer layer, indicated by strong Balmer hydrogen spectral lines; DB, a neutral helium-rich atmosphere indicated by neutral helium spectral lines, (He I lines); DO, an ionized helium-rich atmosphere indicated by ionized helium spectral lines, (He II lines); DC, no strong spectral lines.\\
Objects have been fitted with either a blackbody curve or a power-law slope characterised by $S \propto \nu^{-\alpha}$. The best fit value of T or $\alpha$ is shown. See \S3.2 and \S4.1 for details.}
\end{flushleft}
\end{table*}

\section{Broadband optical and near infrared studies}
\subsection{SDSS Magnitudes}
Prior to obtaining the above VLT spectra we also checked the Sloan Digital Sky Survey for photometry of candidate BL Lac objects.  SDSS Data Release 4 now includes photometric magnitudes\footnote{SDSS uses photometric magnitudes  designated $u'$, $g'$, $r'$, $i'$ and $z'$, with central 
wavelengths of 3500 \AA, 4800 \AA, 6250 \AA, 7700 \AA \, and 9100 \AA \, 
respectively, and AB magnitude system 
with a common zero point for all filters (Fukugita et al.\,1996)} for 19 2BL objects in the equatorial strip of the 2QZ (the only exception being 2QZJ105534.3-012617).\

 Curve fitting was also carried out on the 19 2BL objects with SDSS optical magnitudes; good fits to a blackbody function were found for five objects, while a further three were better fit by a blackbody than a powerlaw slope. Of the remaining 11 objects three were found to be well fit by a powerlaw slope (J113900.5$-$020140, J121834.8$-$011955 and J142526.2$-$011826), one was better fit by a powerlaw than a blackbody function (J103607.4+015658) and seven produced a poor fit in both cases. Fits to optical and NIR photometric magnitudes are listed in table 3 and shown in Appendix B; the unusual SED of J113039.1-004023 is a result of broad absorption and emission features, evident in the SDSS spectrum of this object. 

\subsection{Infrared imaging}
\begin{figure}
\psfig{file=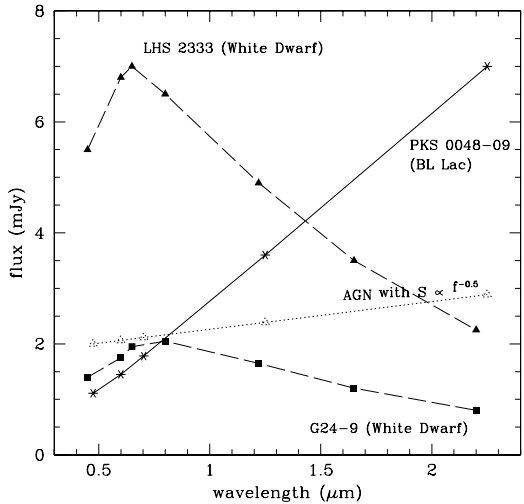,height=2.9in}
\caption{The energy distributions of 2 DC WDs (T$_{eff} \approx $ 7000K) from Bergeron et al.\,(2001) (solid symbols) compared to 
that of a luminous (V=16.4) BL Lac (asterisks) (data from Tanzi et al.\,1988) 
and an $f \propto \nu^{-0.5}$ energy distribution typical of an AGN/QSO. For 
the 2 WDs the $J-K$ colours are $+$0.19 (G24--9) and $+$0.12 (LHS 2333). 
Note the 3--4 leftmost
points are insufficient to differentiate between the  thermal,
blackbody curve of WDs and the  nonthermal continuum of an AGN; thus in some cases near-IR data 
points are essential to determine the fit.}
\label{berg}
\end{figure}

Given that the SEDs of both thermal and nonthermal objects can be similar at optical wavelengths (see Fig.\,\ref{berg}) we searched the Two-Micron All-Sky Survey 
(2MASS)\footnote{ http://www.ipac.caltech.edu/2mass/} for NIR magnitudes of our 2BL sources. We found only one detection of the bright 6QZ object J022044.7--303713 which has a $J-K$ colour of  0.67, indicative of a nonthermal source.  As 2MASS limiting magnitudes for point sources are 
15.8, 15.1 and 14.3 for the $J$- $H$- and $K_s$-bands (1.25, 1.65 and 2.17 
microns respectively), no 2QZ objects are  found in the 2MASS catalogue, although J105534.3$-$012617 has a 3$\sigma$ detection. 
We therefore carried out our own infrared imaging as a further means of distinguishing between cool, featureless DC white dwarfs and genuine candidate BL Lac objects. \

NIR imaging was first carried out in April/May 2002 using the infrared camera, SQIID, on the 2.1m  telescope at Kitt Peak, Arizona. This instrument produces simultaneous images in the $J$, $H$, $K_s$,
and narrowband $L$ passbands.  With 440 $\times$ 460 pixels and 0.69 arcsec per pixel, 
the field of view is 304 $\times$ 317 arcsec ($\sim 5 \times 5$ arcmin).  Candidate BL Lac objects were observed in $J$, $H$ and $K$  (corresponding to an effective central wavelength of 1.267, 1.672 and 
2.225 microns respectively).  \

Sources were observed using 20 co-added 15s exposures and a 5-image dither pattern. Spectroscopic standards were chosen from Persson et al.\,(1998). Objects and standards observed on the first night, which was not photometric, were briefly re-observed at the beginning of the third night. Data reduction was performed using standard IRAF routines. \

Of the six 2BL objects observed, four were subsequently removed from the sample on account of their $>2.5\sigma$ proper motion. All four were found to have thermal NIR magnitudes, consistent with their optical SDSS magnitudes. 
Optical/IR SEDs obtained for the two remaining objects, J113900.5$-$020140 and J123437.6$-$012953 are shown in Table \ref{ca_not}.   Although the SED of these objects cannot be well fit by a single powerlaw slope,
their $J-K$ values are consistent with those found for QSOs by Barkhouse and Hall (2001) using 2MASS sources. Similar NIR magnitudes were found for these objects in later observations at the NOT and Calar Alto (see Fig. B1). 

Infrared  observations of 17 2BL objects were also carried out during April 2003, January 2004 and March 2004 on the 2.56m Nordic Optical Telescope\footnote{observations made with the Nordic Optical Telescope, operated
on the island of La Palma jointly by Denmark, Finland, Iceland,
Norway, and Sweden, in the Spanish Observatorio del Roque de los
Muchachos of the Instituto de Astrofisica de Canarias}, La Palma, Canary Islands, using the wide field camera (NOTCam) providing a field of view of 3.9' $\times$ 3.9' (0.23''/pixel). $J$, $H$ and $K$ filters at 1.235, 1.662 and 2.159 microns were used. Results are presented in Table \ref{ca_not}. Data reduction was carried out with IRAF's APPHOT package using standard NIR techniques (sky-subtraction using a sky-frame evaluated from a 12-frame set of dithered images, followed by flatfield division using twilight flats, and finally shifting and combining). With the exception of one night (Jan 2004) NOT data was taken in non-photometric conditions, hence data quality is not very high. The target  magnitudes were estimated using differential 
photometry against 2MASS field stars. Typically more than two 2MASS stars 
were available. As a check, data from the 1st of January 2004 was
calibrated using standard stars from Hunt et al.\,1998  
and no significant differences were found between the 2MASS and 
Hunt et al. calibrations.

Three 2BL objects now removed from the sample on account of their high proper motion were also observed; two are confirmed to be thermal however from both SDSS and NOT magnitudes the source J140021.0+001956  would appear to be a nonthermal object. A 3-$\sigma$ radio detection of 0.113 mJy was also recorded for this object with the VLA (\S 3, Table 5). \
\begin{table*}
\begin{center} 
\begin{tabular}{|l|c|c|c|c|c|r|} \hline
\multicolumn{1}{|c|} { Object } &
\multicolumn{5}{|c|}{Kitt Peak} &
\multicolumn{1}{|c|}{Classification}\\
\multicolumn{1}{|c|} { } &
\multicolumn{1}{|c|} {$J$ } &
\multicolumn{1}{|c} {$H$} &
\multicolumn{1}{|c} {$K_s$} &
\multicolumn{3}{c|} { } \\ \hline
J113900.5--020140   &  17.78 $\pm0.06$ & 
17.45 $\pm0.07$  & 16.94 $\pm0.07$ & & & ??  \\
J123437.6--012953  & 18.58$\pm0.09$ 
& 17.99 $\pm0.10$ &  17.88 $\pm0.11$ & & & ?? \\
\hline
\multicolumn{1}{|c|} {Object} &
\multicolumn{3}{|c|} {Nordic Opt Telescope } &
\multicolumn{2}{|c|} {Calar Alto} &
\multicolumn{1}{|c|} {Classification}\\ 
\multicolumn{1}{|c|} { } &
\multicolumn{1}{|c|} {$J$} &
\multicolumn{1}{|c|} { $ H$} &
\multicolumn{1}{|c|} {$K$ } &
\multicolumn{1}{|c|} {$J$} &
\multicolumn{1}{|c|} {$K'$} &
\multicolumn{1}{|c|} { }\\ \hline
J105355.1$-$005538 & 19.111 $\pm$0.092 & & $>$ 19.0 & & &stellar\\
J105534.3$-$012617 & 16.884 $\pm$0.087 & 16.237  $\pm$0.067 & 15.540 $\pm$0.130 & 16.97 $\pm$0.22 & 15.70 $\pm$0.16& nonthermal\\
 & 16.739  $\pm$0.094 &  16.069  $\pm$0.055 & 15.280  $\pm$0.118 & & &\\
J110644.5+000717 & $>$ 20.0 & $>$ 19.2 & $>$ 17.7 & &  $>$ 17& ??\\
J113039.1$-$004023 & 18.451$\pm$0.081 &18.340 $\pm$0.124 &17.734 $\pm$0.111& & 18.27 $\pm$0.31& ??\\
J113413.4+001041& &&& 17.57 $\pm$0.17 & 16.80 $\pm$0.17& ??\\
J113900.5$-$020140 & 17.848 $\pm$0.082 & 17.771 $\pm$0.119 & 17.241 $\pm$0.119 &18.24 $\pm$0.20 & 17.20 $\pm$0.21& variable?\\
 & 17.617 $\pm$0.099 & 17.191 $\pm$0.080 & 16.955 $\pm$0.150 & &&\\
J114137.1$-$002730 &18.199 $\pm$0.055 & 17.837$\pm$0.085& 17.768 $\pm$0.174   &17.66 $\pm$0.15 & 16.65 $\pm$0.15 & variable?\\
J114221.4$-$014812 &17.739 $\pm$0.122 &17.353 $\pm$0.086& 16.884 $\pm$0.118 & 17.70 $\pm$0.26 & 17.26 $\pm$0.24& stellar\\
J114327.3$-$005050 &18.294 $\pm$0.098 & 17.480 $\pm$0.131 &&18.48 $\pm$0.34 & 17.49 $\pm$0.22& nonthermal\\
& 18.112 $\pm$0.128 & 17.511 $\pm$0.131 &17.137 $\pm$0.116 & &&\\
J114554.8+001023  & 19.216 $\pm$0.077 & &19.683 $\pm$0.270 & & $>$ 18 & stellar\\
J115909.6$-$024534 & 18.193 $\pm$0.079 && 17.467 $\pm$0.106& 18.76 $\pm$0.46 & 17.32 $\pm$0.20& variable?\\
& 18.305 $\pm$0.083 & & 17.538 $\pm$0.106 &&&\\
J120015.3+000552 & 17.973 $\pm$0.058 & 17.268 $\pm$0.093 & 17.145 $\pm$0.054 &&& ??\\
J121834.8$-$011955 &  15.496 $\pm$0.136 & 14.543 $\pm$0.139 & 13.574  $\pm$0.049 & & & nonthermal\\
J123437.6$-$012953 &18.521 $\pm$0.098 & & 17.470 $\pm$0.114 & & 17.58 $\pm$0.25& ??\\
& & 17.990 $\pm$0.067 & 17.517 $\pm$0.103 &&&\\
J125435.7$-$011822 &18.733 $\pm$0.087 &18.557 $\pm$0.198 & 18.167 $\pm$0.168 & & 17.63 $\pm$0.54 & stellar\\
J140207.7$-$013033 & $>\,19.3$ $\pm$0.5  & &  $>\,18.2$ $\pm$0.5 & &   $>\,17.3$\\
J141040.2$-$023020 &19.239 +-0.232 & & & && stellar\\
J142526.2$-$011826 & 17.656 $\pm$0.164 & & 15.972 $\pm$0.127 & 18.13 $\pm$0.25 & 16.15 $\pm$0.08 & nonthermal\\
 & 17.793 $\pm$0.093 & & & &&\\
\hline
\multicolumn{7}{|c|} {Objects with high proper motion now removed from the 2BL sample }\\ \hline
J102615.3$-$000629 &   18.988 $\pm$0.132 & & 19.097 $\pm$0.210 & & & stellar\\
J130009.9$-$022600 & 18.129 $\pm$0.176 & & 18.427 $\pm$0.211 & & & stellar\\
J140021.0+001956 &18.494  $\pm$0.082 & &17.508 $\pm$0.055 & & & nonthermal\\
\hline
\end{tabular}
\end{center}
\caption{$J$, $H$ and $K/K'$ magnitudes of objects observed at Kitt Peak, the Nordic Optical Telescope and Calar Alto}
\label{ca_not}
\end{table*}

Further NIR images in $J$ and/or $K'$ (1.250 and 2.160 microns respectively) of 17 2BL objects were taken at the 
Calar Alto 2.2m telescope\footnote
{observations collected at the Centro Astron{\'o}mico Hispano Alem{\'a}n (CAHA) at Calar Alto, operated jointly by the Max-Planck Institut f\"ur Astronomie and the Instituto de Astrof{\'i}sica de Andaluc{\'i}a (CSIC)} during the period May, 19-27 2005. The NIR 
camera MAGIC in high-resolution mode was used, which provides a field 
of view of $\sim 2.7' \times 2.7'$ (0.636"/pixel) by means of a 256 
$\times$ 256 pixel HgCdTe array. For the observations, we adopted a 3 
$\times$ 3 dither pattern, with the individual raster points separated 
by 10". Depending on the filter, at each of the individual 
raster points 15 images 2sec each in $J$ or 
30 images 1sec each in $K'$ were recorded, summed and saved. 
The total integration time for one dither pattern was thus 
450 sec. Since our aim was to detect the source and measure its 
(unknown) flux, the integration times in total per filter and object 
were adopted adaptively. If the  quick online reduction at the 
telescope shows the object to be detected  after one dither pattern, 
the observations of this particular object were either repeated using  
the other filter or stopped. Otherwise the observations in a particular filter
were repeated until the object was detected on the 
final summed image. If the object was not detected after 4 dither patterns
($\sim$ 30min), the observations of this source were terminated as well. 
Observing conditions varied greatly during our 8 night run (4 full 
and 4 half nights), and observations taken during poor seeing 
conditions ($>$2" FWHM) were repeated. Only two of the nights were 
photometric during which we regularly observed standard stars from the 
lists of Elias et al.\,(1982) and Hawarden et al.\,(2001) to set the zero points. The data reduction was standard for NIR-observations (subtracting 
a scaled sky derived from adjacent observations of the same field, 
flatfielding by means of a dome flat taken as the difference of 
exposures of the illuminated and non-illuminated dome). Bad pixels were 
identified from the flatfield exposures and subsequently flagged on the 
individual images. The reduced images were then aligned to integer pixels 
and averaged by a clipping algorithm to get rid of the bad 
pixels. Photometry was done using SExtractor (Bertin \& Arnouts 1996)), where we used the 
photometry provided by Mag\_Best. As only a few nights were photometric 
calibration for all sources was performed via 2MASS stars in the same frame. The zero points derived via 2MASS 
observations and those derived from our own standard star measurements 
agree very well. The resulting magnitudes and errors for our 2BL objects 
are given in Table \ref{ca_not}. As can be seen, the errors are partly very large. 
This is due to the fact that for the majority of our sources only 
1-2 2MASS stars were present in the same frame, and these stars typically had 
large measurement errors. Thus our errors are dominated by systematic errors. 
The internal errors are much smaller -- of the order of 0.02-0.05mag. 
\begin{table}
\label{trav}
\begin{center}
\begin{tabular}{|l|r|l|} \hline
\multicolumn{1}{|c|} {Object} &
\multicolumn{1}{|c|} {$S_{8.4}$ (mJy) } &
\multicolumn{1}{|c|} {rms} \\ \hline
J114137.1--002730 & 3.580 & 0.052 \\ 
J115909.6--024534 & 0.296 & 0.037 \\
J140021.0+001955$\dag$  & 0.336 & 0.034 \\
J031056.9--305901 & 0.129 & 0.035 \\
J140207.7--013033$\dag\dag$ & 0.113 & 0.038 \\ \hline
J004950.6-284907 & $<$0.13 &0.042\\
J023405.5-301519 & $<$0.10 &0.032\\
J103607.4+015658 & $<$0.11 &0.038\\
J105355.1-005538 & $<$0.14 &0.048\\
J110644.5+000717 & $<$0.12 &0.038\\
J113413.4+000411 & $<$0.21 &0.042\\
J113900.5-020140 & $<$0.10 &0.033\\
J114221.4-014812 & $<$0.11 &0.035\\
J114327.3-005050 & $<$0.11 &0.037\\
J120015.3+000552 & $<$0.14 &0.045\\
J120558.1-004215$^*$ & $<$0.12 &0.039\\
J122338.0-015619$^*$ & $<$0.11 &0.036\\
J123437.6-012953 & $<$0.08 &0.026\\
J125435.7-011822 & $<$0.11 &0.035\\
J130009.8-022600$^*$ & $<$0.22 &0.073\\
J132811.5+000227$^*$ & $<$0.09 &0.031\\
J131635.1-002810$^*$ & $<$0.11 &0.035\\
J223233.5-272859 & $<$0.14 &0.047\\
J224559.1-312223 & $<$0.13 &0.044\\
J230443.6-311107 & $<$0.13 &0.043\\
J231749.0-285350 & $<$0.12 &0.039\\
J232531.3-313136 & $<$0.14 &0.048\\
\hline
\end{tabular}
\caption{8.4 GHz flux densities (or upper limits) of 23 2BL sources. $\dag$ This object is assigned a proper motion of 3.87$\sigma$ in the updated SuperCosmos catalogue and has been removed from the sample. However IR observations conducted at the NOT identify this object as a nonthermal source, consistent with SDSS optical photometry. $^*$These objects have now been removed from the sample on account of their $>\,2.5\sigma$ proper motion. $\dag\dag$ This object is identified from the VLT spectrum as an unambiguous WD. }
\end{center}
\end{table}

\section{Radio properties}

Only nine of the 2QZ BL Lac candidates have catalogued FIRST and/or NVSS detections.  However, because of their faint optical flux densities ($m_{b_J} > 18.25$), it is possible that some of these objects are moderately radio-loud BL Lacs at higher redshift that simply fall below the flux limits of these surveys.  Prior surveys have found that BL Lacs, and other radio-loud AGN, tend to lie above $\alpha_{ro} > 0.2$ whereas radio-quiet AGN lie below $\alpha_{ro} < 0.2$ (Rector et al.\,2000).  The NVSS and FIRST survey detection limits are insufficient to determine whether or not the 2QZ BL Lac candidates are radio-loud objects at higher redshift or if they are indeed radio-quiet BL Lacs that occupy a distinct region of the $\alpha_{ox}$, $\alpha_{ro}$ plane.

Twenty-seven objects from the 2QZ sample that lacked detection in the FIRST  and NVSS  surveys were observed with the NRAO\footnote{The National Radio Astronomy Observatory is a facility of the National Science Foundation operated by Associated Universities, Inc. under a cooperative agreement.} Very Large Array.  The D~configuration was chosen to maximize phase stability.  The targets were observed at a frequency of 8.4~GHz on 20 and 21~April 2003, with all 27 antennas operating.  Each object was observed for 14 to 16 minutes. The data were reduced using the NRAO AIPS software package.   Flux density calibrations were bootstrapped from 3C~48 on April 20th and 3C~286 on April 21st respectively.  All objects within the primary beam were imaged and cleaned to remove residual sidelobes from nearby bright radio sources.  The noise levels in the resultant maps were $\sigma_{s}$ $\simeq$ 30-50 $\mu$Jy~beam$^{-1}$.  

Of the 27 objects observed, only four detections were made coincident with the optical positions of the BL Lac candidates.  All other sources were undetected to within flux densities of $S_{peak}>3\sigma$ above the background noise level.  The VLA observations are summarized in Table 5.  Column 1 is the 2QZ object name.  Columns 2 and 3 give the peak detection and noise level for each object in milli-Janskys.  For non-detections, the peak value is calculated to be a $3\sigma$ upper limit.

Two objects were also observed at 1.4 GHz using Director's time at the Australia Telescope Compact Array (ATCA), Narrabri NSW, in November 2003 and May 2004. No detection for either J215454.3--305654 or J003058.2--275629 was found, placing a $3 \sigma$ upper limit on the flux density of these sources of 0.21mJy and 0.15mJy respectively. 

Radio detections have thus been found for 15 of the candidate BL Lacs.

\section{Summary and Conclusions}
From VLT spectra of 35 2BL objects we identify 11 WDs with weak or absent absorption features over a temperature range of 6800 -- 13500~K, one unambiguous DA at 14500~K, plus a further unusual WD with split absorption features at a temperature of $\sim 25000$K. This represents contamination of the 2BL sample by stellar objects of $\sim$40 per cent. NIR observations at the NOT and Calar Alto identify a further two stellar objects, as well as three objects classified as stellar from the VLT spectra.
Two other objects (J110644.5+000717 and J125501.2+015513) have SDSS optical magnitudes that are well fit by a thermal blackbody curve. J110644.5+000717 was also observed at the NOT and Calar Alto but was too faint for a detection. NIR upper limits are consistent with the SDSS magnitudes that identify it as a thermal object. Note however that Nesci et al.\,(2005) report this object as showing evidence of variability.\

 Twelve objects with VLT spectra (four of which are radio-detected) are identified as weak-lined AGN; eleven of these are in the redshift range $z=$ 1.23--1.91 where the MgII line has fallen in the noisy red end of the 2QZ spectra.  A further X-ray- and radio-loud object (J014310.1-320056) at $z=$0.375, has clearly varied since the first 2QZ spectrum was taken.  Nine 2BL objects have SDSS spectra (Fig A1); three are of objects without VLT spectra, and all three appear to be AGN.  \

 Ten 2BL sources remain without a clear classification. For three of these objects with SDSS and NIR photometry it would appear that either the source has varied considerably between the NIR observations  and those carried out for SDSS or these are a new class of object with an unusual kink in the spectrum between the optical and infrared. Note that within the errors the NIR observations carried out at the NOT in 2003 and 2004 and at Calar Alto in 2005 are generally in agreement. One of these three objects, J114137.1$-$002730, is quoted by Nesci et al.\,(2005) as having high proper motion, but would appear from our observations to be a 3.58mJy radio source.  For the other seven objects only the 2dF/6dF spectra are available.\

\begin{figure}
\psfig{file=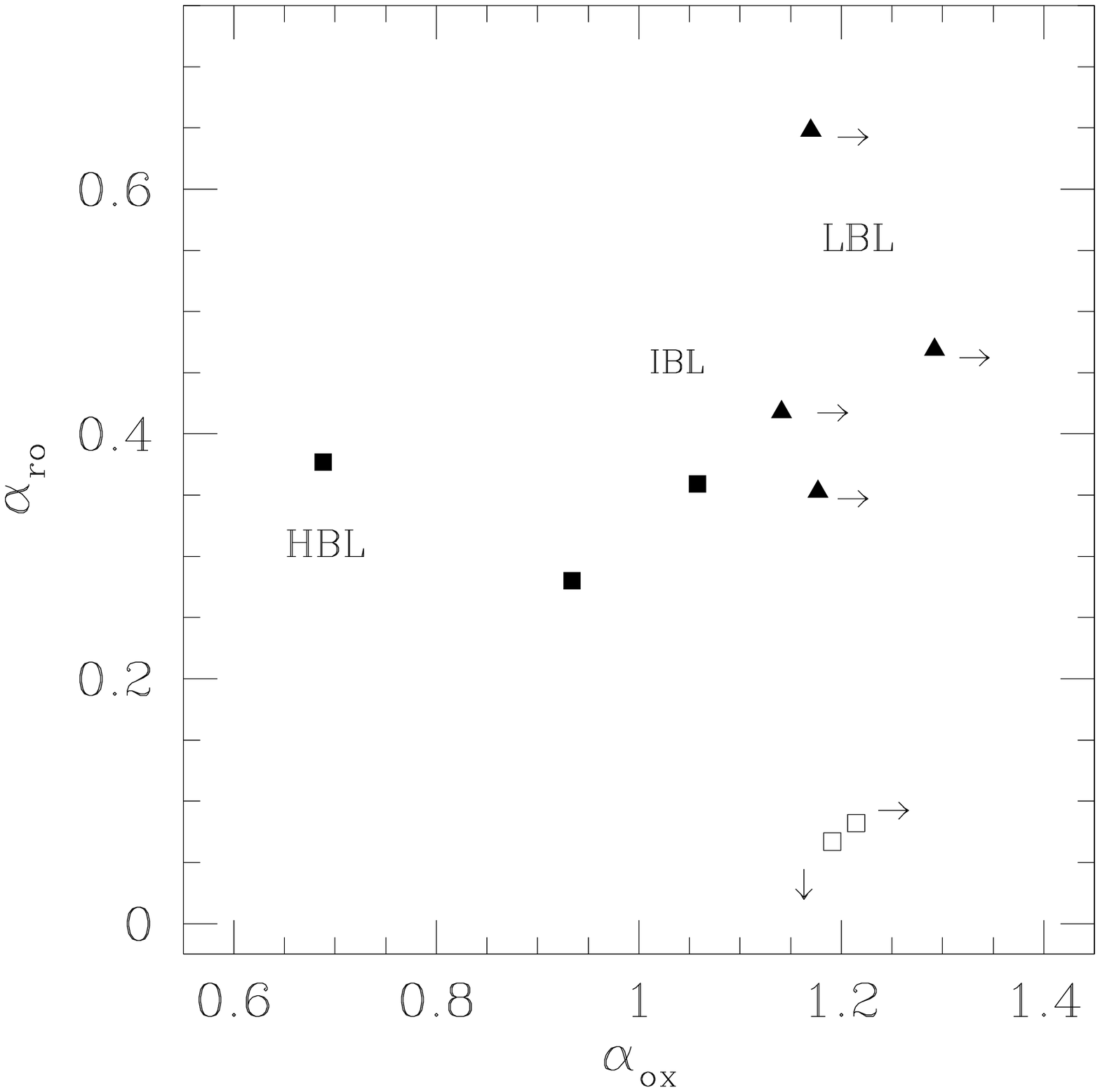,height=2.8in}
\psfig{file=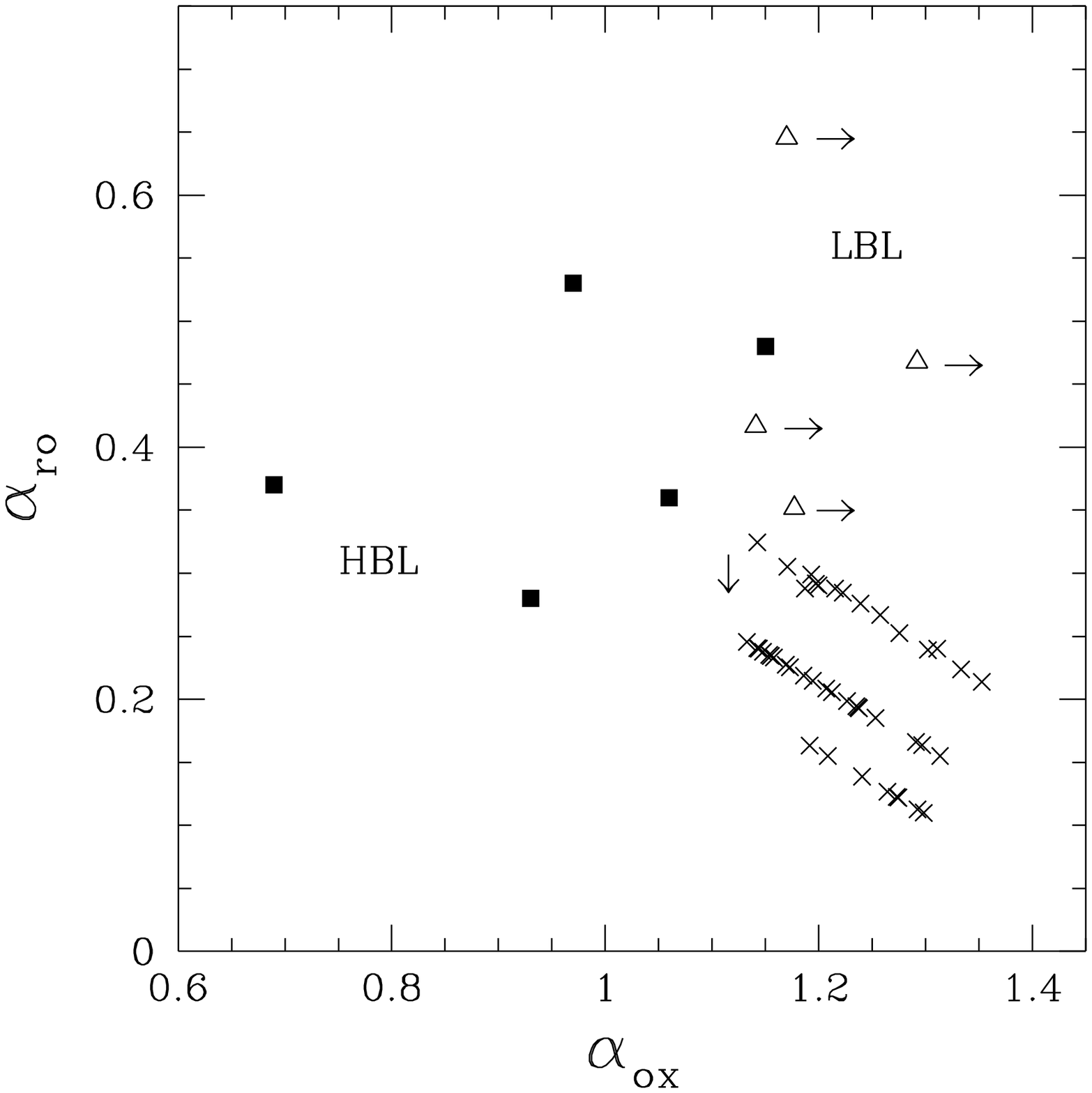,height=2.8in}
\caption{Top: $\alpha_{\rm ro}-\alpha_{\rm ox}$ diagram of 2BL BL Lacs;  Solid squares denote objects with both radio and X-ray fluxes, while triangles are those with no X-ray flux. Arrows indicate upper limits of $f_x = 8 \times 10^{-14}$ erg s$^{-1}$ cm$^{-2}$. The open squares are the two radio-quiet candidate BL Lacs.  Bottom: $\alpha_{\rm ro}-\alpha_{\rm ox}$ diagram from Paper 1 (Fig 14) showing 47 radio-quiet candidate BL Lacs.}
\label{newalf}
\end{figure}

A total of eight 2BL objects are classed as continuum objects. Seven are radio-loud and three of these are detected at both X-ray and radio frequencies.  NIR observations  confirm the nonthermal nature of three of these eight continuum objects.  J215454.3$-$305654 is undetected at radio frequencies. The VLT spectrum confirms that it is extragalactic, at a redshift of 0.494. As the spectrum does show weak, narrow [OIII] emission this object could belong to the recently discovered population of low luminosity AGN with no broad line region (Nicastro et al.\,2003; Elitzur \& Shlosman 2006). This object has been included in a Chandra proposal to obtain  X-ray data for a sample of radio-weak/radio-quiet featureless continuum objects (PI J. Gelbord, MIT). A further object, J224559.1-312223, is potentially a radio-quiet BL Lac; the VLT spectrum shows a weak feature that could be MgII at $z=$1.324, however the shape of the spectrum is also well fit by a blackbody curve of T\,=\,7100\,K, and the feature could then be H$\alpha$ at zero redshift. 
\begin{table}
\caption{Classification of 2BL sources from VLT spectra, NIR imaging and SDSS spectra and photometry}
\begin{center}
\begin{tabular}{c c}
\multicolumn{1}{c} {number} &
\multicolumn{1}{c} {classification}\\
\hline
17 & stellar \\
16 & AGN \\
10 & ?? \\
7  & continuum (radio-detected)\\
2 & continuum (no radio detection) \\
\hline
\end{tabular}
\end{center}
\end{table}
A colour-colour diagram of these nine 2BL candidate BL Lacs is shown in Fig \ref{newalf} (Top); the two radio quiet candidates lie well outside the region of colourspace populated by hitherto discovered BL Lacs. These two objects are now the only remaining radio-quiet candidate BL Lacs of 47 objects included in Paper 1 (see Fig. \ref{newalf} (Bottom)).

A similar search for optically selected candidate BL Lac objects (or QFOs, quasi-featureless objects) was carried out using the SDSS survey (Collinge et al.\,2005).  This search made use of colour-selection to separate extragalactic objects from galactic QFOs -- mostly DC white dwarfs -- where no reliable proper motion measurements were available. Of 240 probable QFOs identified from the 2860 sq deg of the SDSS spectroscopic survey, 27 objects are currently undetected at the 5$\sigma$ level of the FIRST and NVSS surveys, however Collinge et al.\,(2005) indicate that more sensitive radio observations are needed before deeming these objects to be radio-quiet. In comparison the 2BL sample, from a survey area of $\sim$650 sq deg and magnitude range $18.25 \leq \bj \leq 20.0$ (i.e. not including the 6QZ survey as all three candidates are now classified as AGN), identifies seven BL Lac objects plus two radio-quiet candidate BL Lacs.  Ten 2BL objects remain unclassified;  based on contamination levels evident from the VLT spectra we may expect one or two of these unclassified objects to be a candidate BL Lac, giving a sample of $\sim$10 optically selected BL Lac objects. Further high signal-to-noise, high resolution spectroscopy and/or NIR imaging is needed for these ten objects to determine their identity.\
 
Optical selection of BL Lac candidates ${can}$ be an effective method of identifying  candidate objects, providing the spectra are flux calibrated and of high enough signal-to-noise and resolution to eliminate contamination by weak-lined AGN. Overcoming contamination by cool, featureless WDs is more difficult (and hence time-consuming), and relies on accurate proper motion data together with colour selection strategies in order to eliminate the bulk of thermal objects in the initial selection procedure. However some overlap in colourspace between WDs and QSOs remains in the optical regime. NIR imaging or spectroscopy of radio undetected candidate BL Lacs is thus necessary to distinguish between nonthermal objects and cool DC WDs. \

For the 2BL sample the selection criteria of the original 2QZ and 6QZ catalogues were less than optimal for finding candidate BL Lac objects -  the bright limit for 2QZ sources of $\bj > 18.25$, and the requirement that objects be classed as point sources, will have eliminated many nearby BL Lac objects from the 2QZ catalogue (indeed six BL Lacs were independently identified in the 2dF Galaxy Redshift Survey (Colless et al.\,2001), Lewis, I.J., private communication). However, from a search of the online V\'eron \& V\'eron-Cetty (2000) catalogue, conducted once the original 2BL sample had been selected, we determined that only two possible BL Lac objects (denoted BL? by V\'eron \& V\'eron-Cetty) were missed as a result of the 2QZ's initial colour (UV excess) selection criteria.

While there is evidently not a large population of radio-quiet BL Lacs, optical selection does probe to lower $\alpha_{ro}$ values than do X-ray and radio-selected samples; inclusion of these faint candidate BL Lacs is clearly necessary if evolutionary trends for BL Lac objects are to be correctly interpreted. \

\section*{Acknowledgments} 
The 2QZ/6QZ is based on observations made with the Anglo-Australian
Telescope and the UK Schmidt Telescope; we would like to thank our
colleagues on the 2dF and 6dF survey teams and all the staff at
the AAT who have helped make this survey possible. We also thank
the staff at the Mt Stromlo Observatory and at the Australia Telescope Compact Array, which is funded by
 The Commonwealth of Australia for operation as a national Facility 
managed by CSIRO.\\
This research has made use of data obtained from the SuperCOSMOS Science Archive, prepared and hosted by the Wide Field Astronomy Unit, Institute for Astronomy, University of Edinburgh, which is funded by the UK Particle Physics and Astronomy Research Council.\\
Data has also been obtained from SDSS; Funding for the SDSS and SDSS-II has been provided by the Alfred P. Sloan Foundation, the Participating Institutions, the National Science Foundation, the U.S. Department of Energy, the National Aeronautics and Space Administration, the Japanese Monbukagakusho, the Max Planck Society, and the Higher Education Funding Council for England. The SDSS Web Site is http://www.sdss.org\\
TP thanks Amanda Djupvik for providing the NIR data reduction script. JH and DL thank Johannes Ohlert for assisting in the CA observations.
  DL wishes to thank Hermine Landt for interesting discussions and the School of Physics at the
University of Sydney for a Postgraduate (Mature Age) Scholarship.\\

\newpage
\appendix
\section{SDSS spectroscopy of 2BL objects}
\begin{figure*}
\centerline{\psfig{file=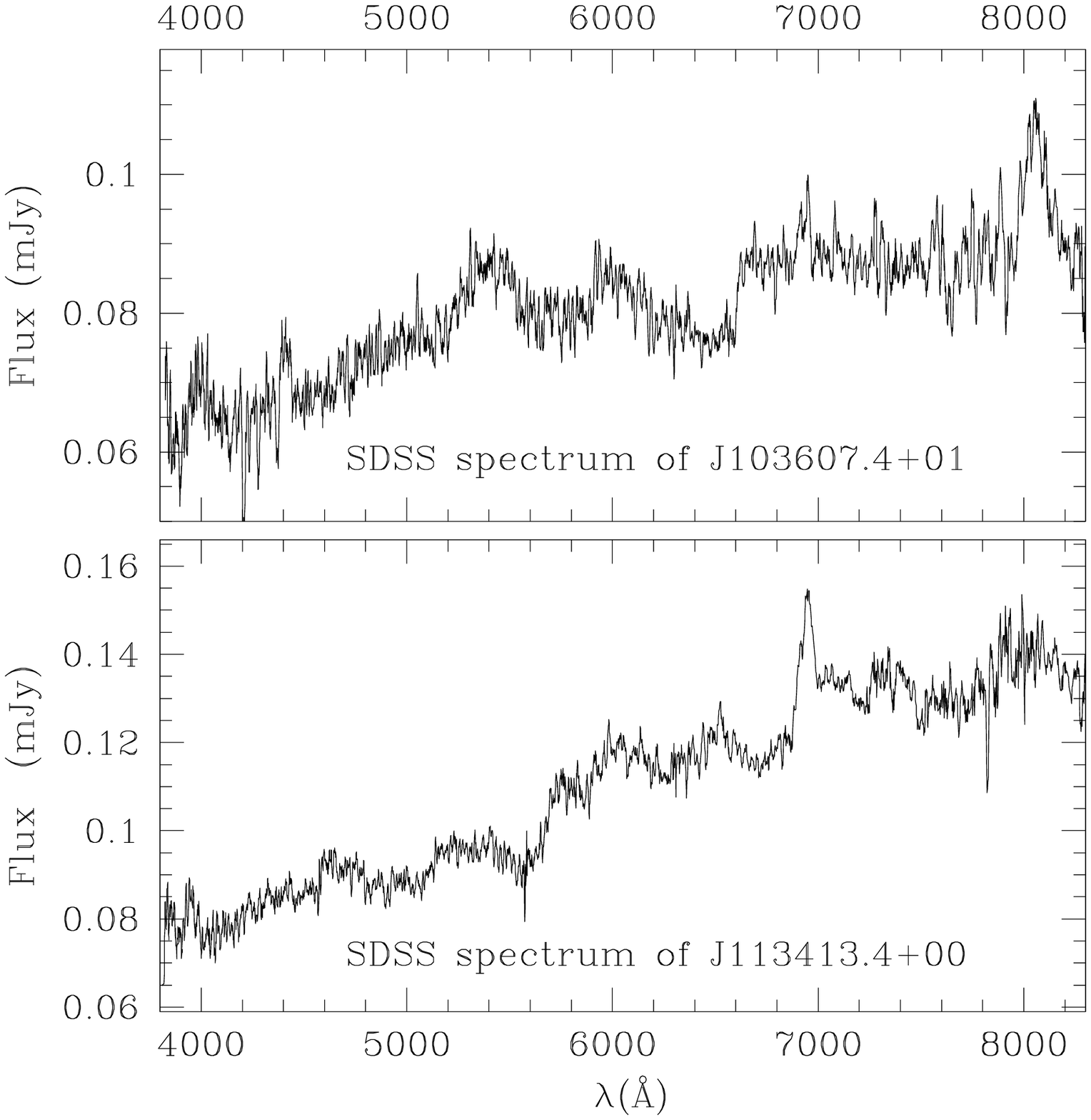,height=3.in}\psfig{file=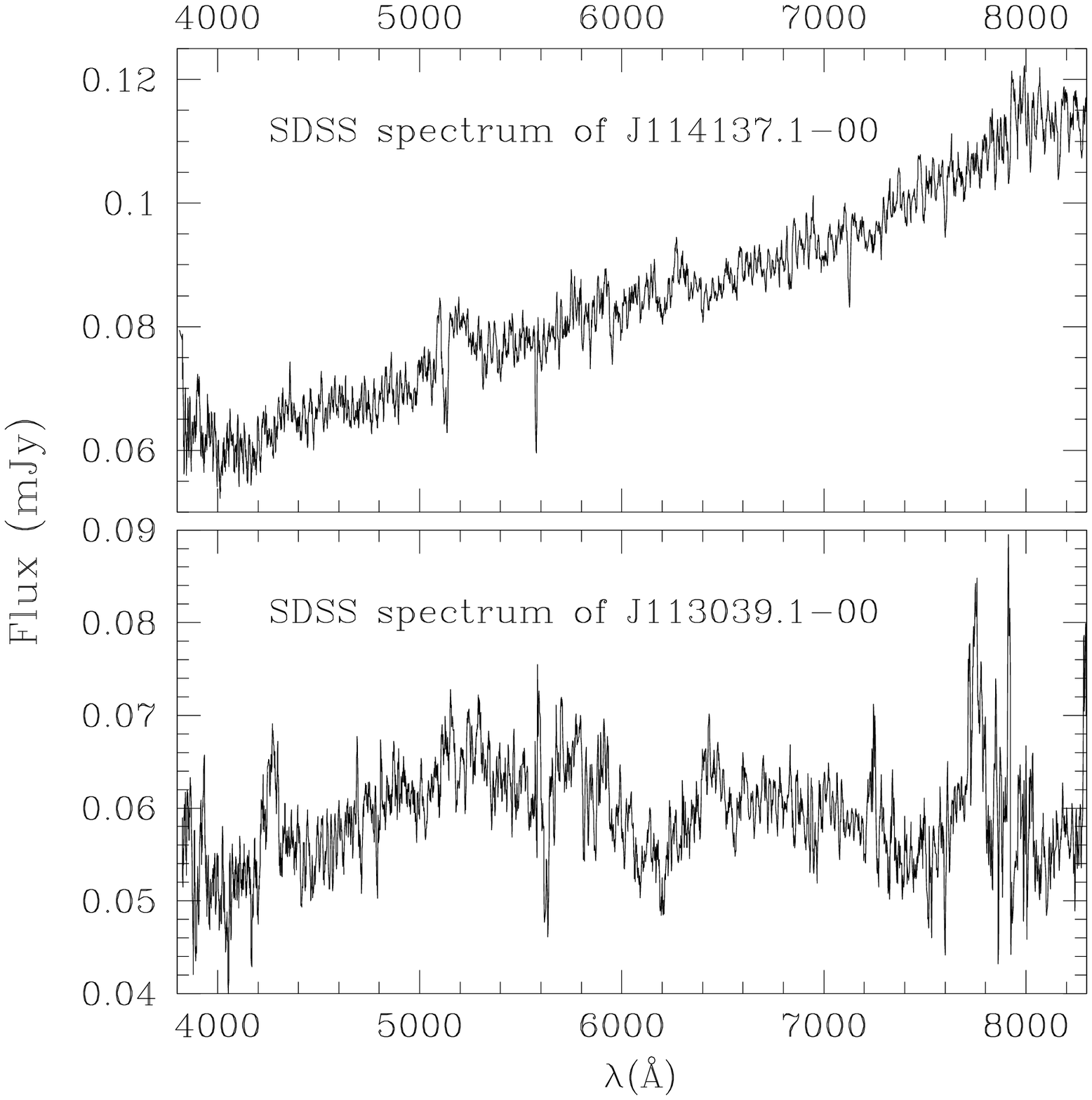,height=3.in}}
\centerline{\psfig{file=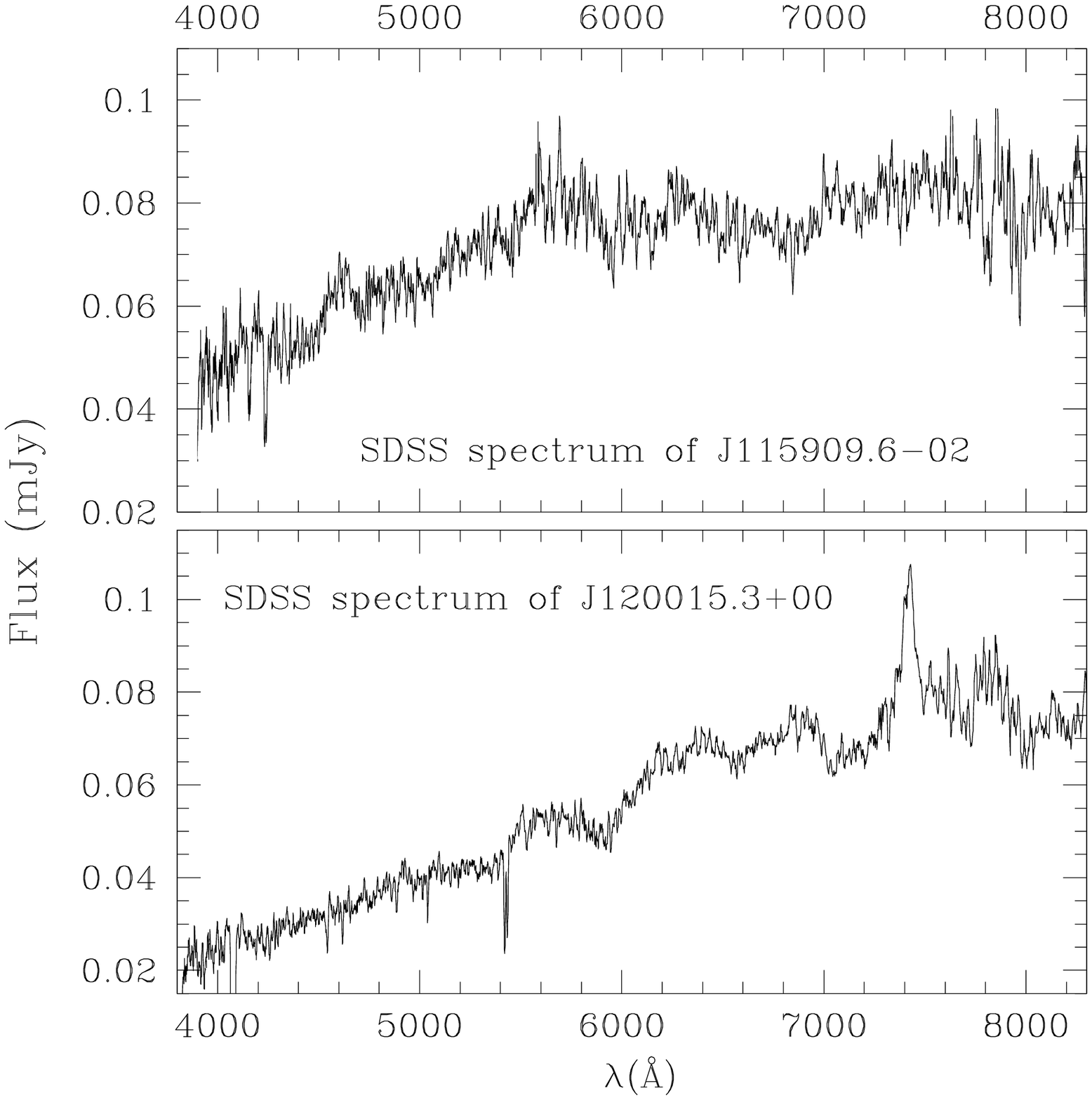,height=3.in}\psfig{file=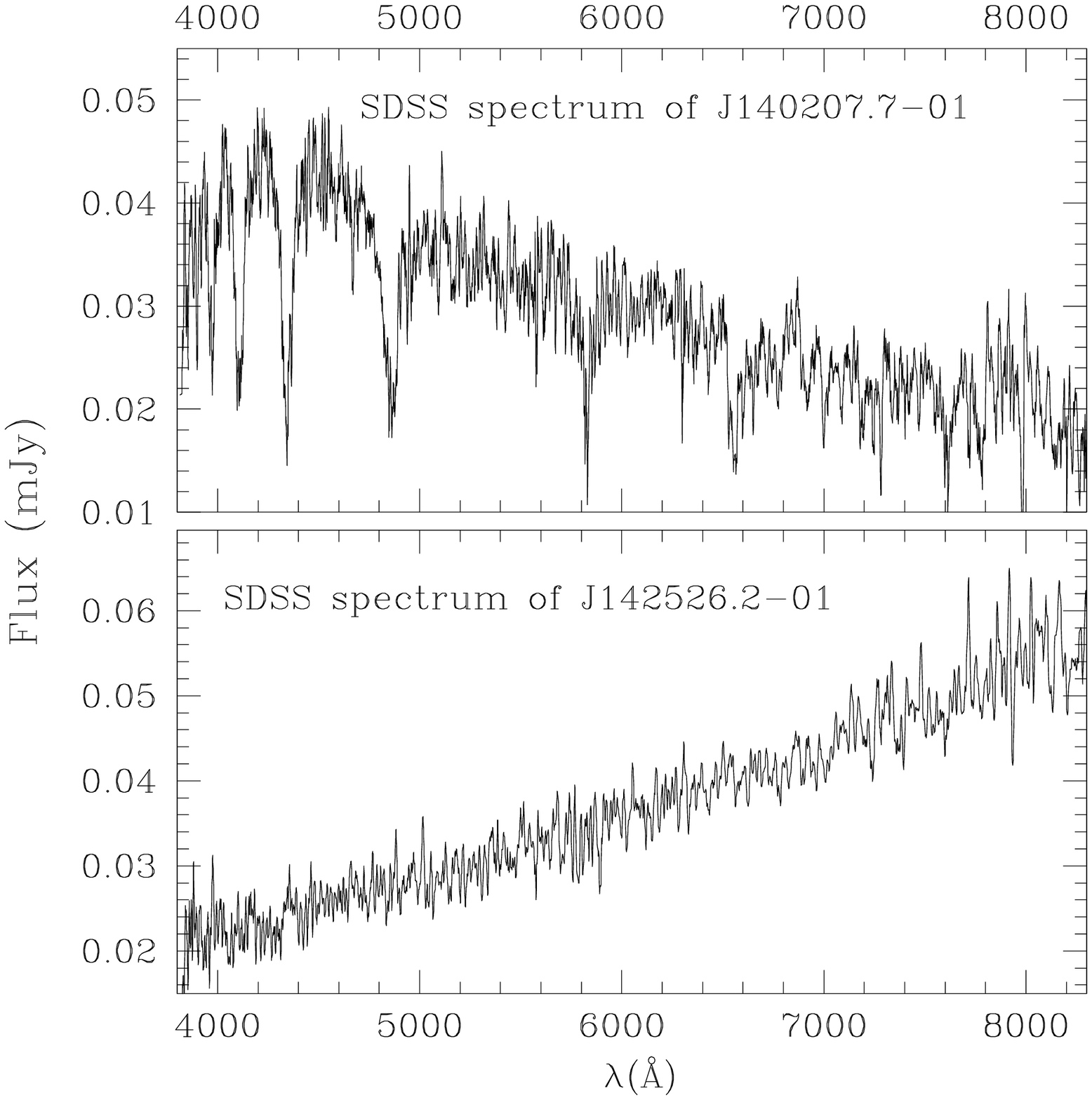,height=3.in}}
\centerline{\psfig{file=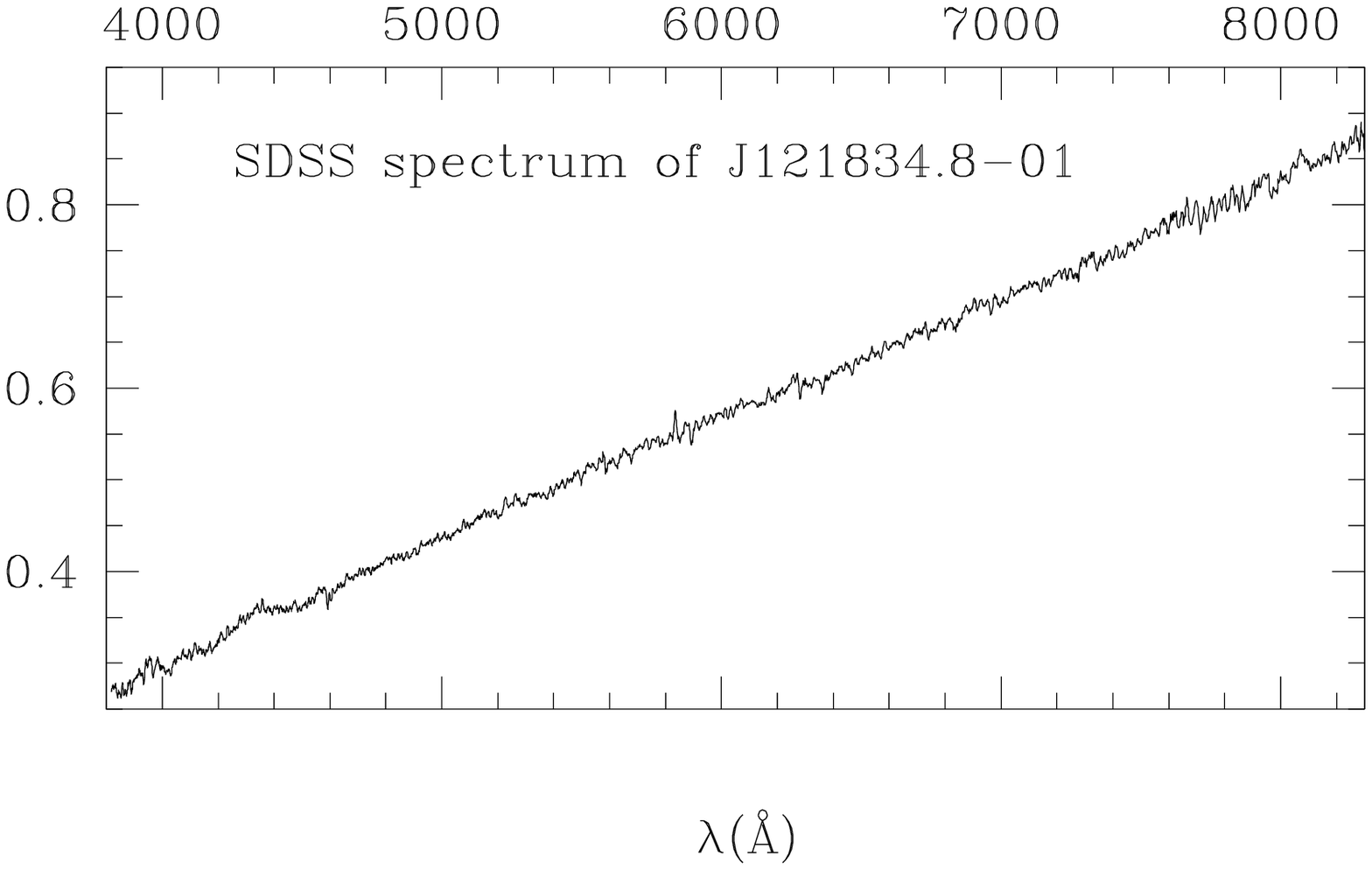,height=3.in}}
\caption{SDSS spectra of 2BL objects}
\label{sdss4}
\end{figure*}
\section{Optical and NIR photometry of 2BL objects}
\begin{figure*}
\caption{Plots of 2BL objects with SDSS optical magnitudes and NIR magnitudes from Kitt Peak, NOT and Calar Alto }
\centerline{\psfig{file=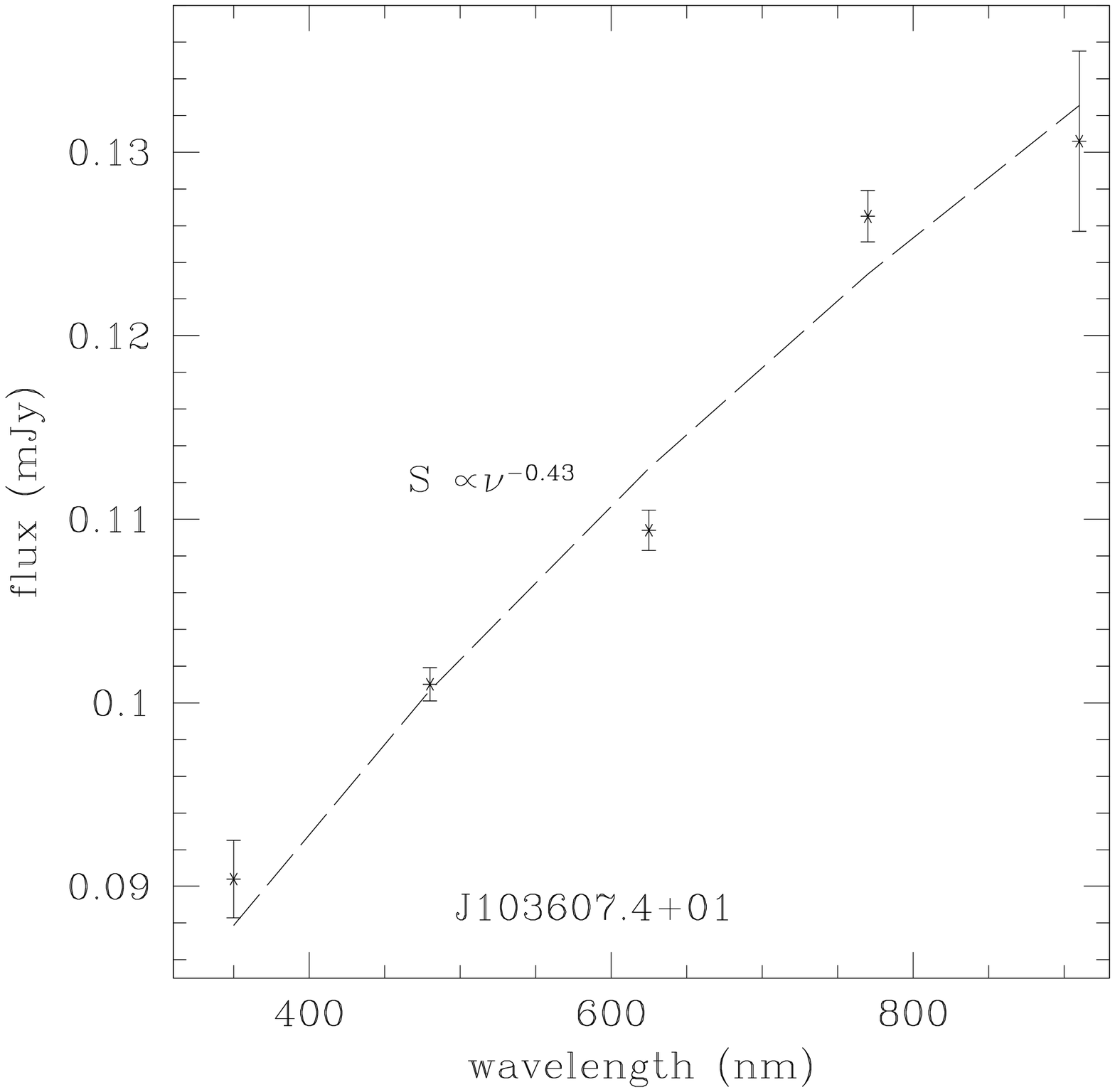,height=3.0in}\psfig{file=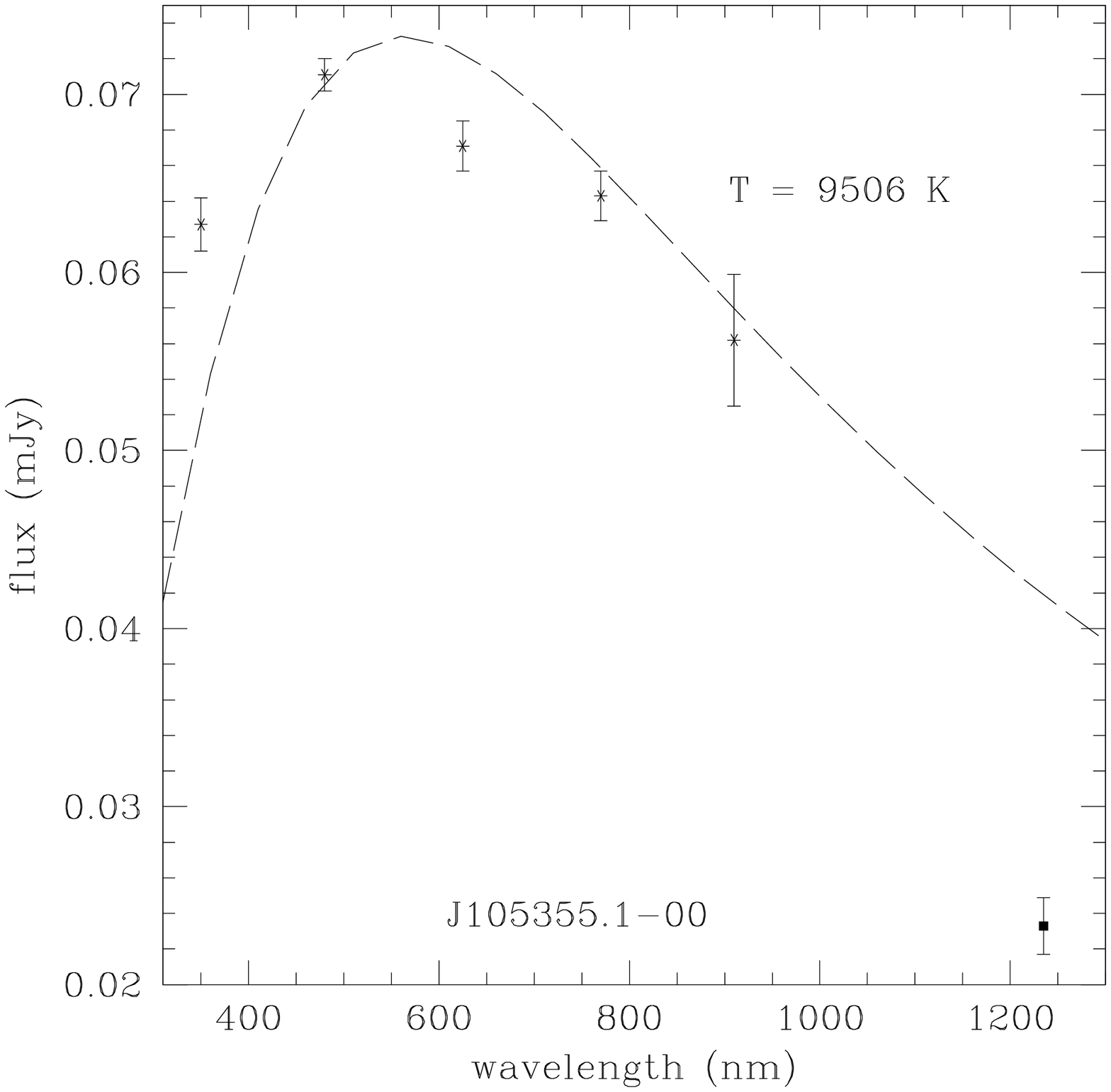,height=3.in}}
\centerline{\psfig{file=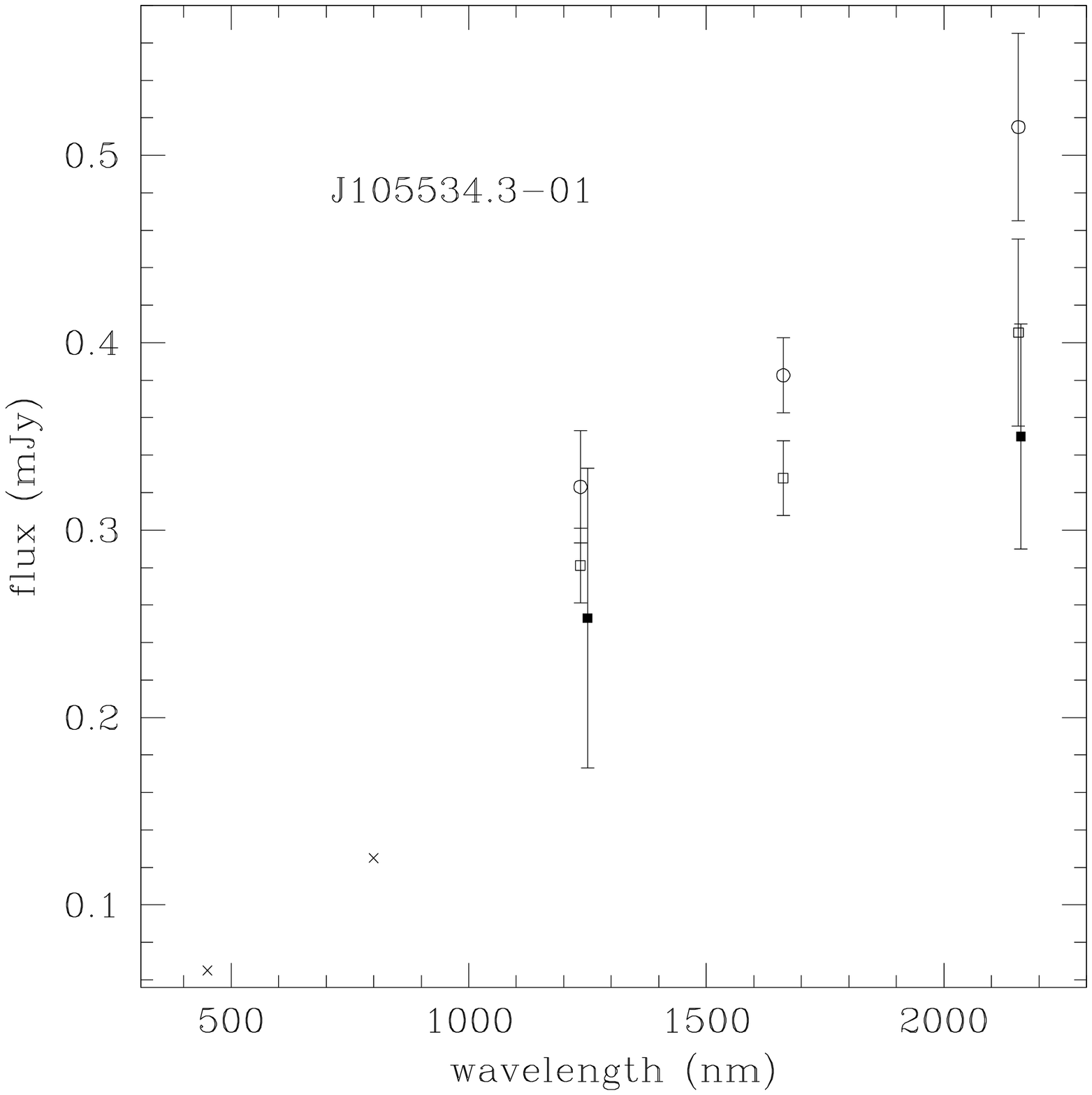,height=3.0in}\psfig{file=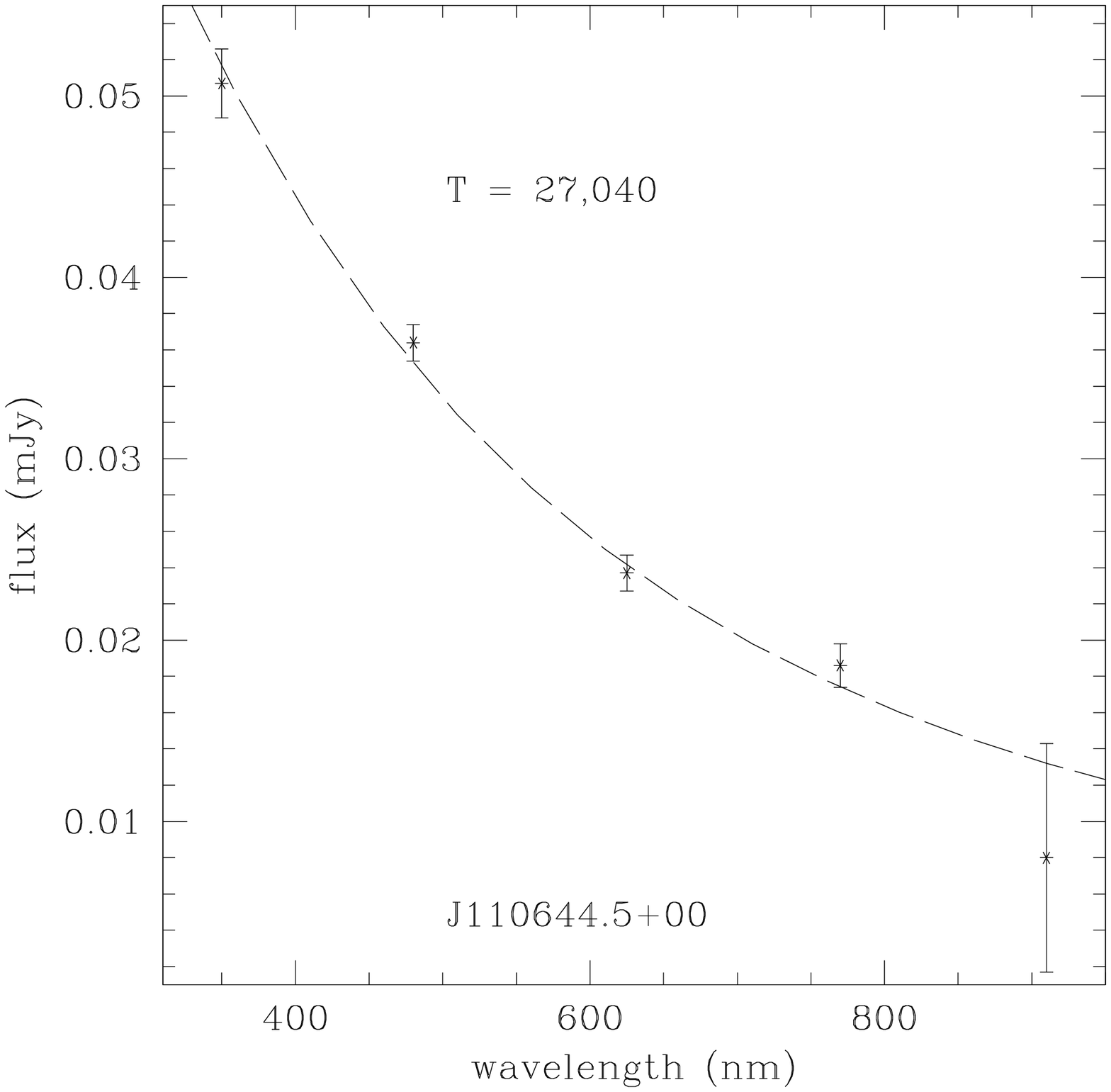,height=3.in}}
\centerline{\psfig{file=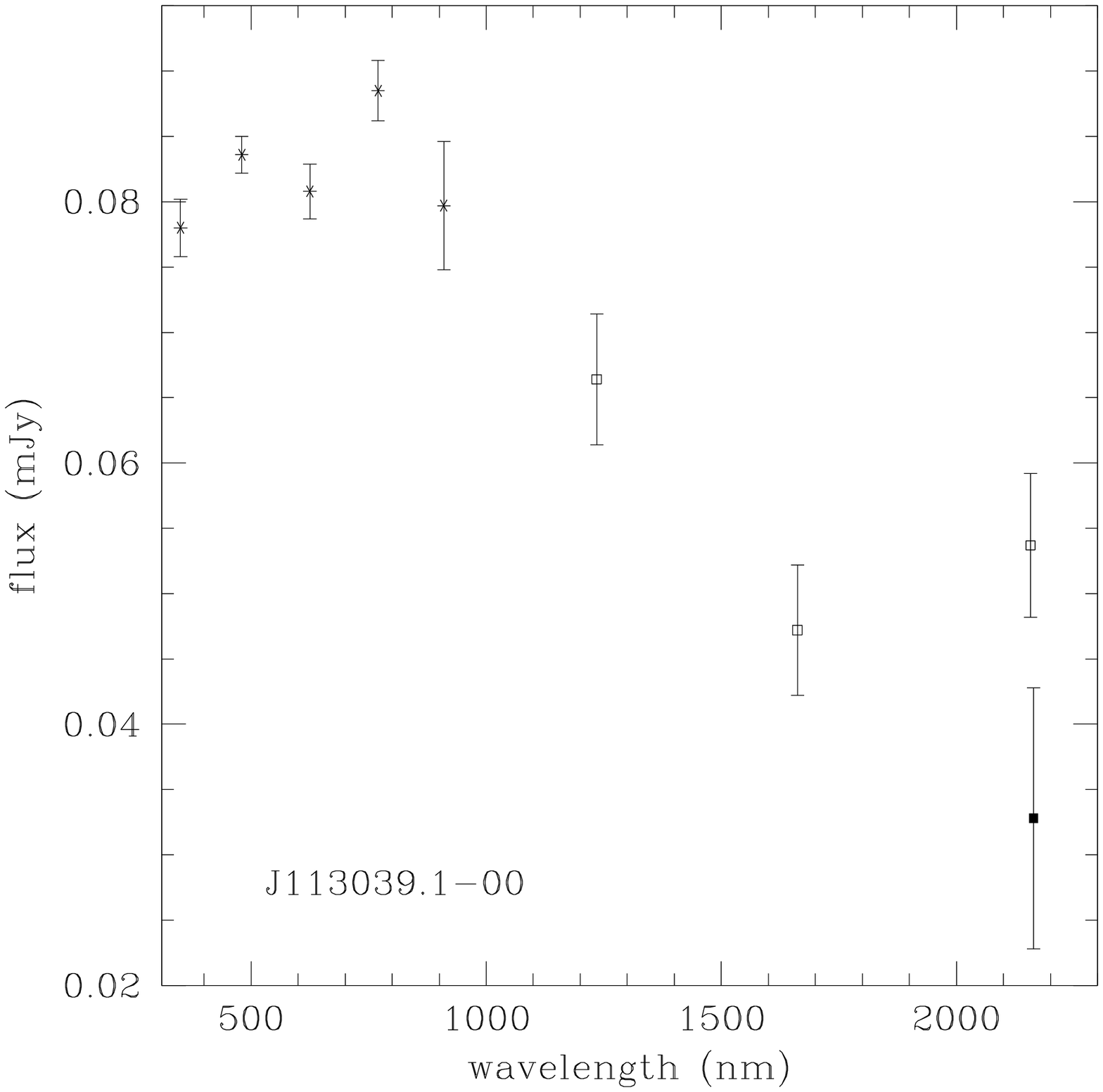,height=3.0in}}
\end{figure*}
\begin{figure*}
\centerline{\psfig{file=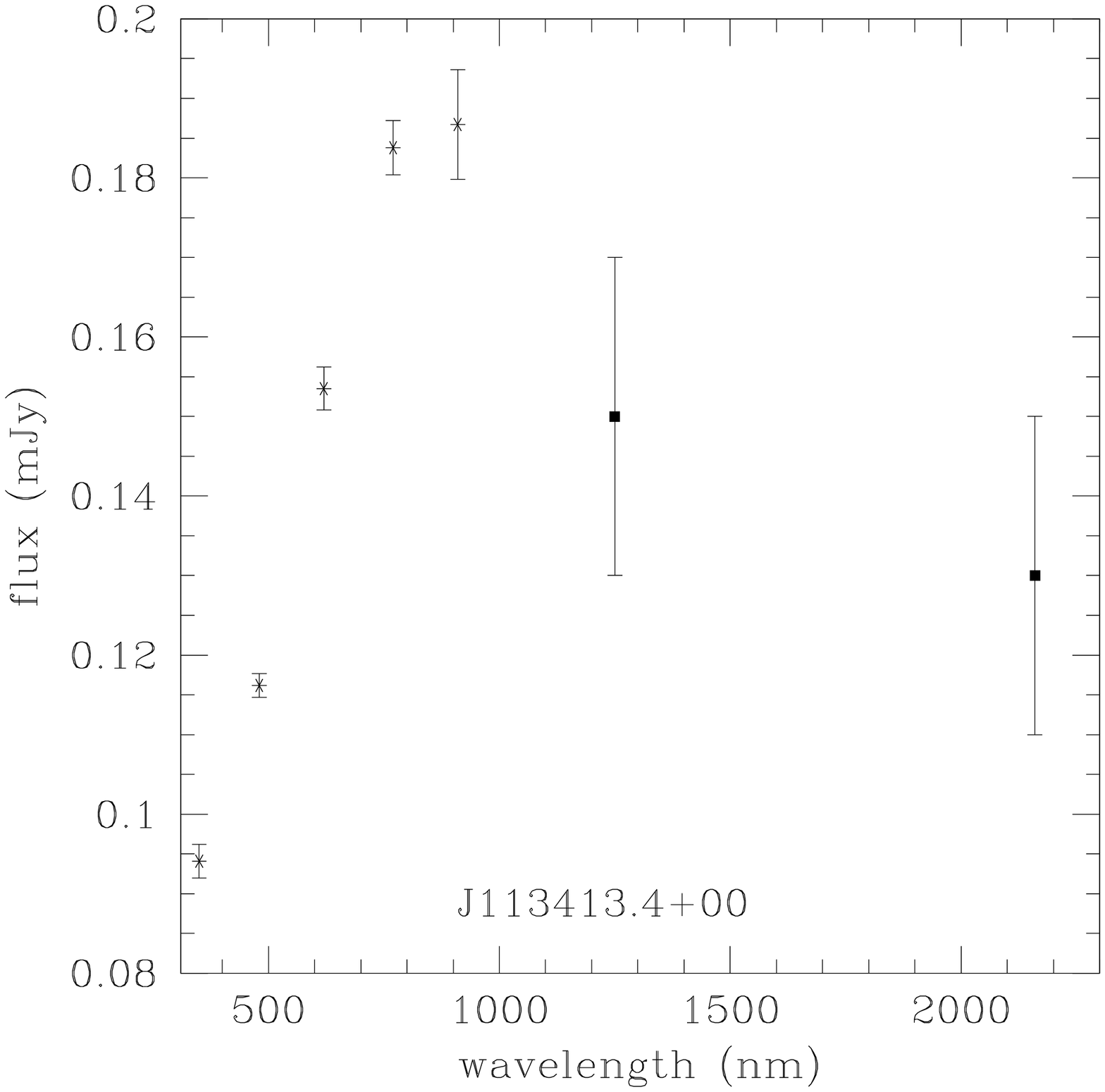,height=3.0in}}
\centerline{\psfig{file=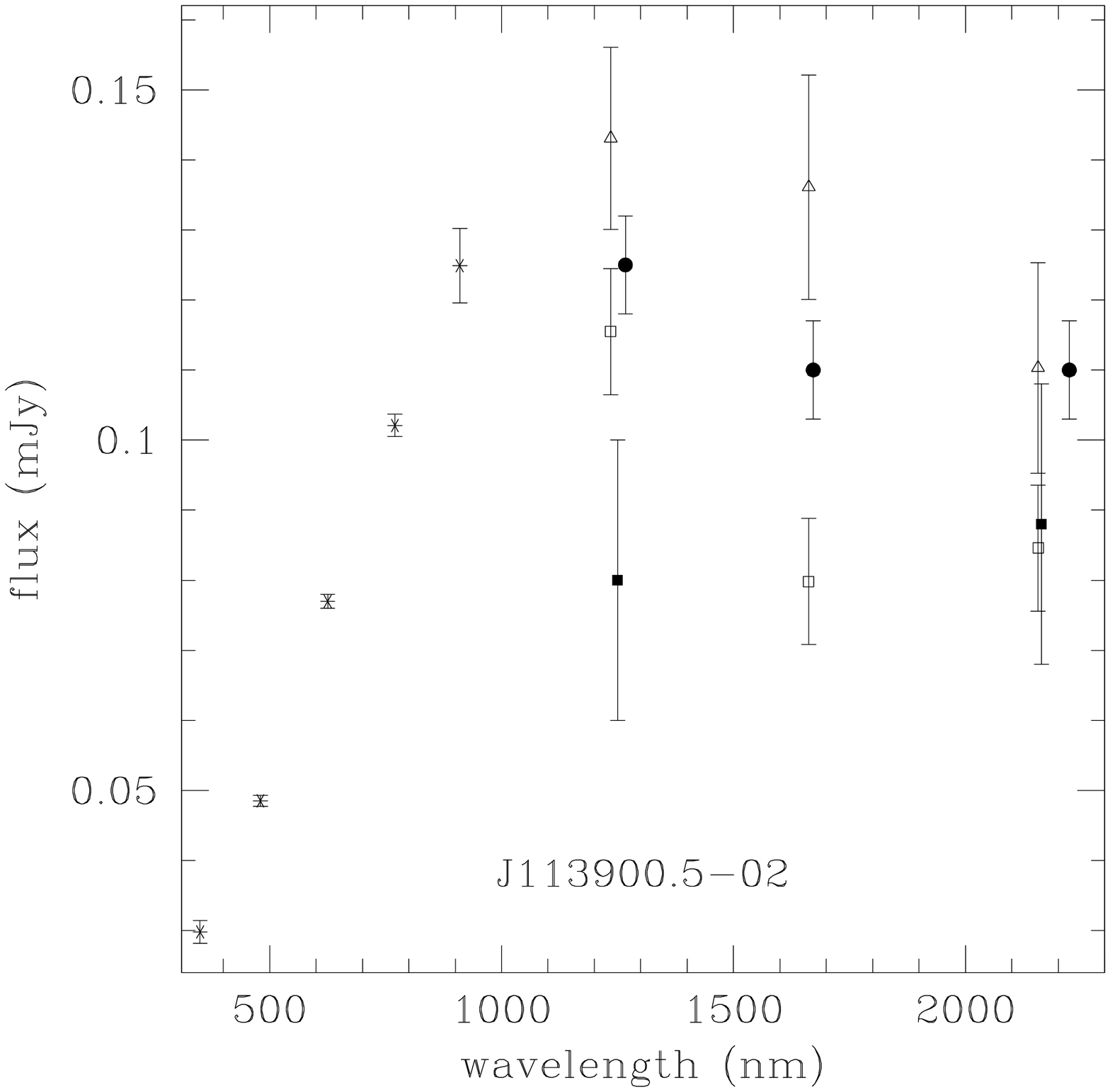,height=3.0in}\psfig{file=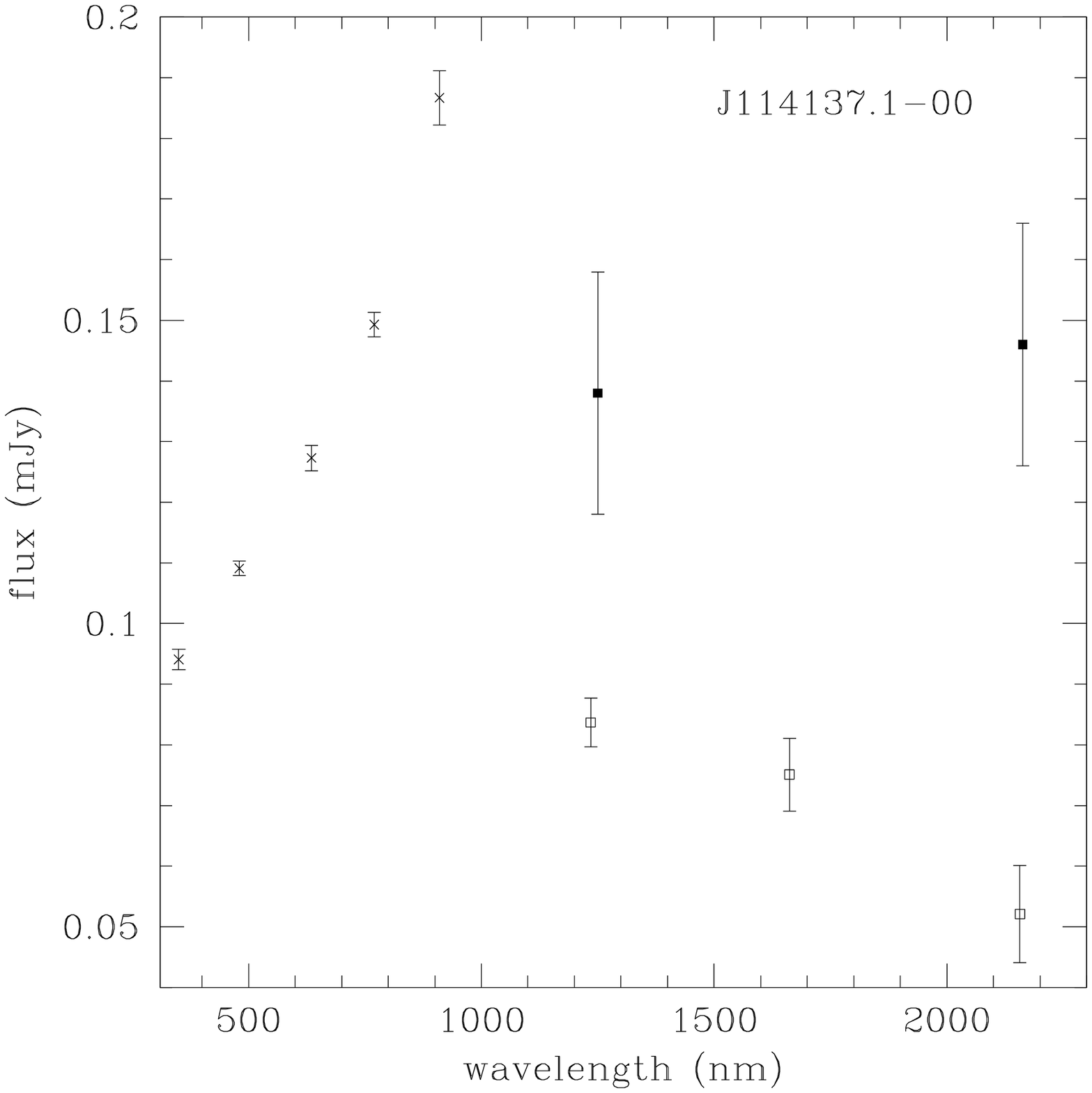,height=3.0in}}
\centerline{\psfig{file=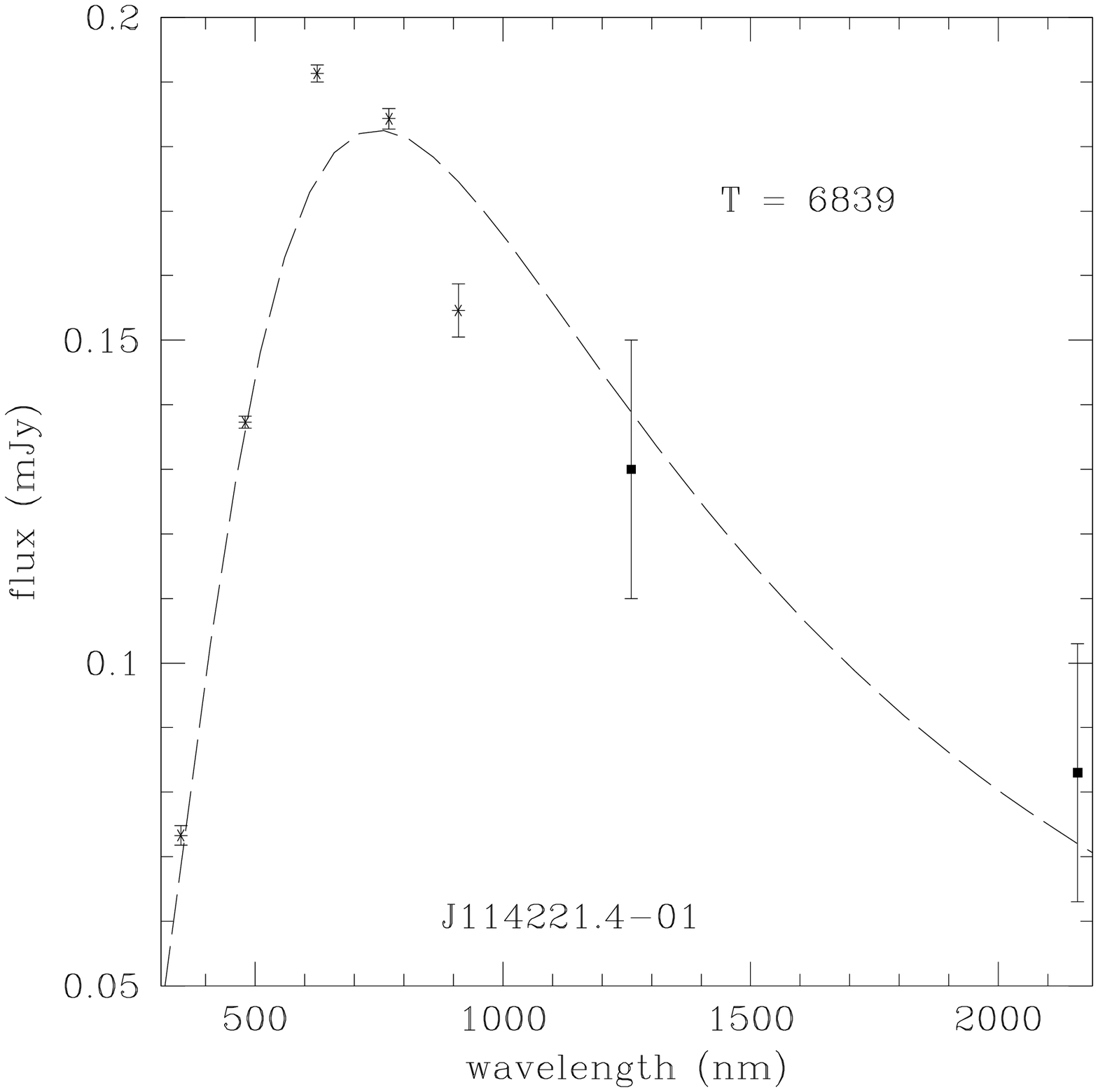,height=3.in}\psfig{file=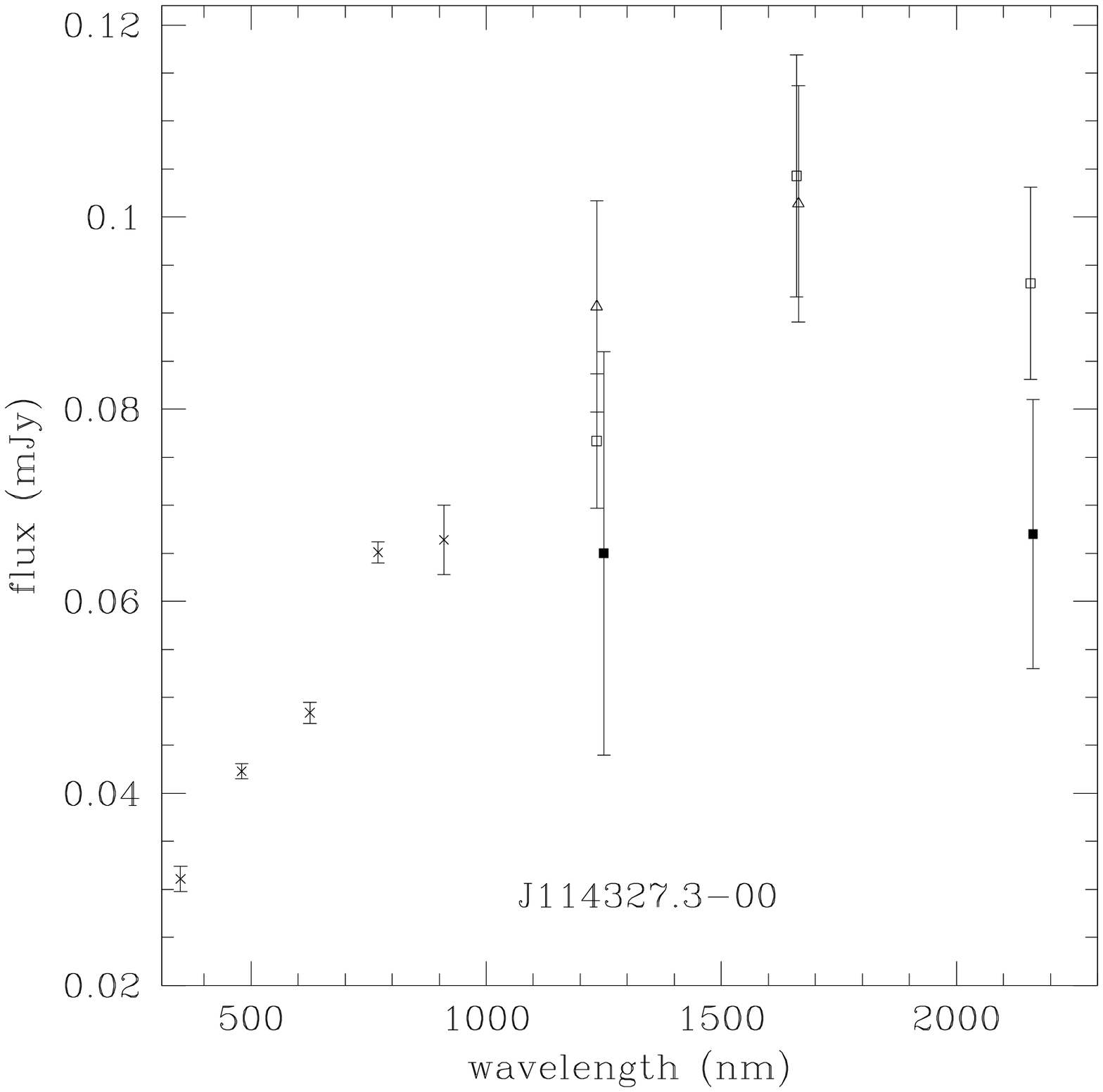,height=3.0in}}
\end{figure*}
\begin{figure*}
\centerline{\psfig{file=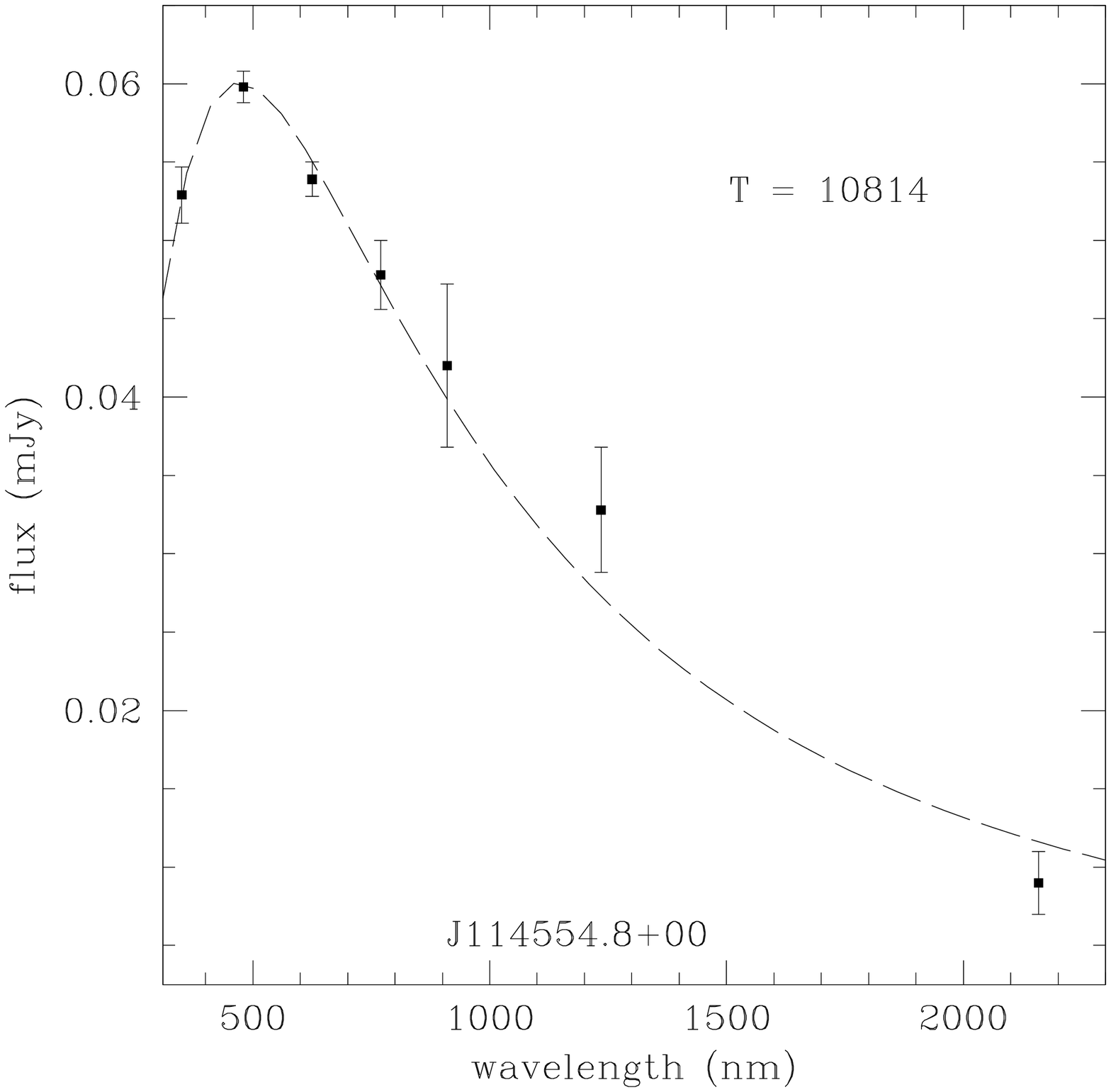,height=3.in}\psfig{file=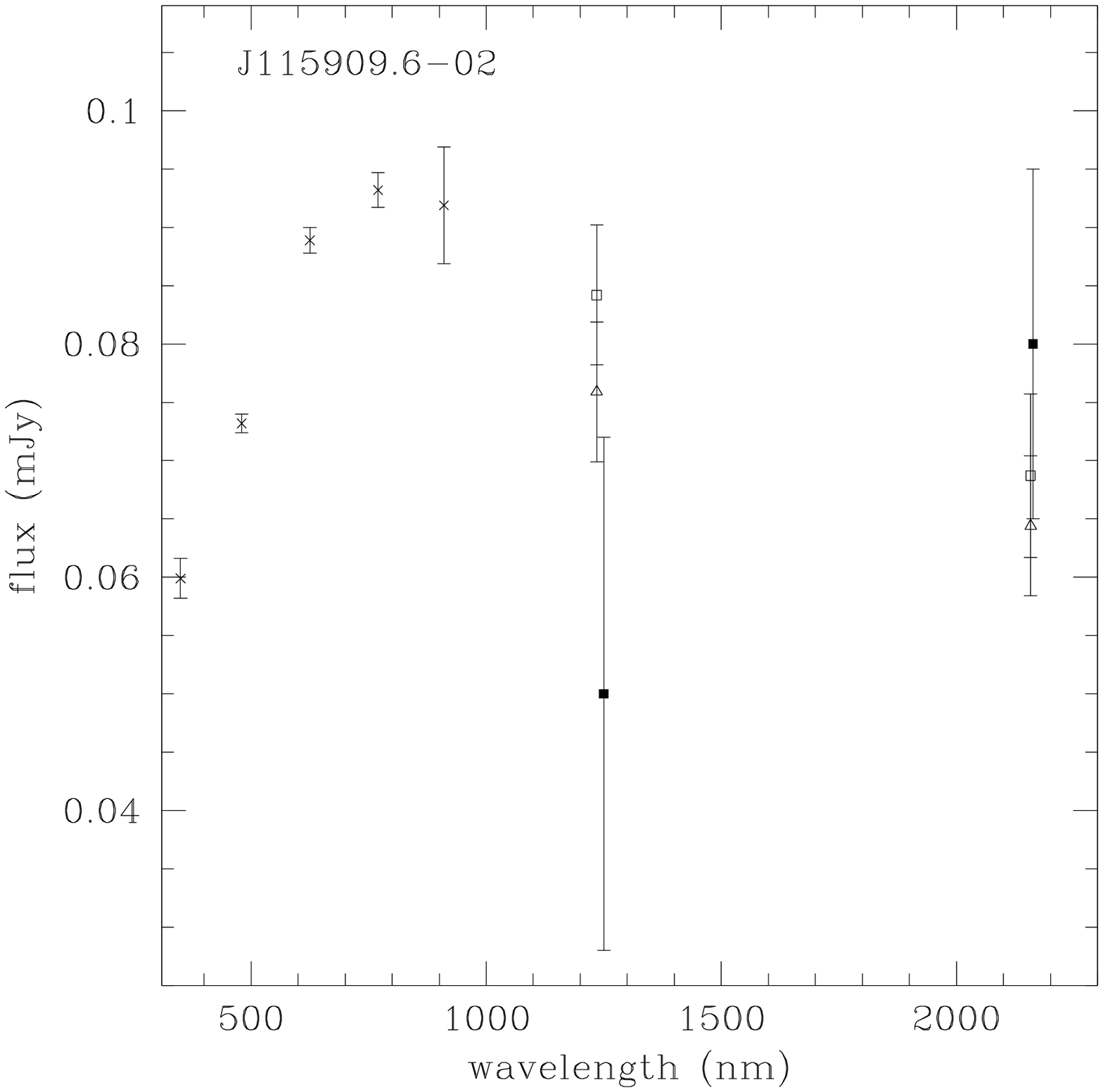,height=3.0in}}
\centerline{\psfig{file=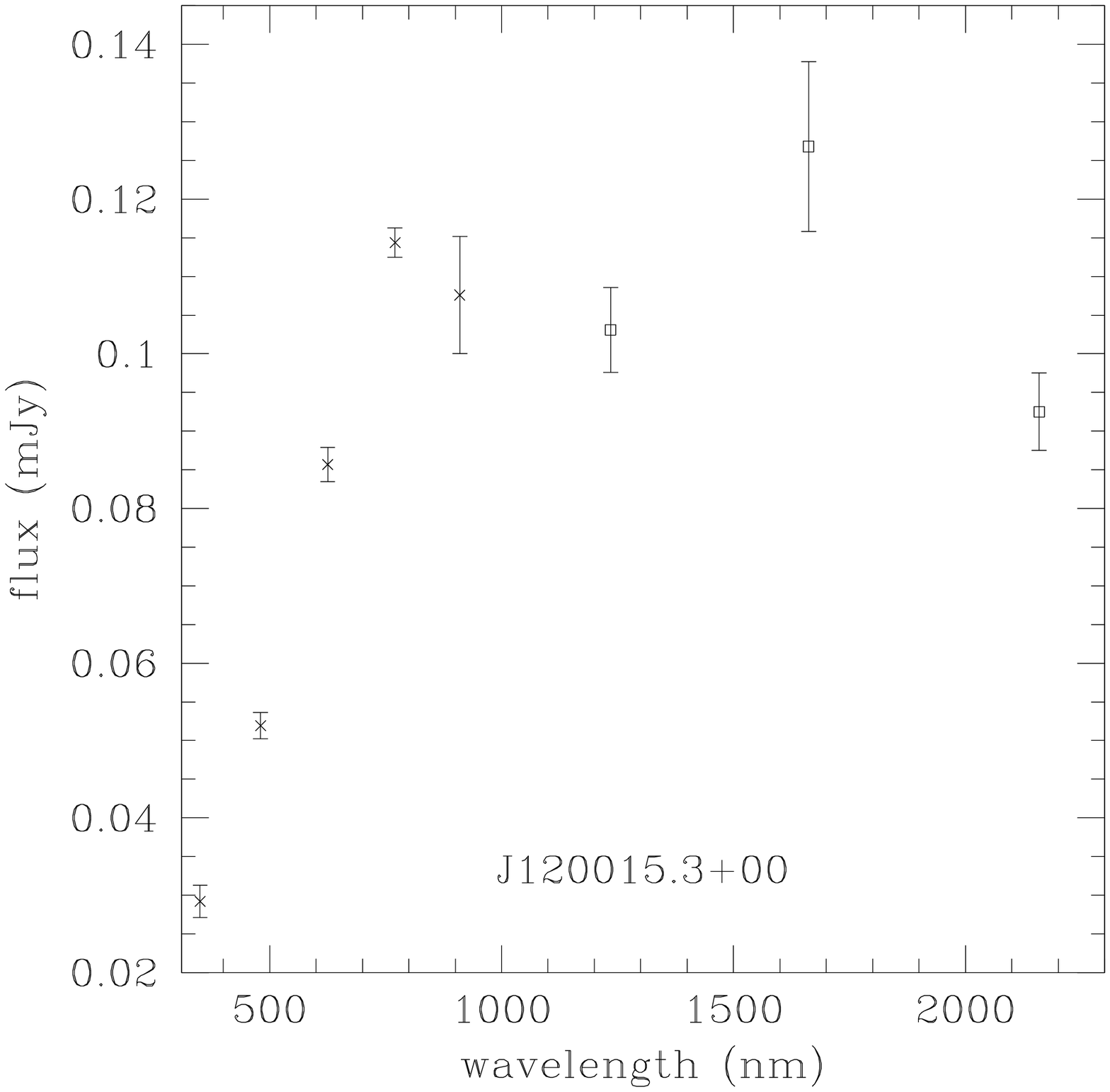,height=3.0in}\psfig{file=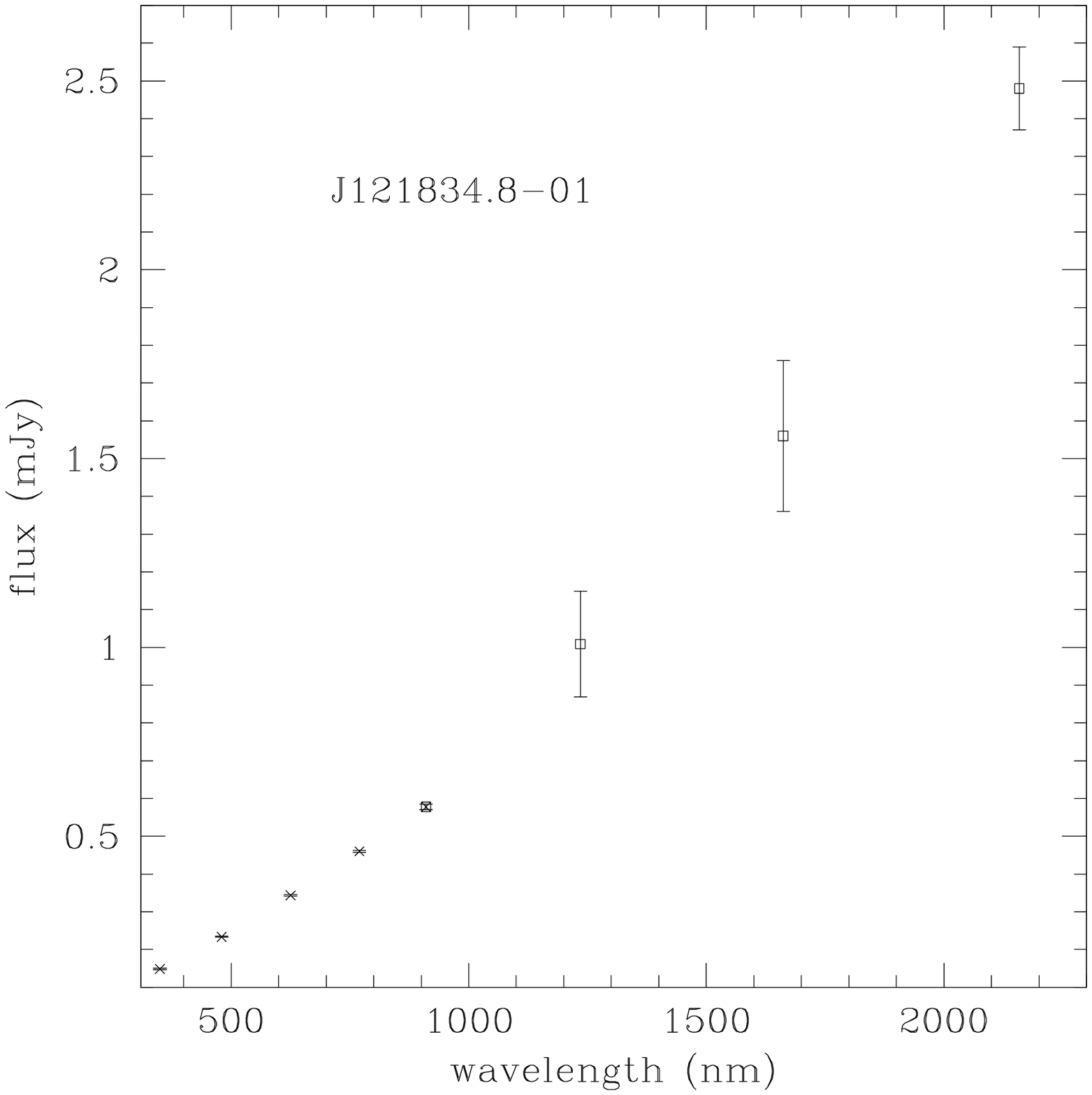,height=3.0in}}
\centerline{\psfig{file=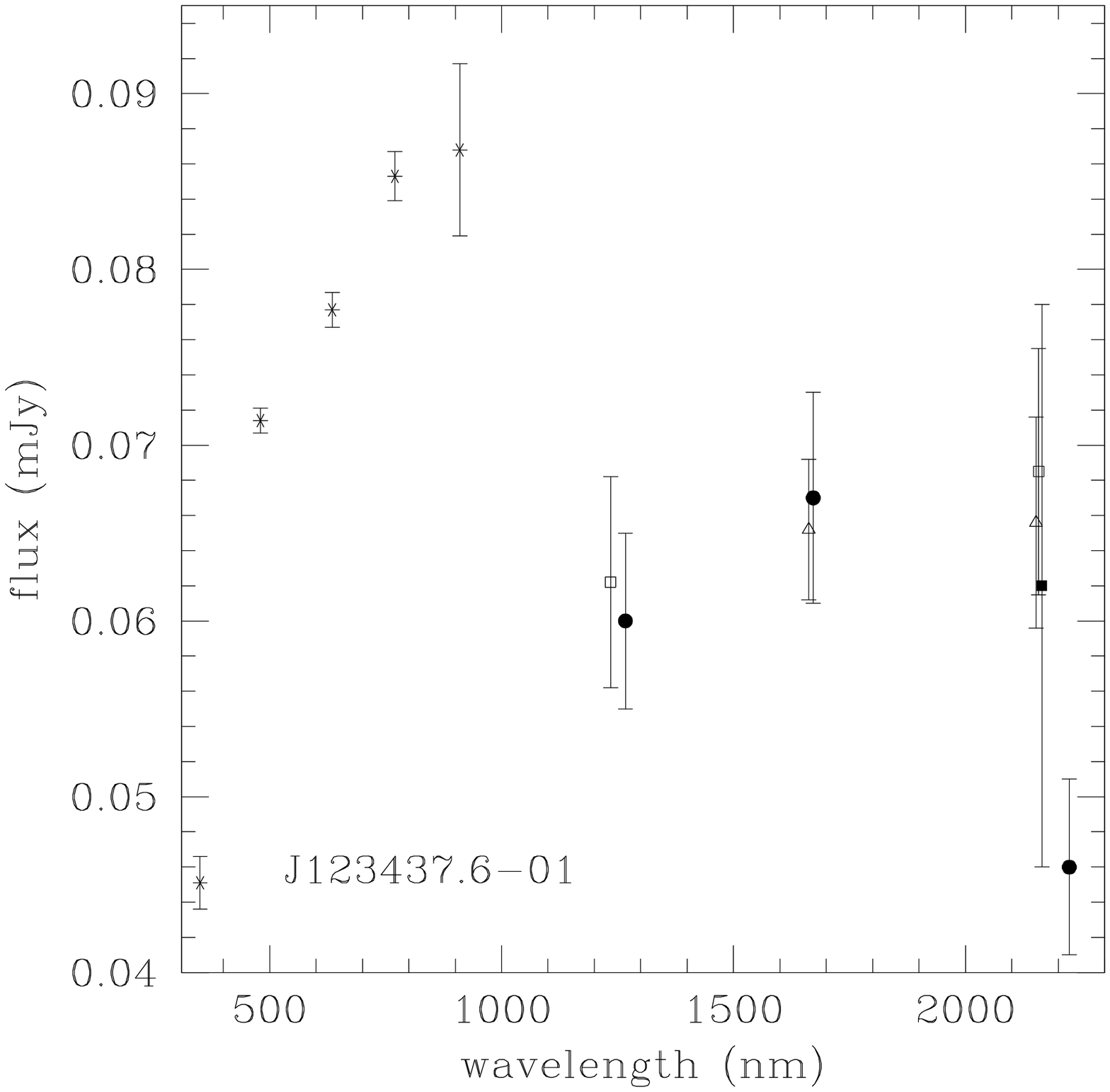,height=3.0in}}
\end{figure*}
\begin{figure*}
\centerline{\psfig{file=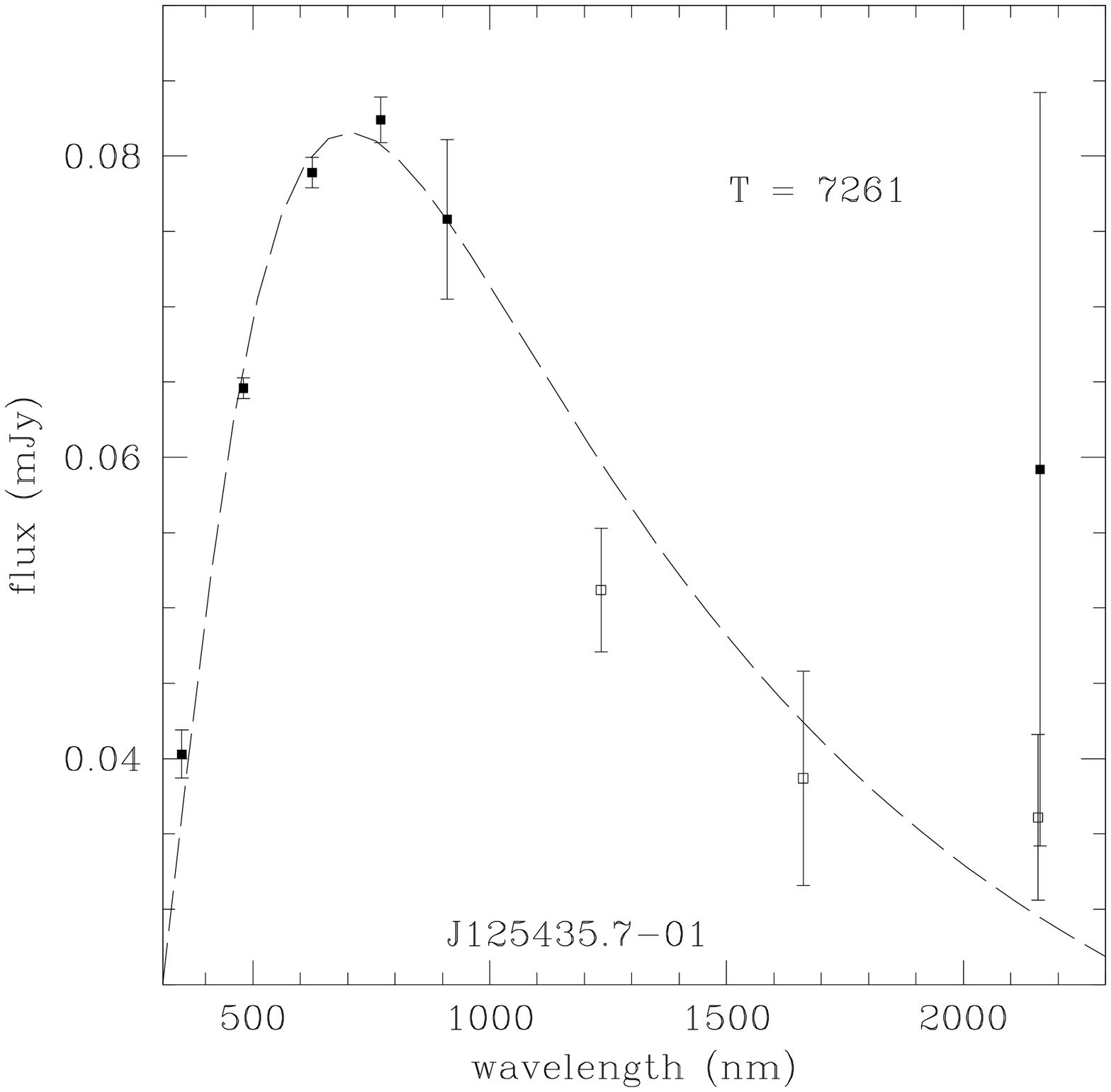,height=3.in}}
\centerline{\psfig{file=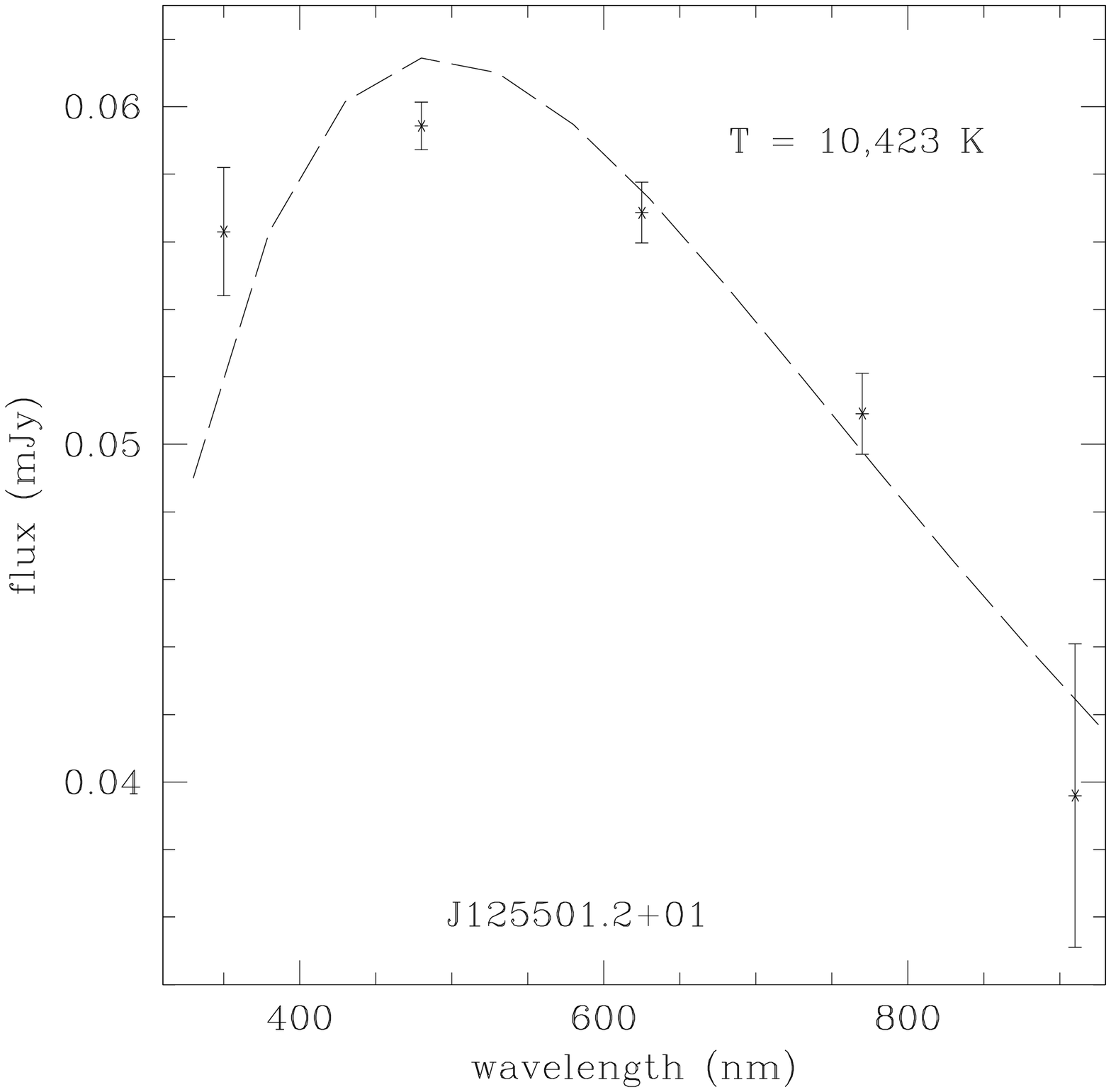,height=3.in}\psfig{file=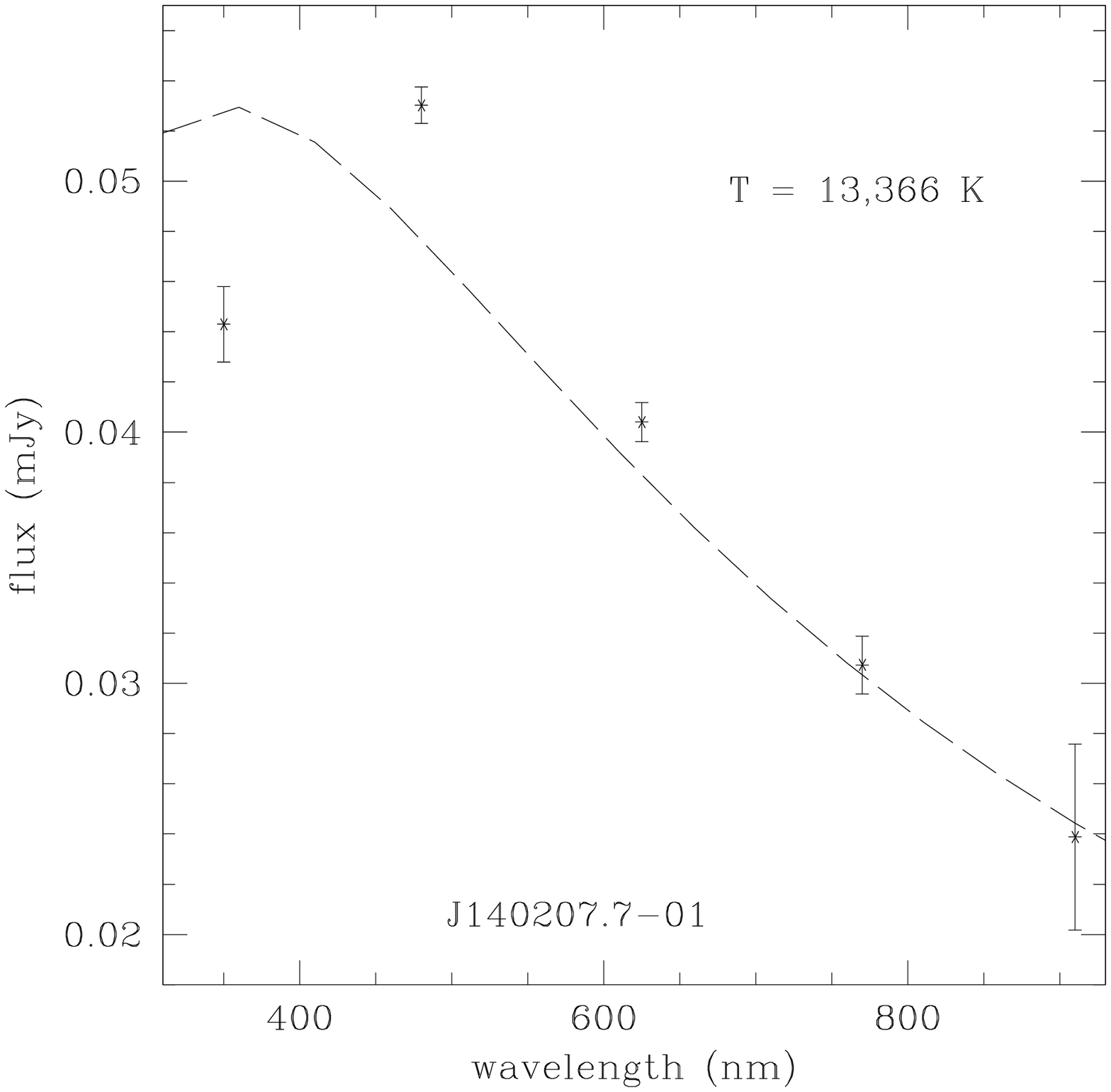,height=3.in}}
\centerline{\psfig{file=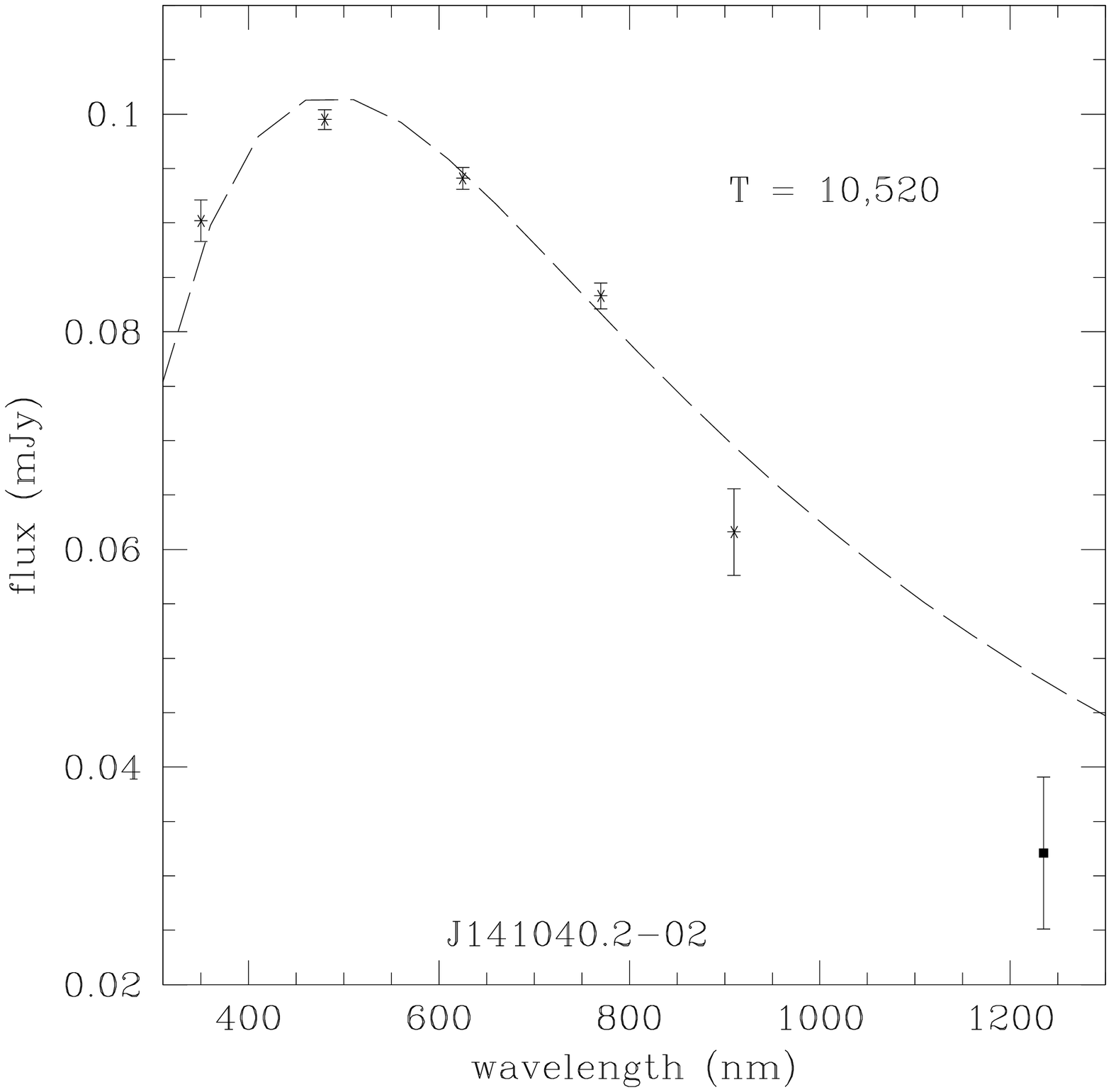,height=3.in}\psfig{file=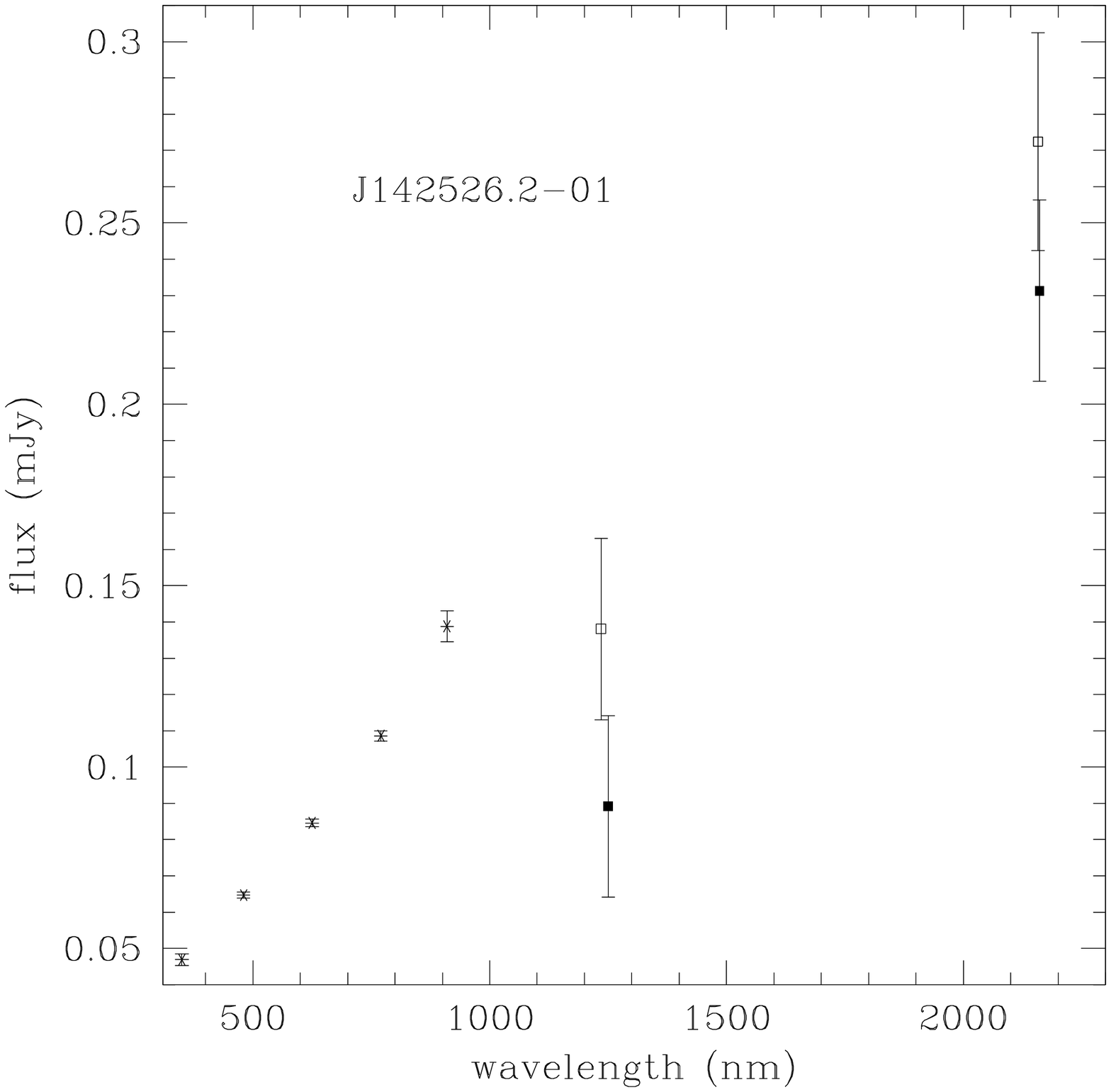,height=3.0in}}
\end{figure*}

\section{Comparison of 2dF/6dF and VLT spectra}
\begin{figure*}
\centerline{\psfig{file=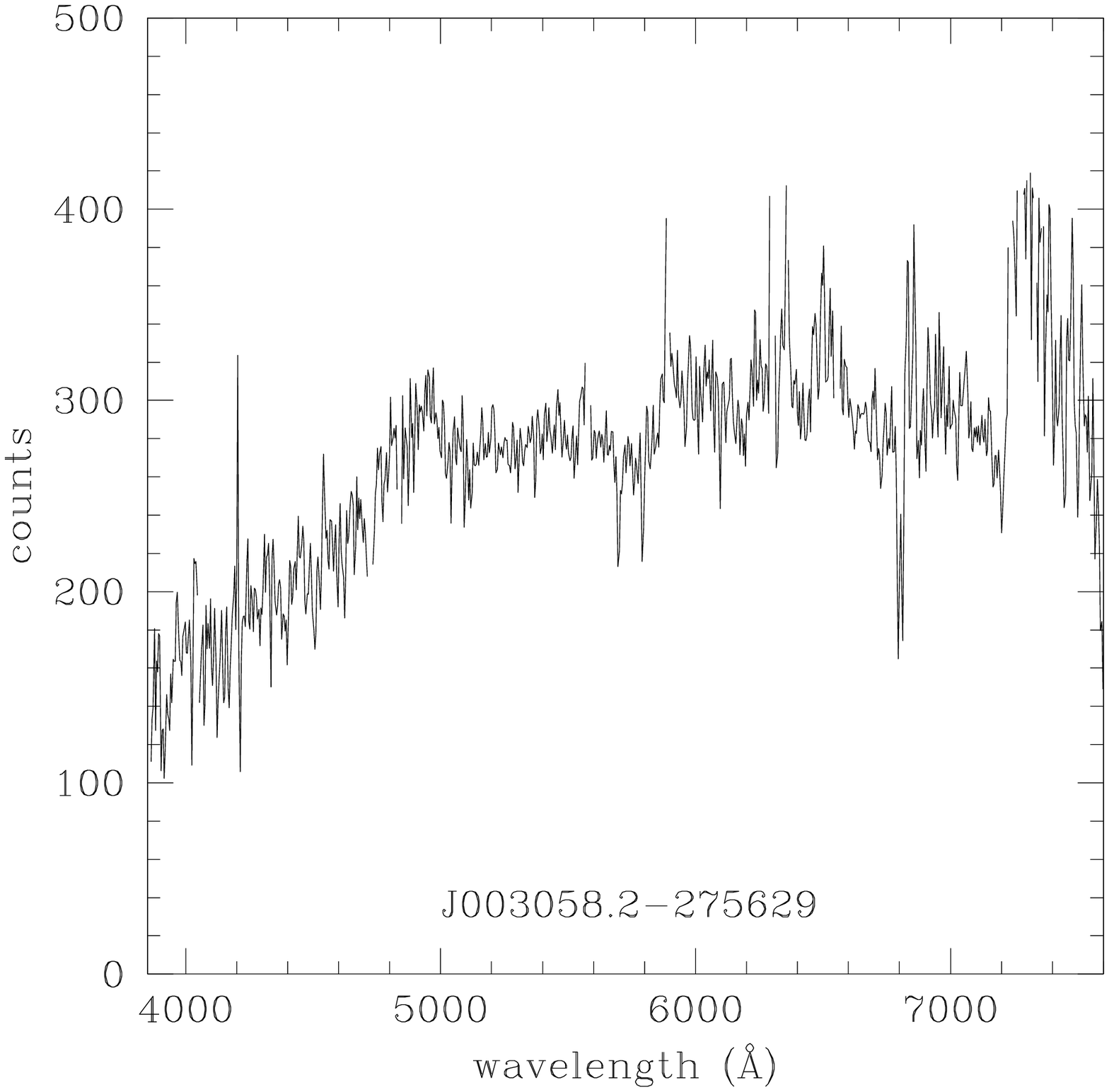,width=2.5in}\psfig{file=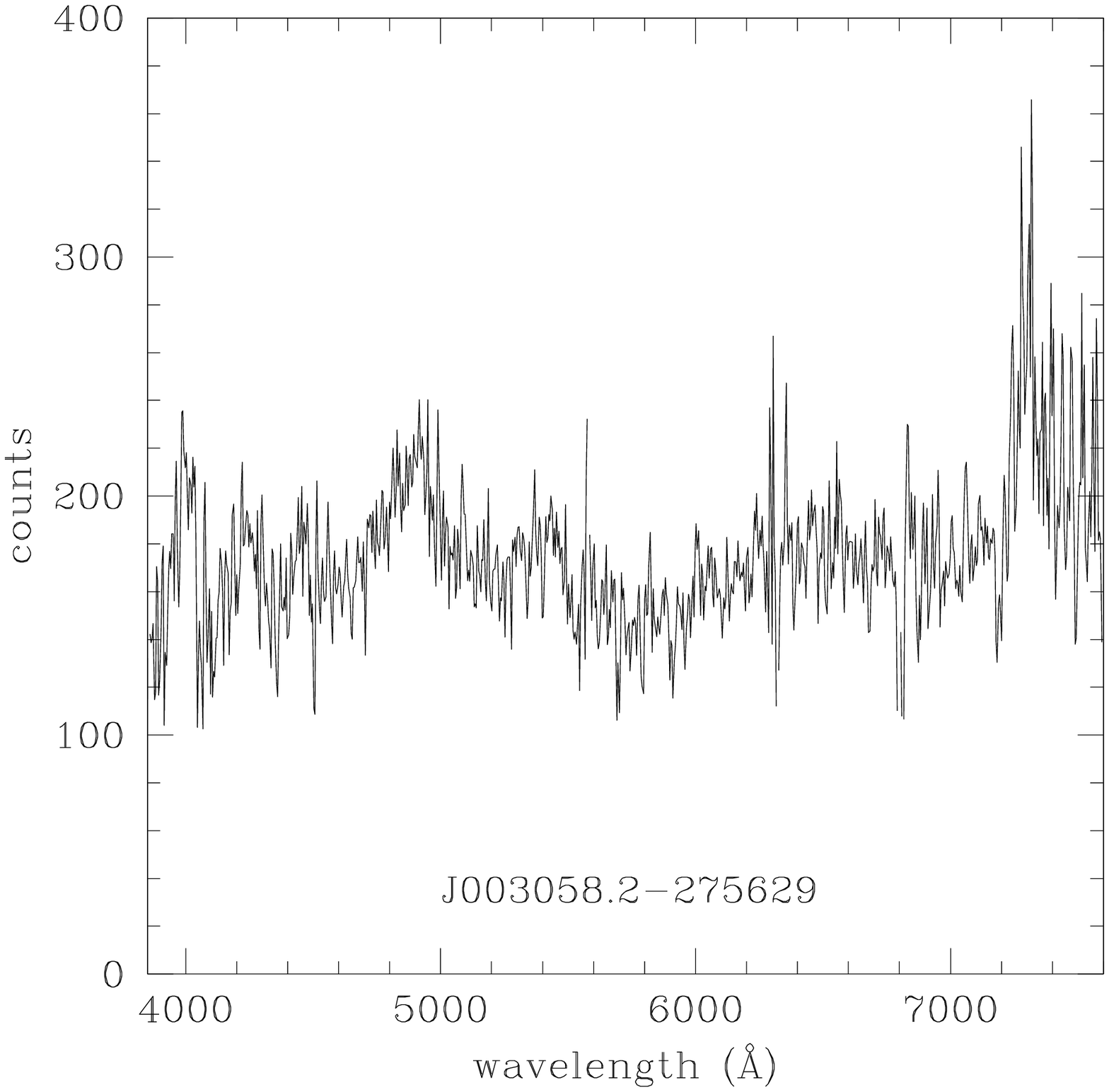,width=2.5in}}
\centerline{\psfig{file=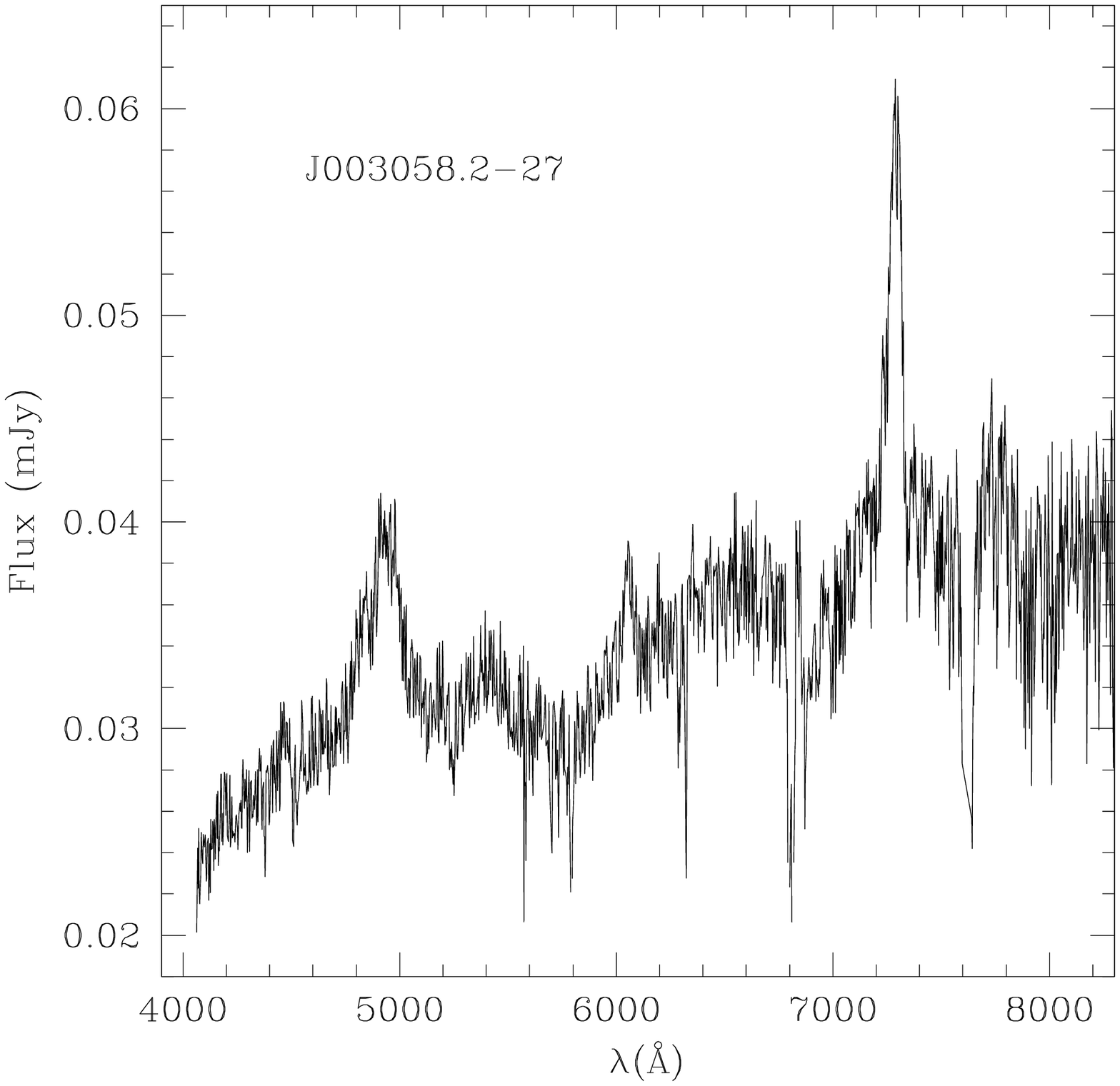,width=2.5in}}
\caption{2dF spectra (top) and VLT spectrum of the radio-quiet object J003058.2$-$275629.}
\end{figure*}
\begin{figure*}
\centerline{\psfig{file=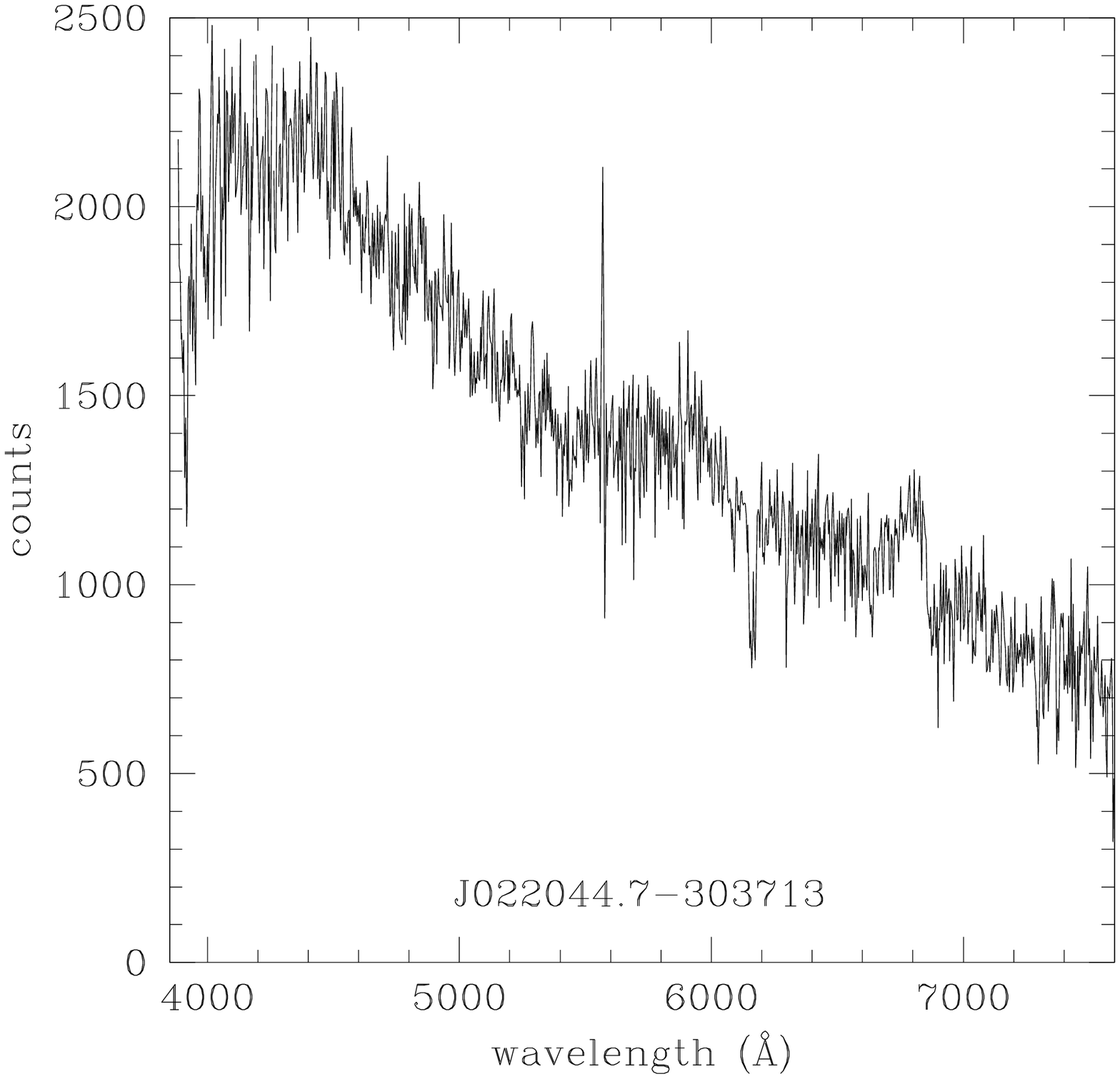,width=2.5in}\psfig{file=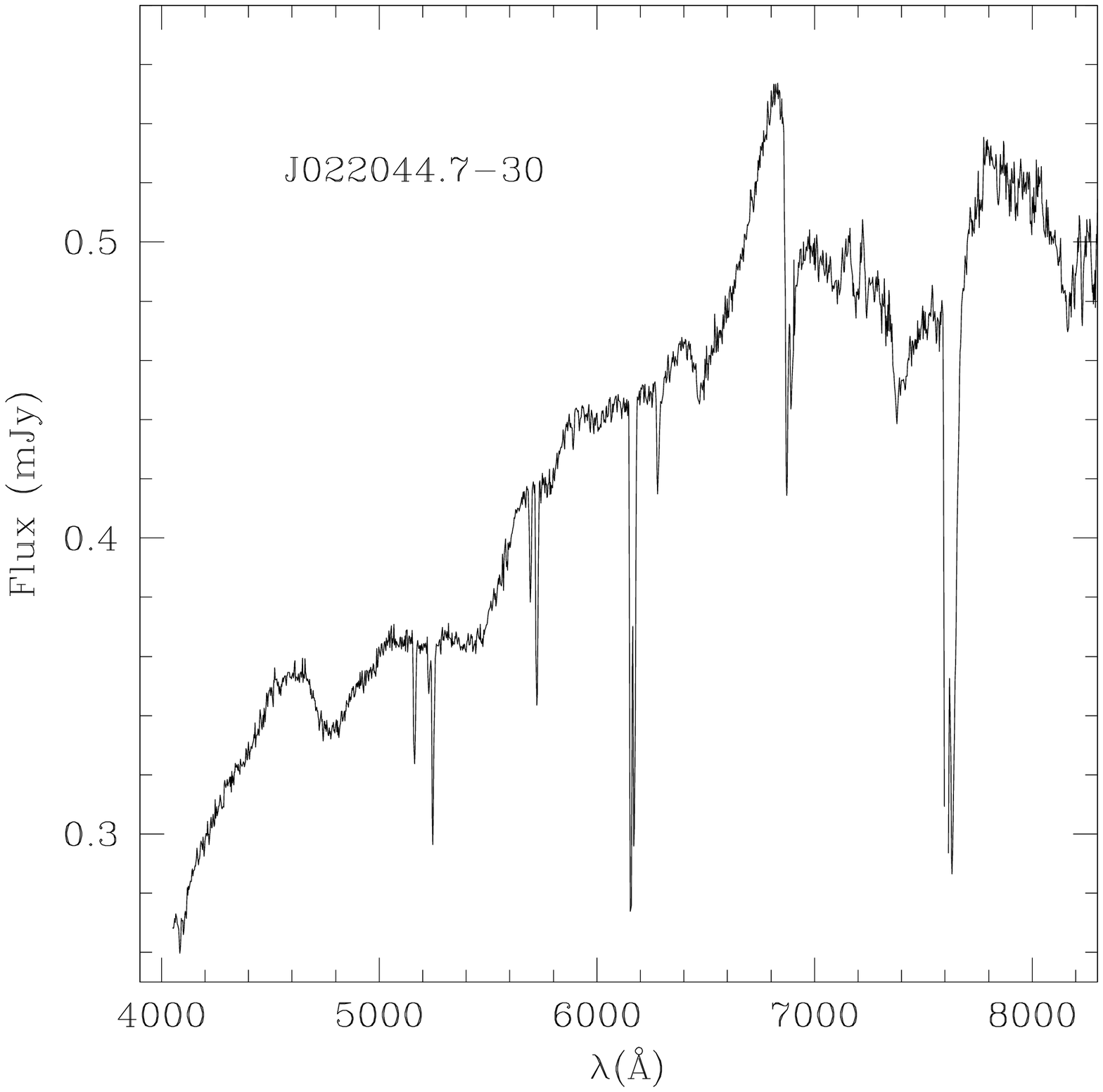,width=2.5in}}
\centerline{\psfig{file=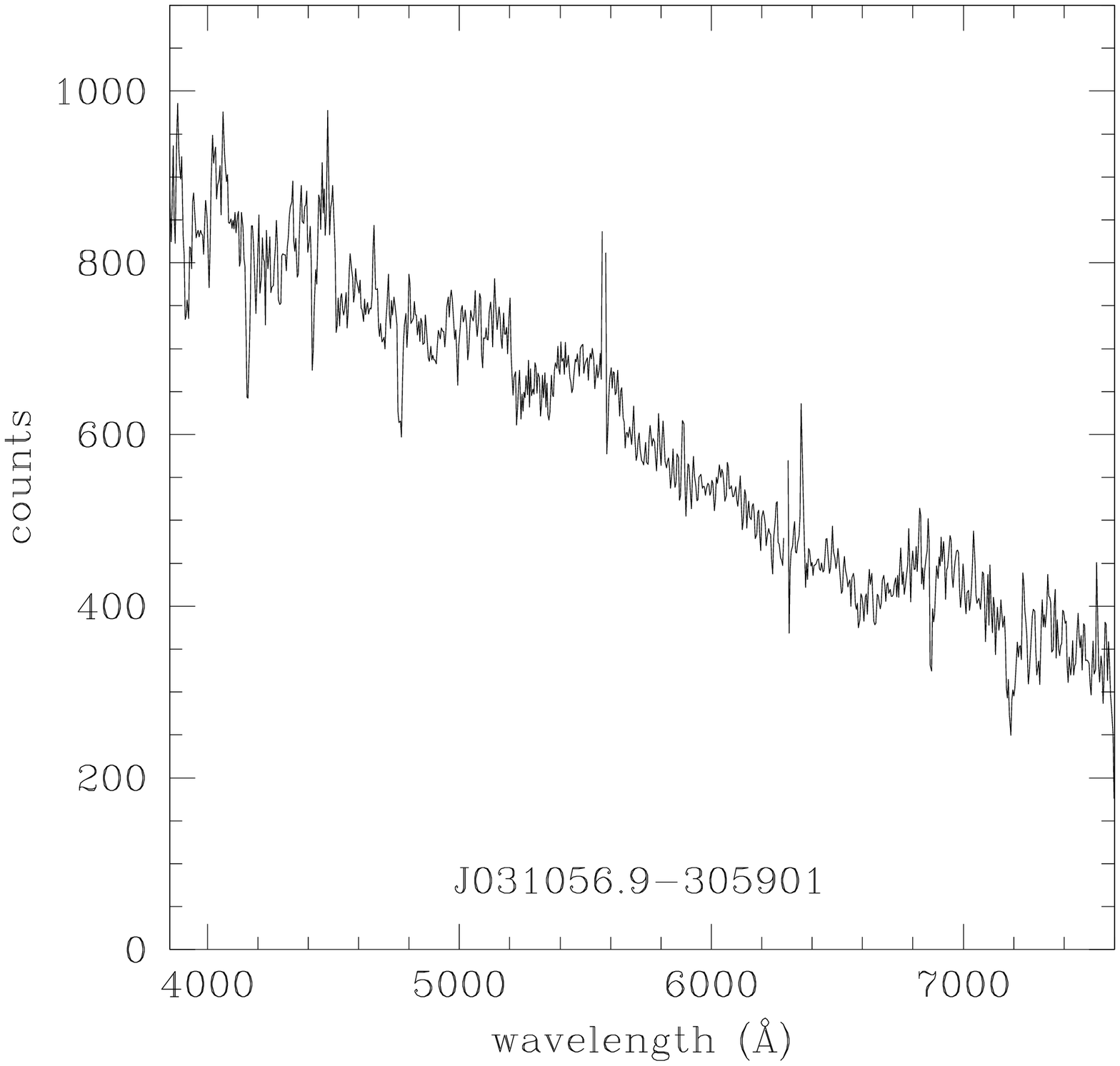,width=2.5in}\psfig{file=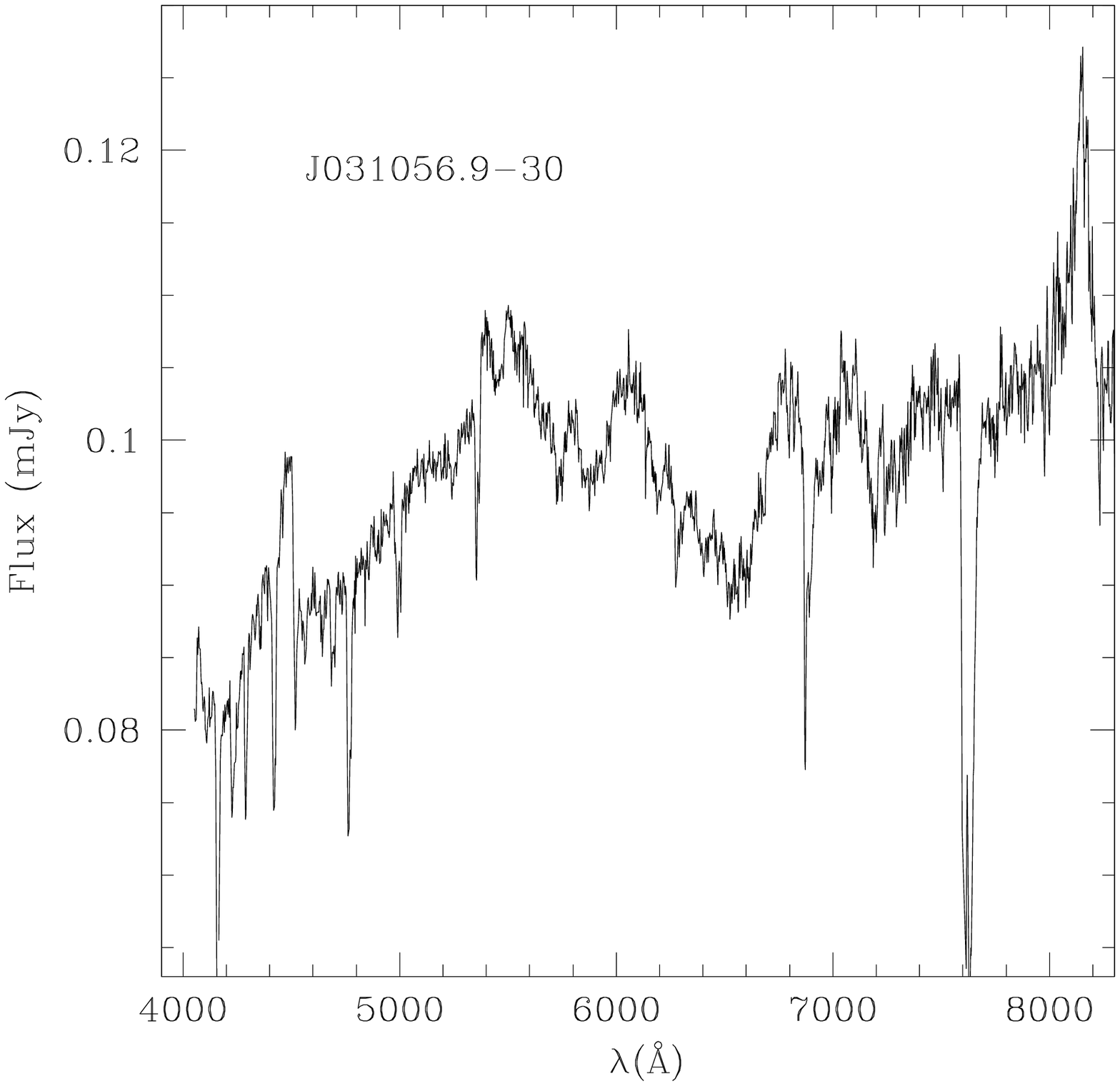,width=2.3in}}
\centerline{\psfig{file=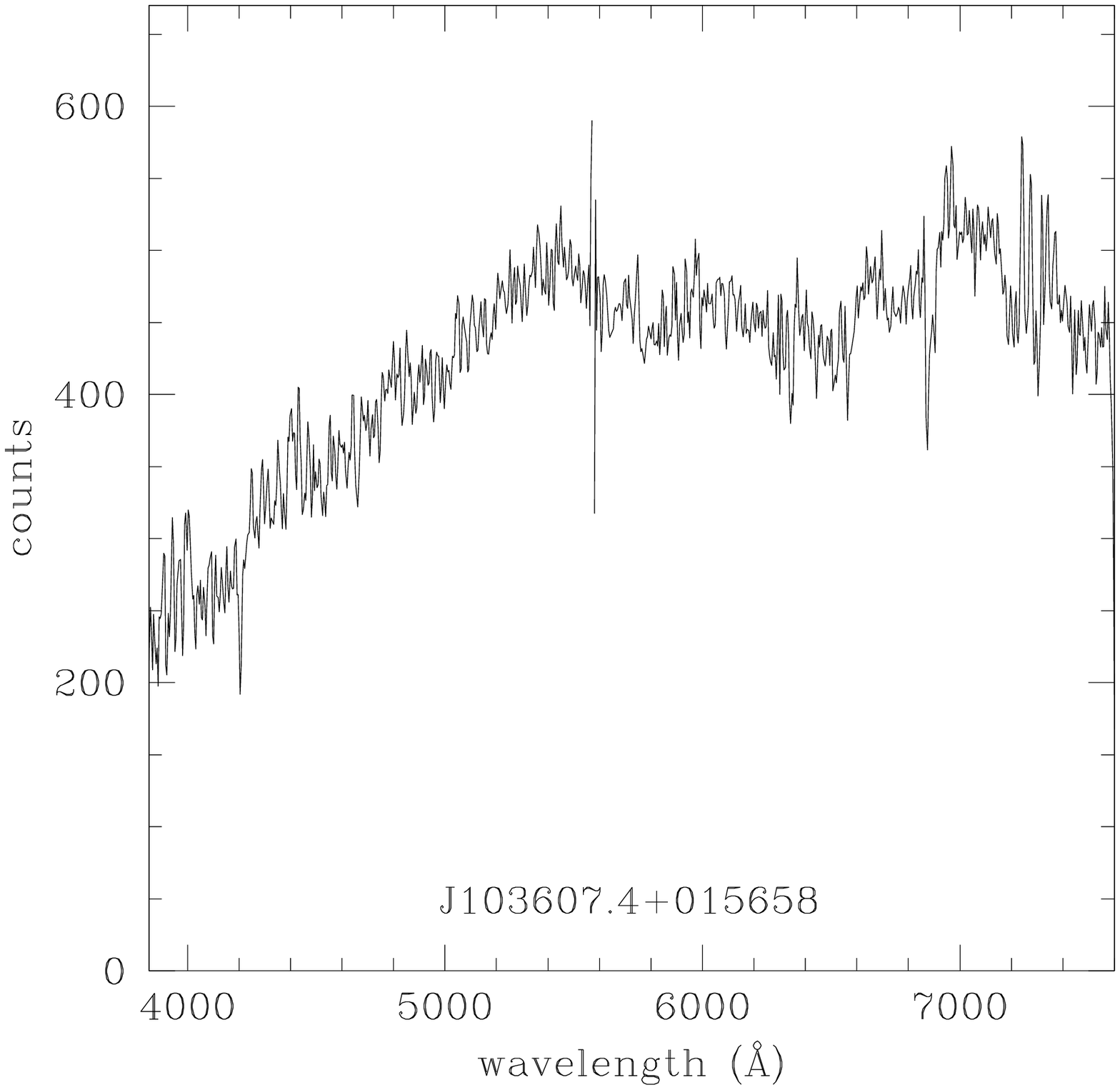,width=2.5in}\psfig{file=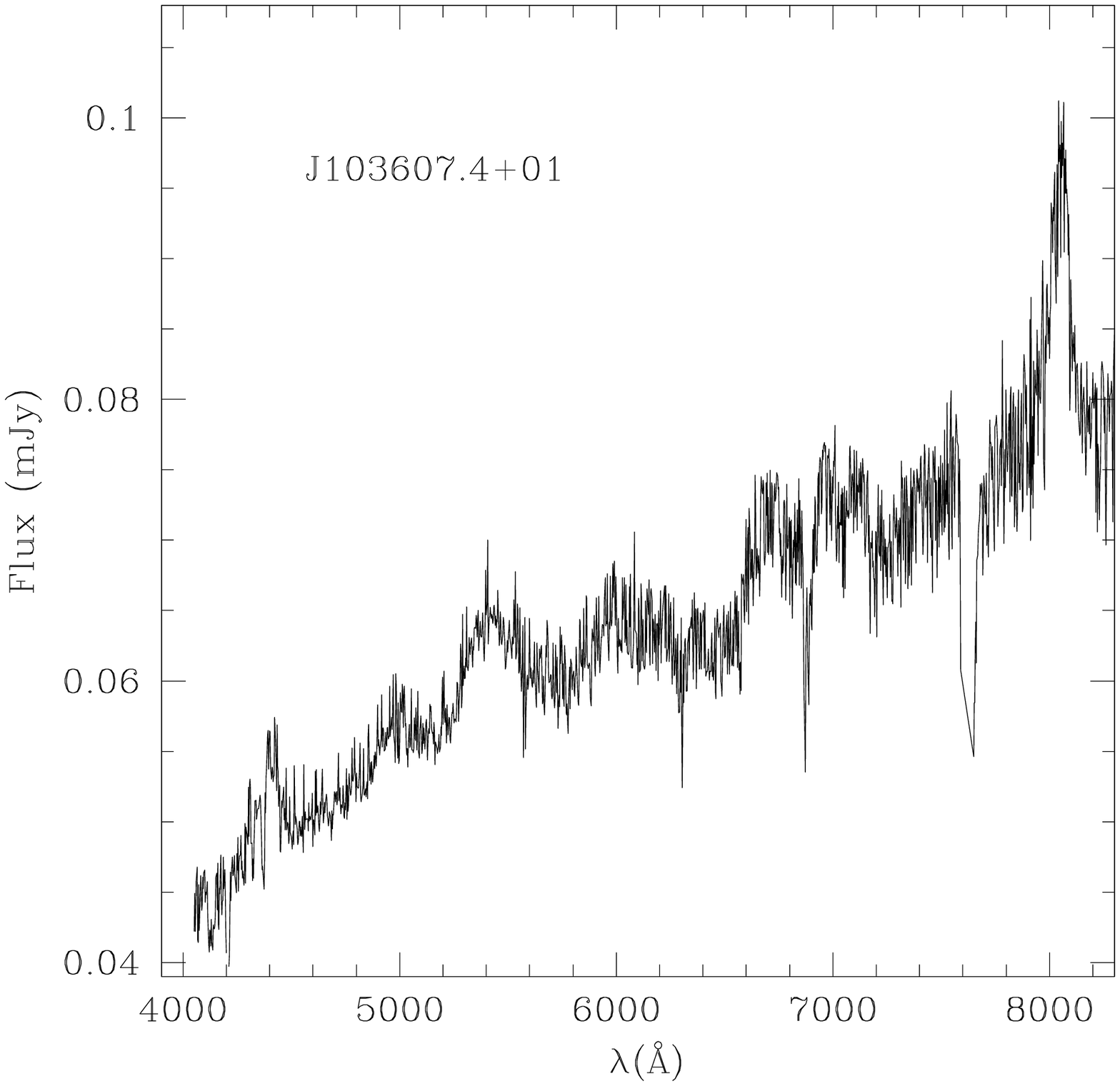,width=2.4in}}
\caption{Top: 6dF and VLT spectrum of the radio-quiet object J022044.7$-$30. Centre: 2dF and VLT spectrum of the radio-detected object J031056.9$-$30. Bottom: 2dF and VLT spectrum of the radio-quiet object J103607.4+01.}
\end{figure*}
\begin{figure*}
\centerline{\psfig{file=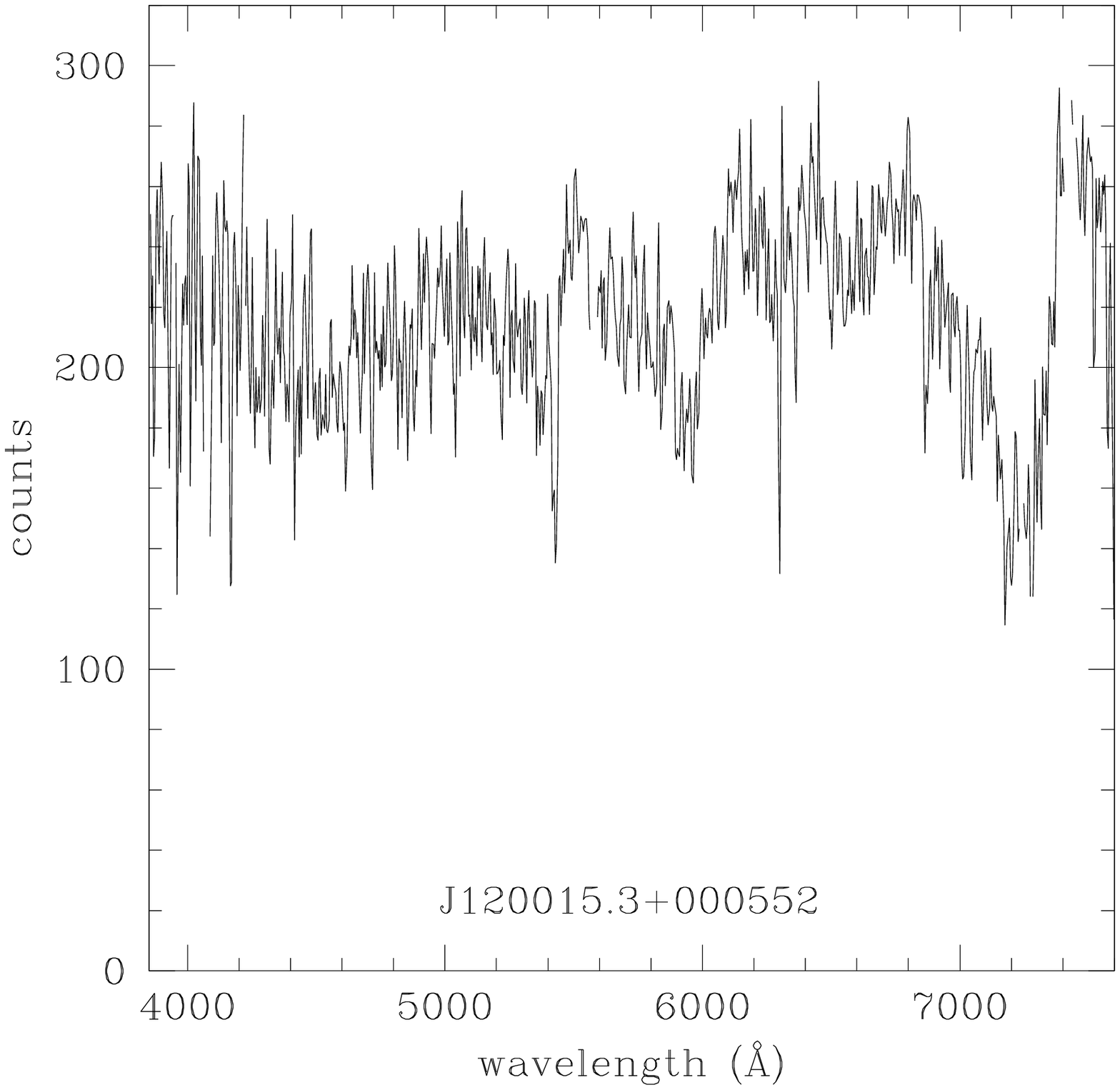,width=2.5in}\psfig{file=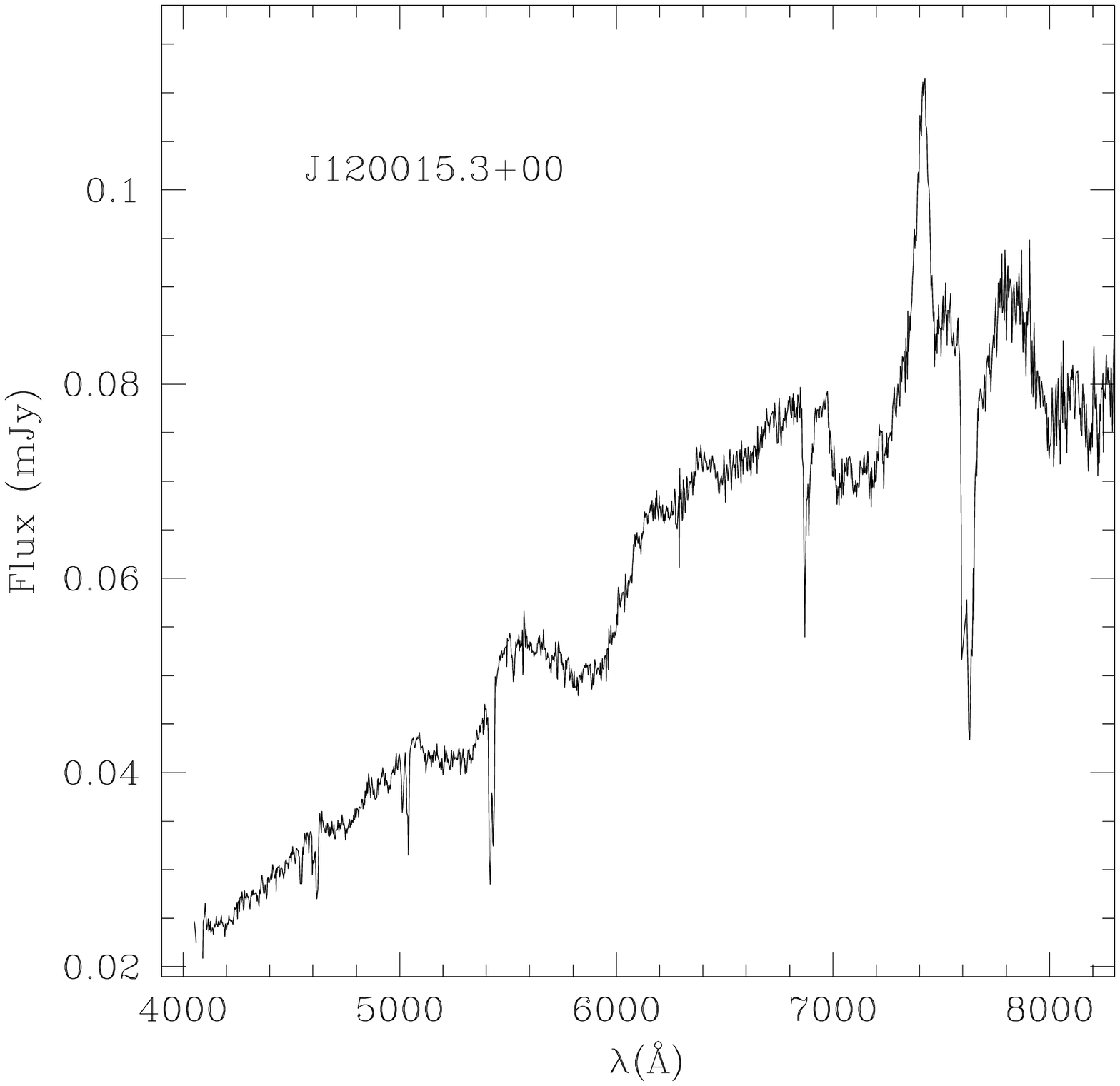,width=2.4in}}
\centerline{\psfig{file=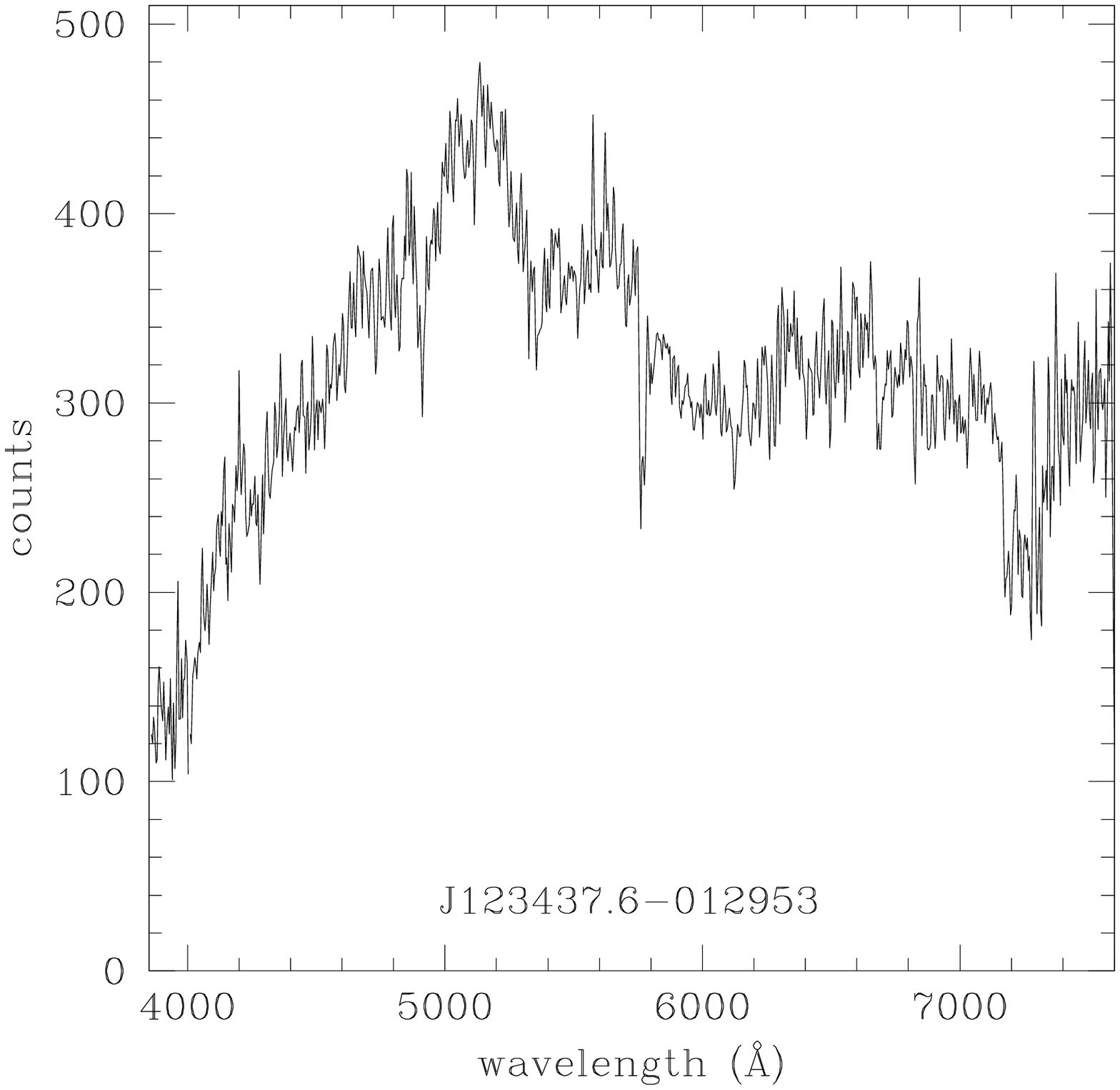,width=2.5in}\psfig{file=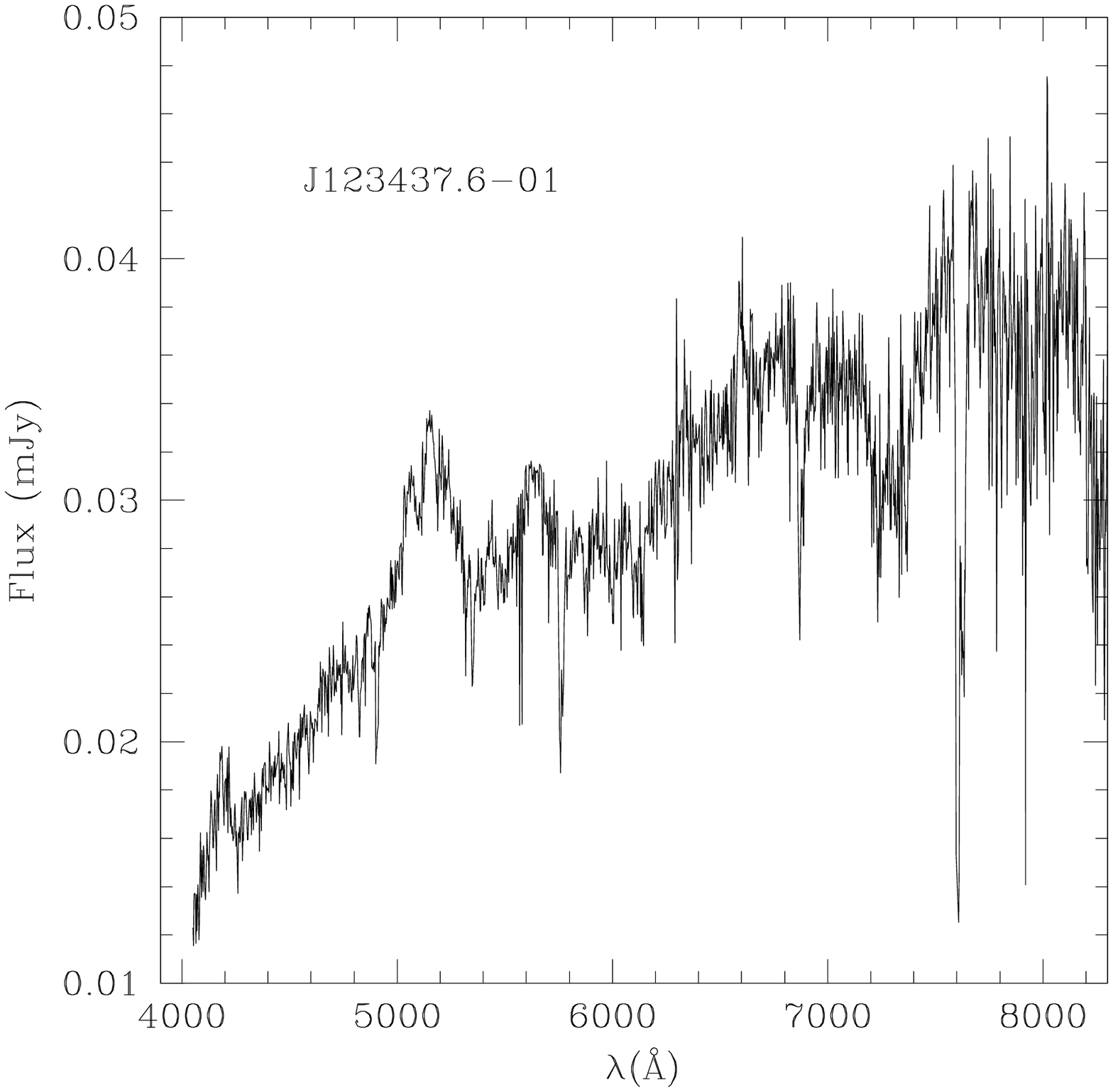,width=2.4in}}
\centerline{\psfig{file=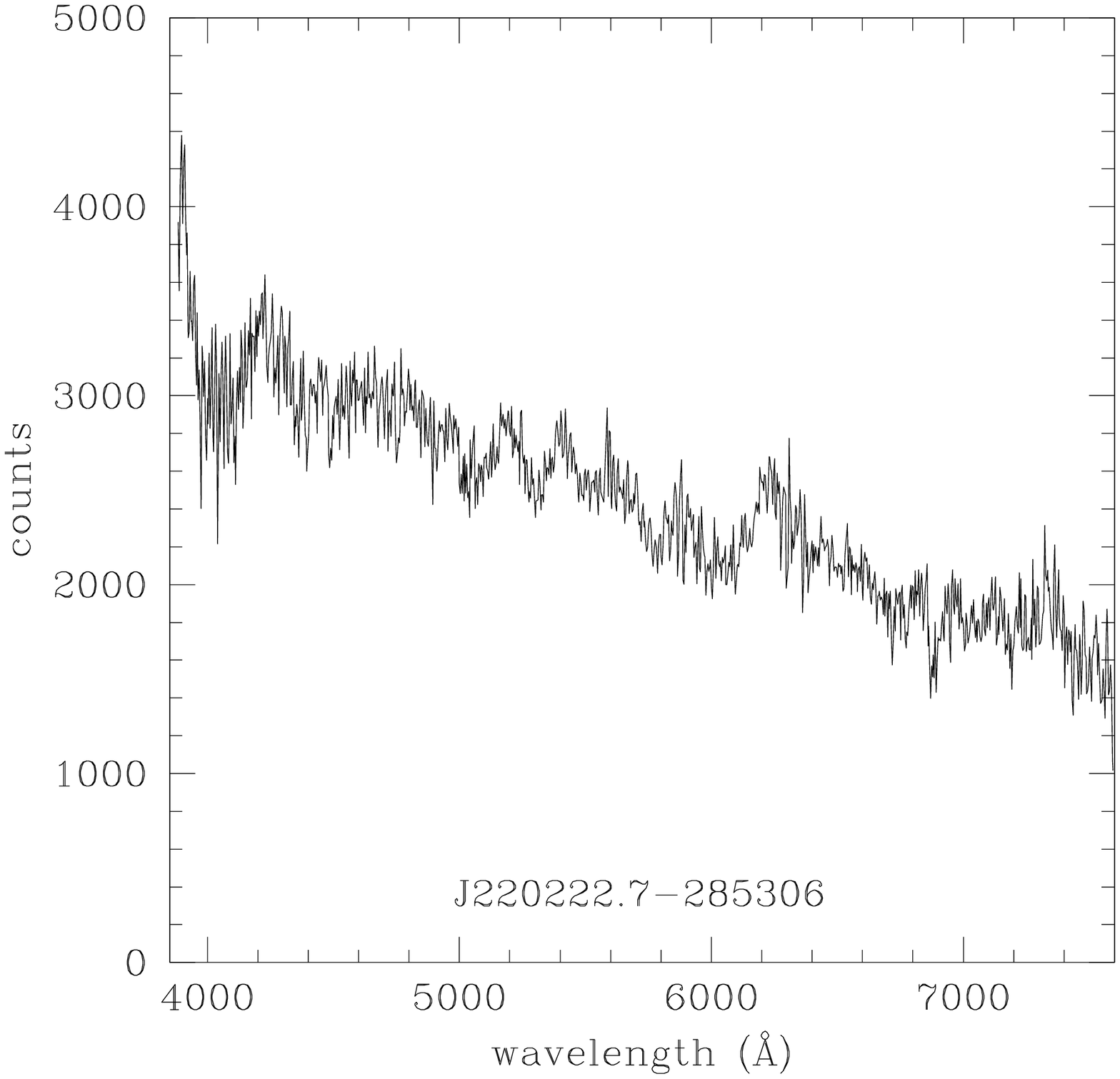,width=2.5in}\psfig{file=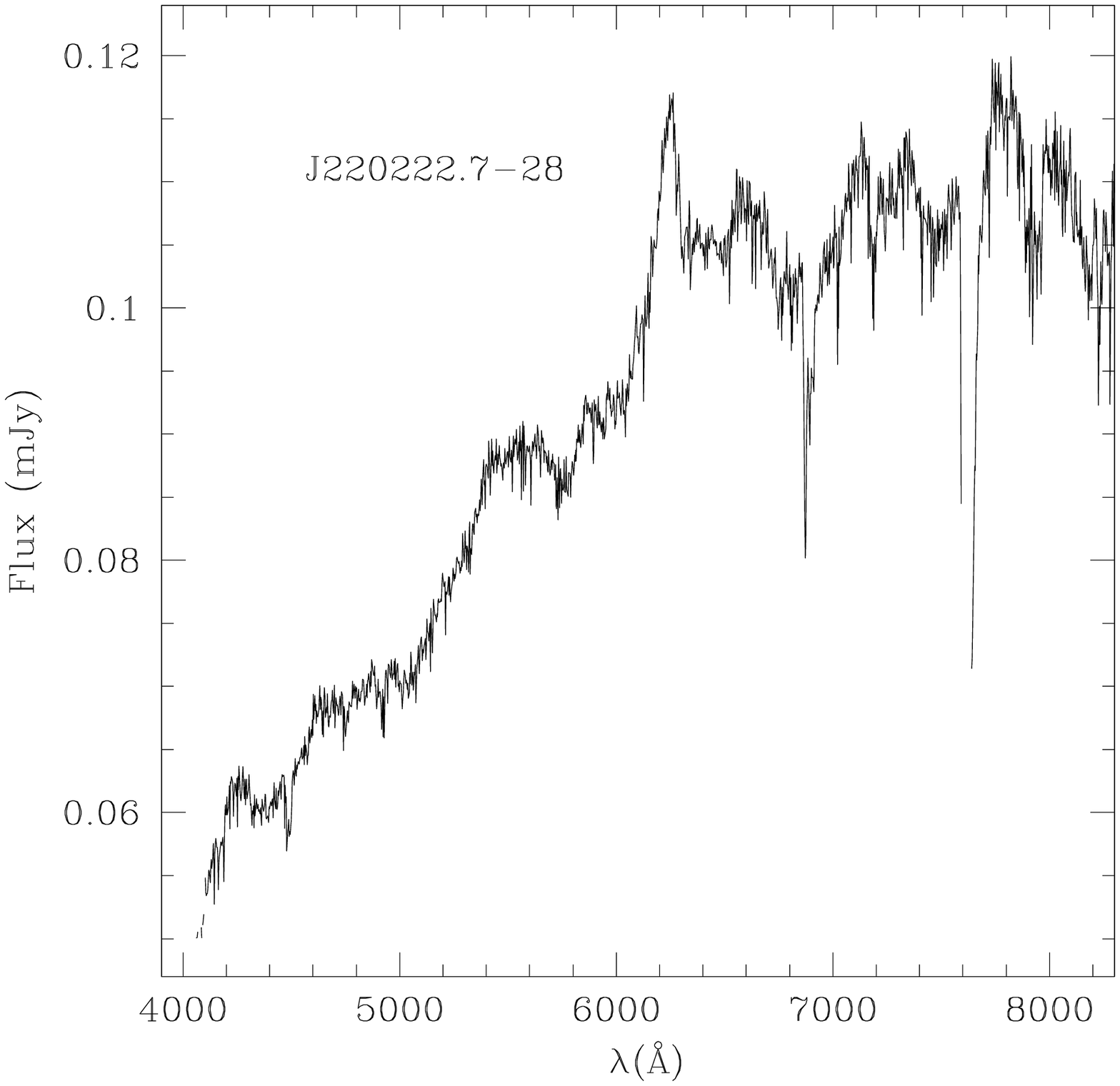,width=2.4in}}
\caption{Top: 2dF and VLT spectrum of the radio-quiet object J120015.3+00. Centre: 2dF and VLT spectrum of the radio-quiet object J123437.6$-$01. Bottom: 6dF spectra and VLT spectrum of the radio-quiet object J220222.7$-$28.}
\end{figure*}

\begin{figure*}
\centerline{\psfig{file=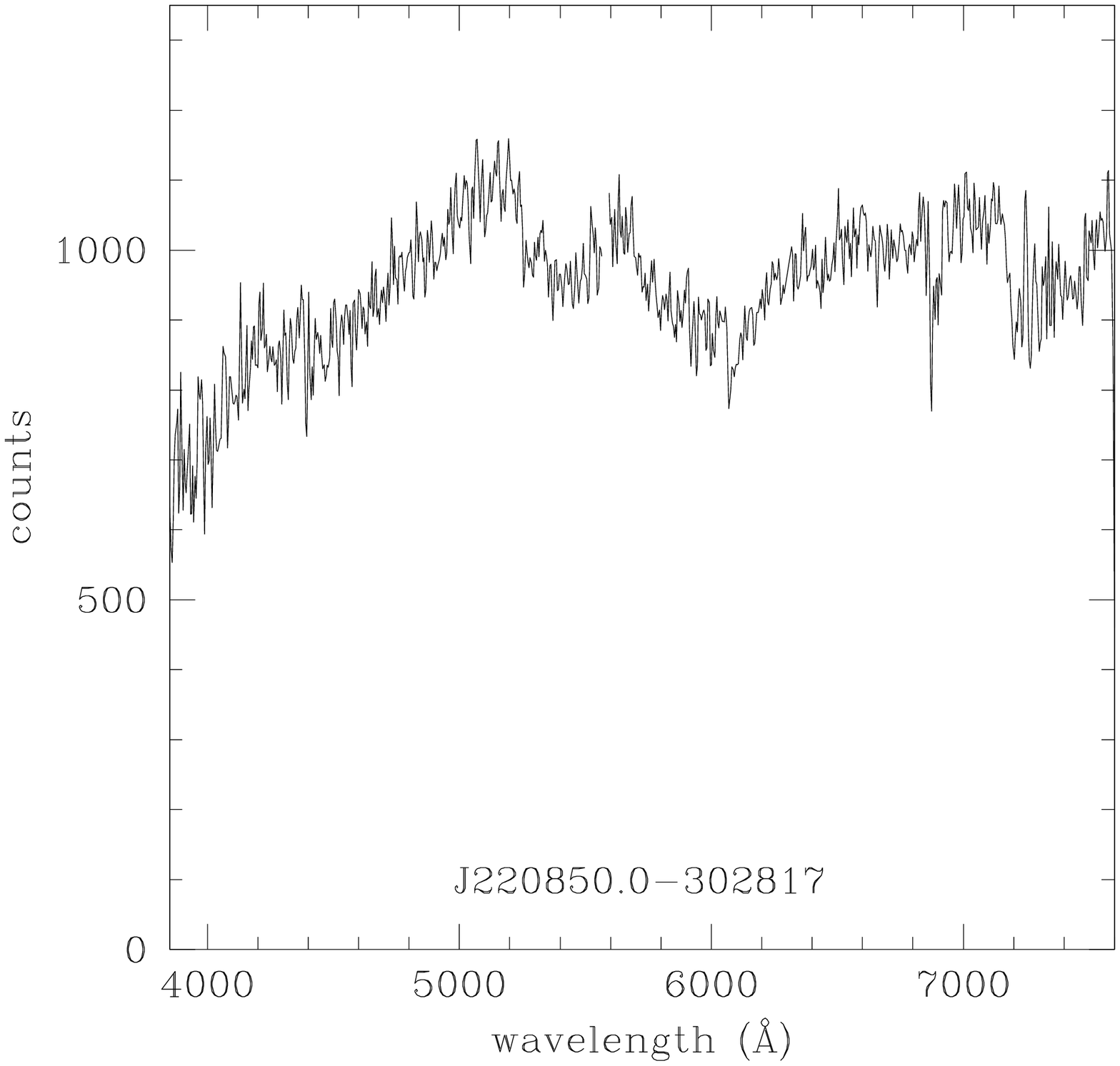,width=2.5in}\psfig{file=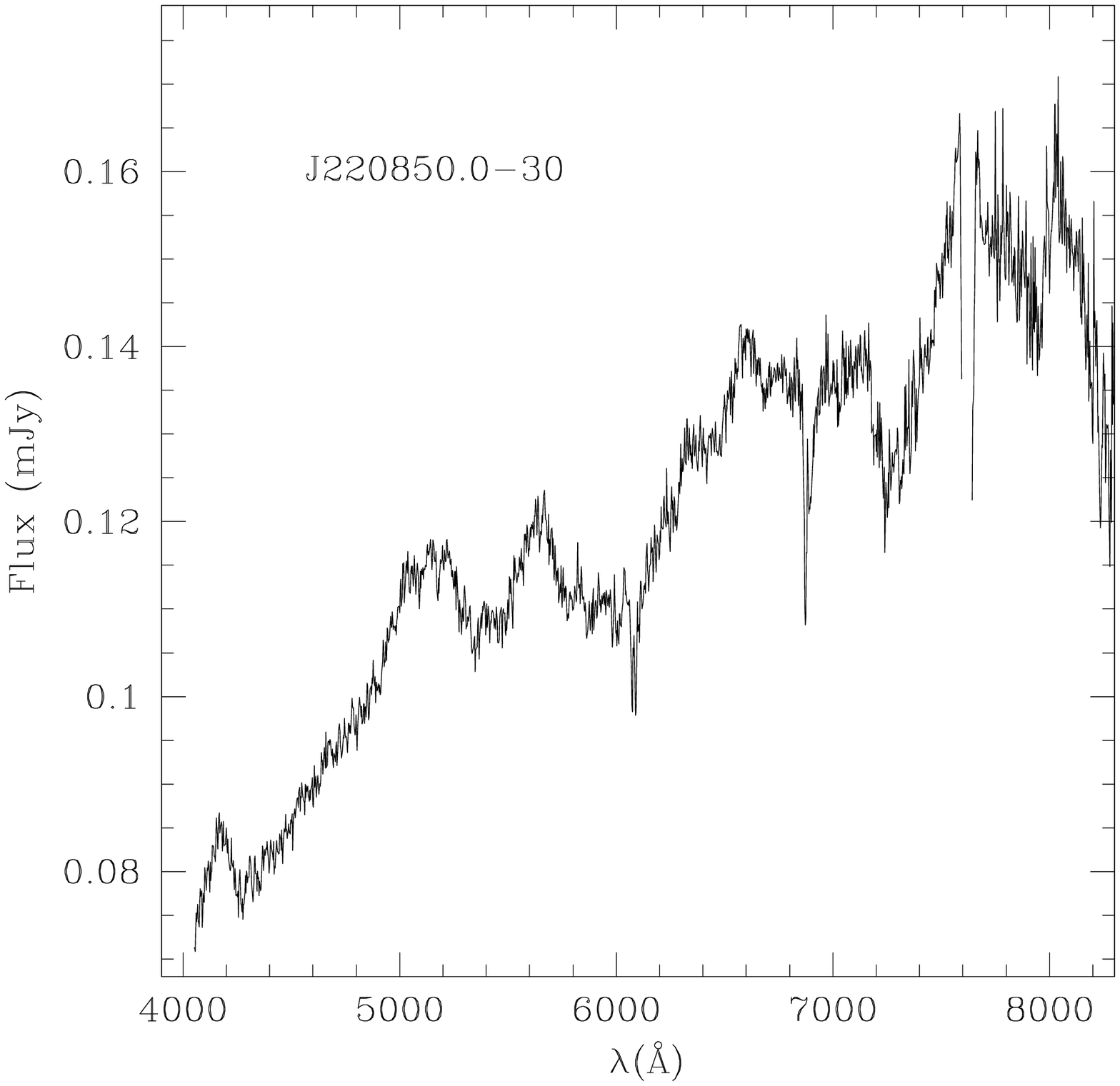,width=2.4in}}
\centerline{\psfig{file=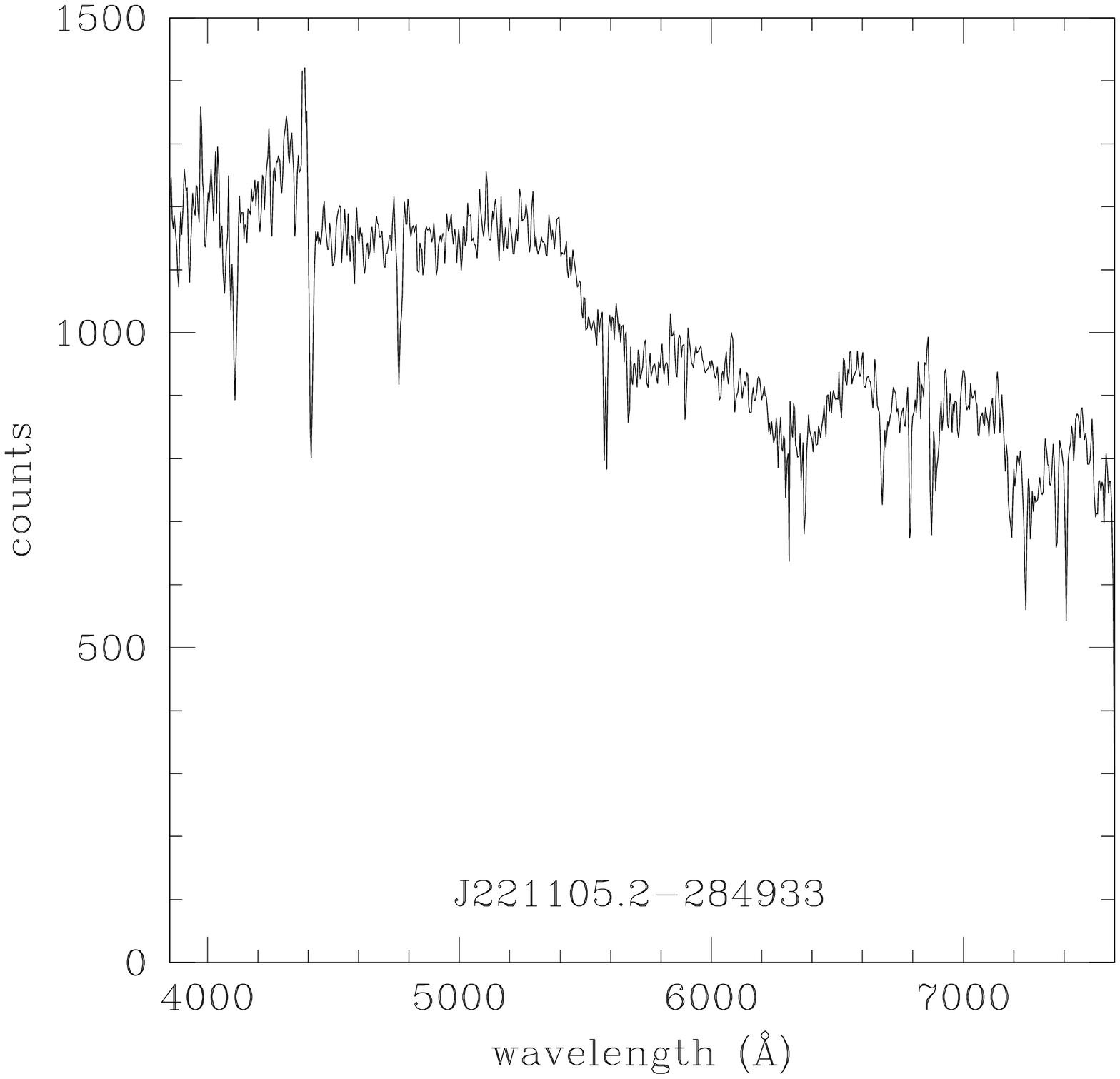,width=2.5in}\psfig{file=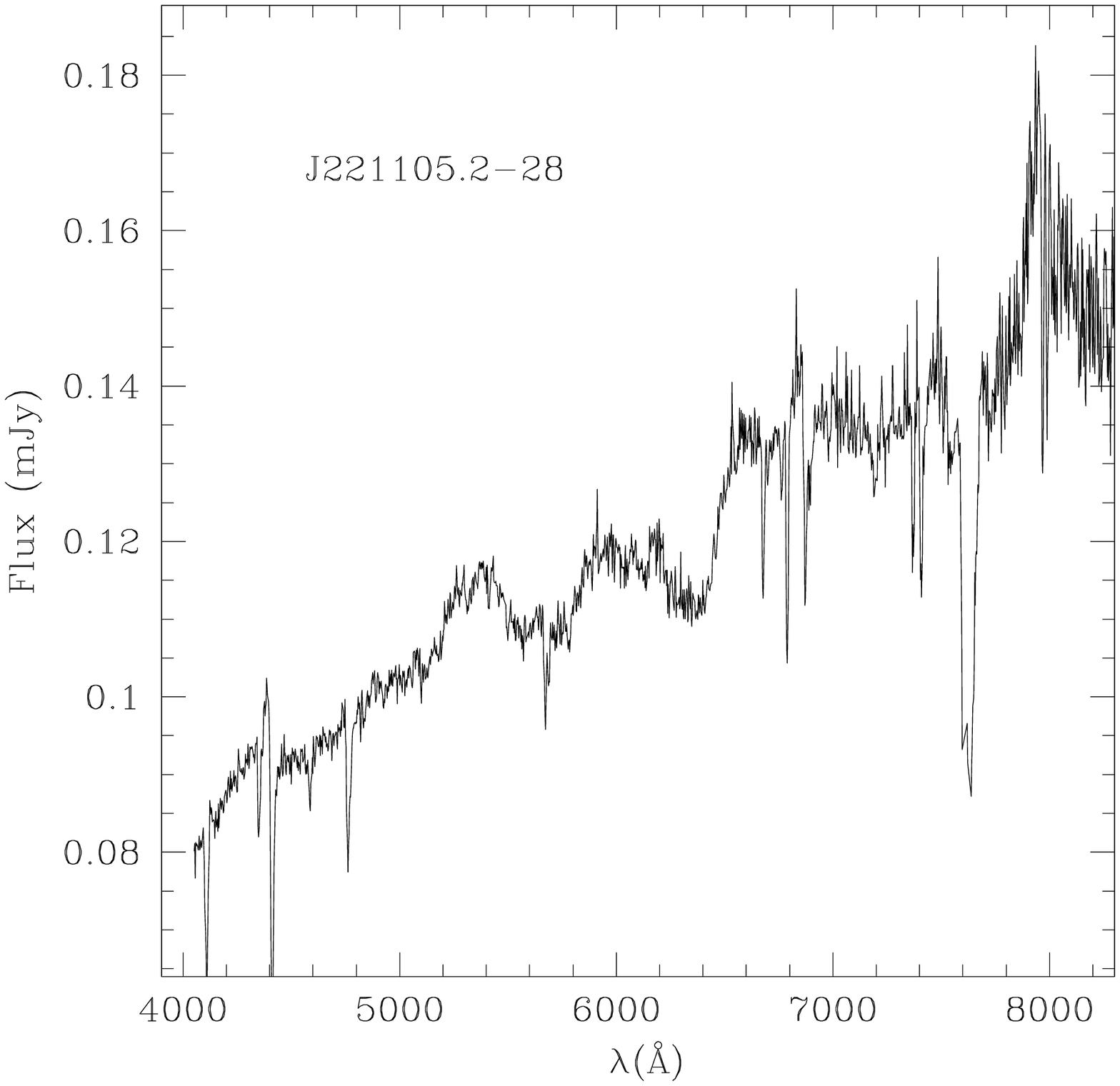,width=2.4in}}
\centerline{\psfig{file=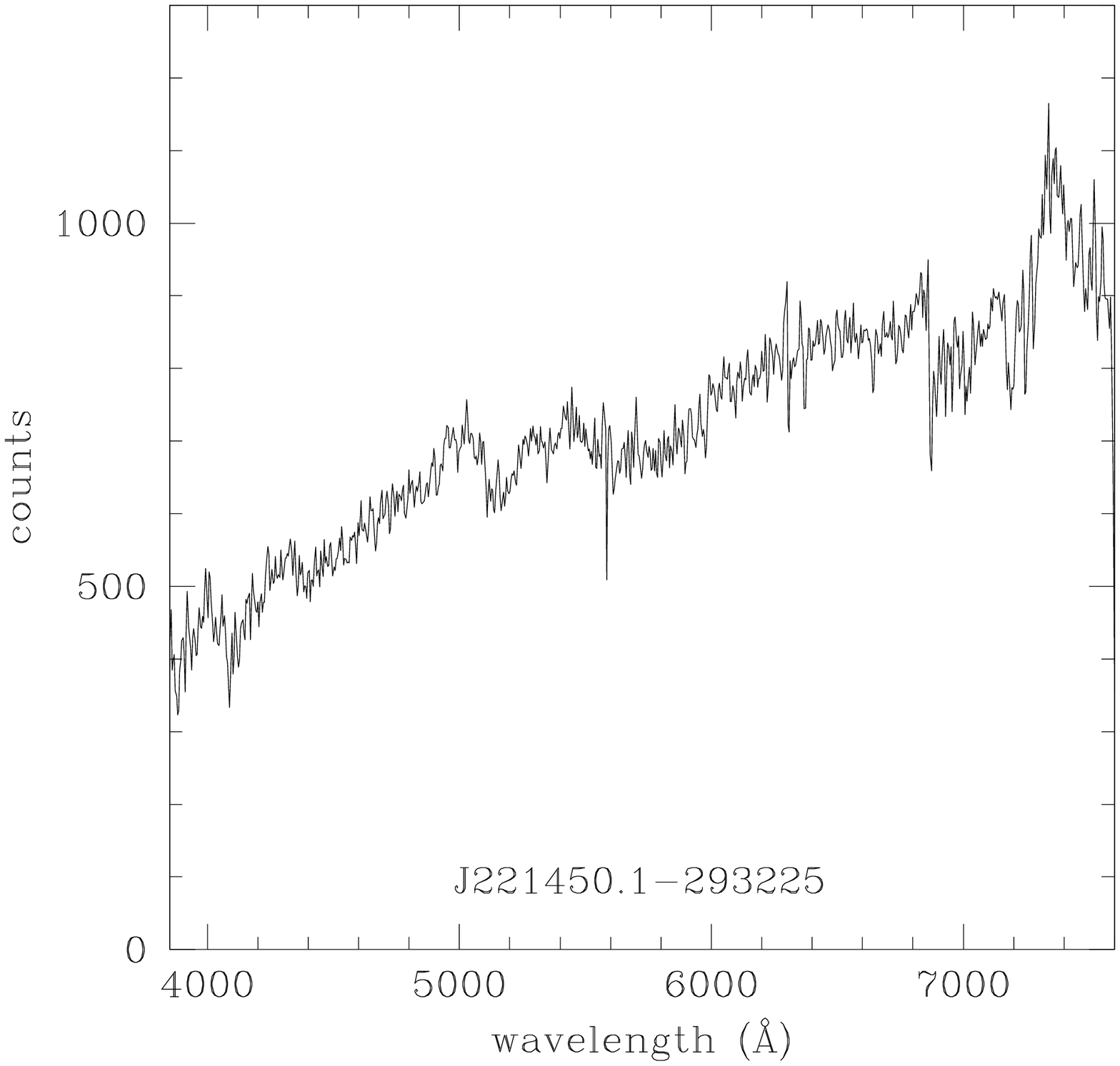,width=2.5in}\psfig{file=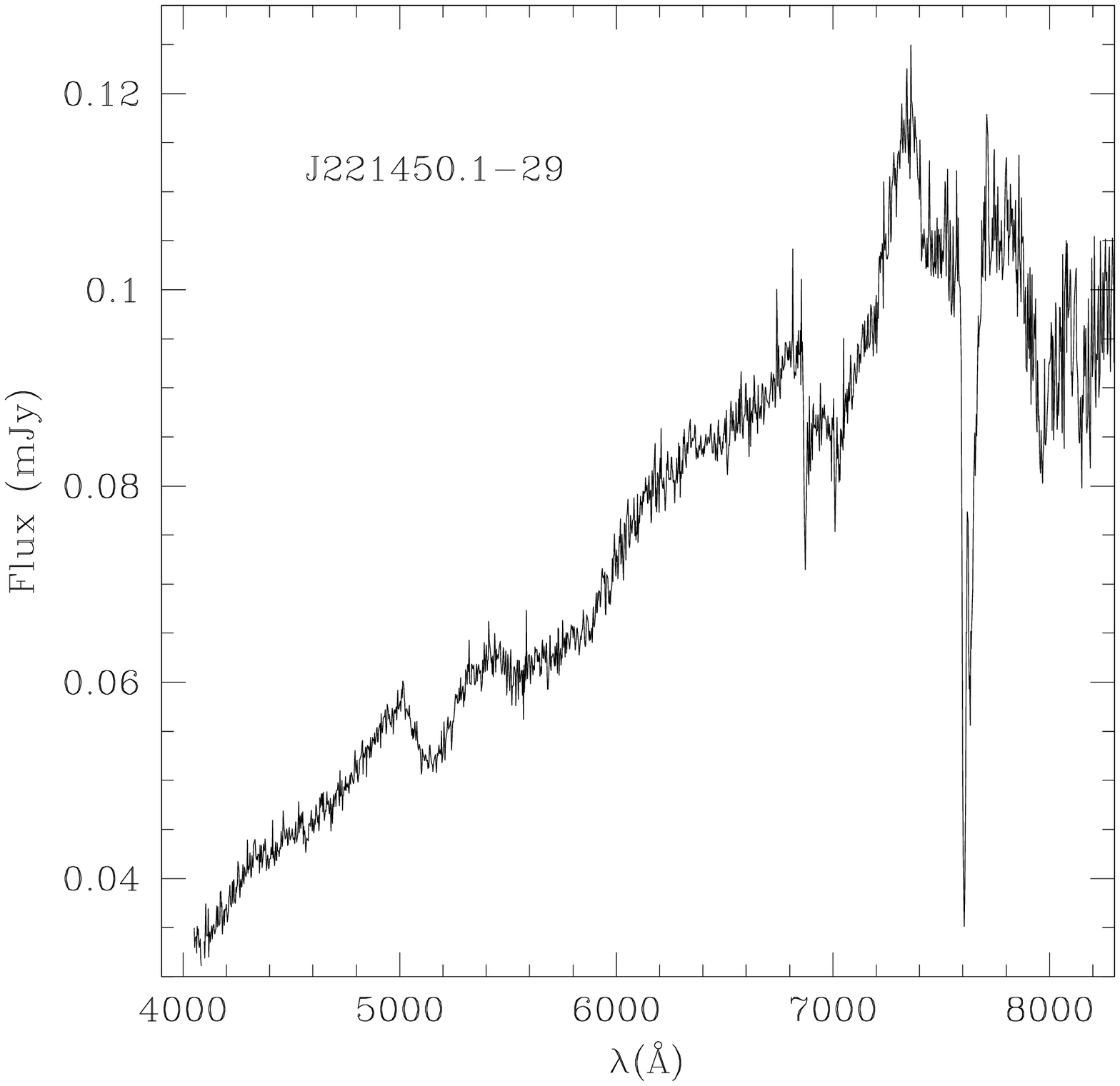,width=2.4in}}
\caption{Top: 2dF and VLT spectrum of the radio-quiet object J220850.0$-$30. Centre: 2dF and VLT spectrum of the radio-quiet object J221105.2$-$28. Bottom: 2dF and VLT spectrum of the radio-quiet object J221450.1$-$29. }
\end{figure*}
\begin{figure*}
\centerline{\psfig{file=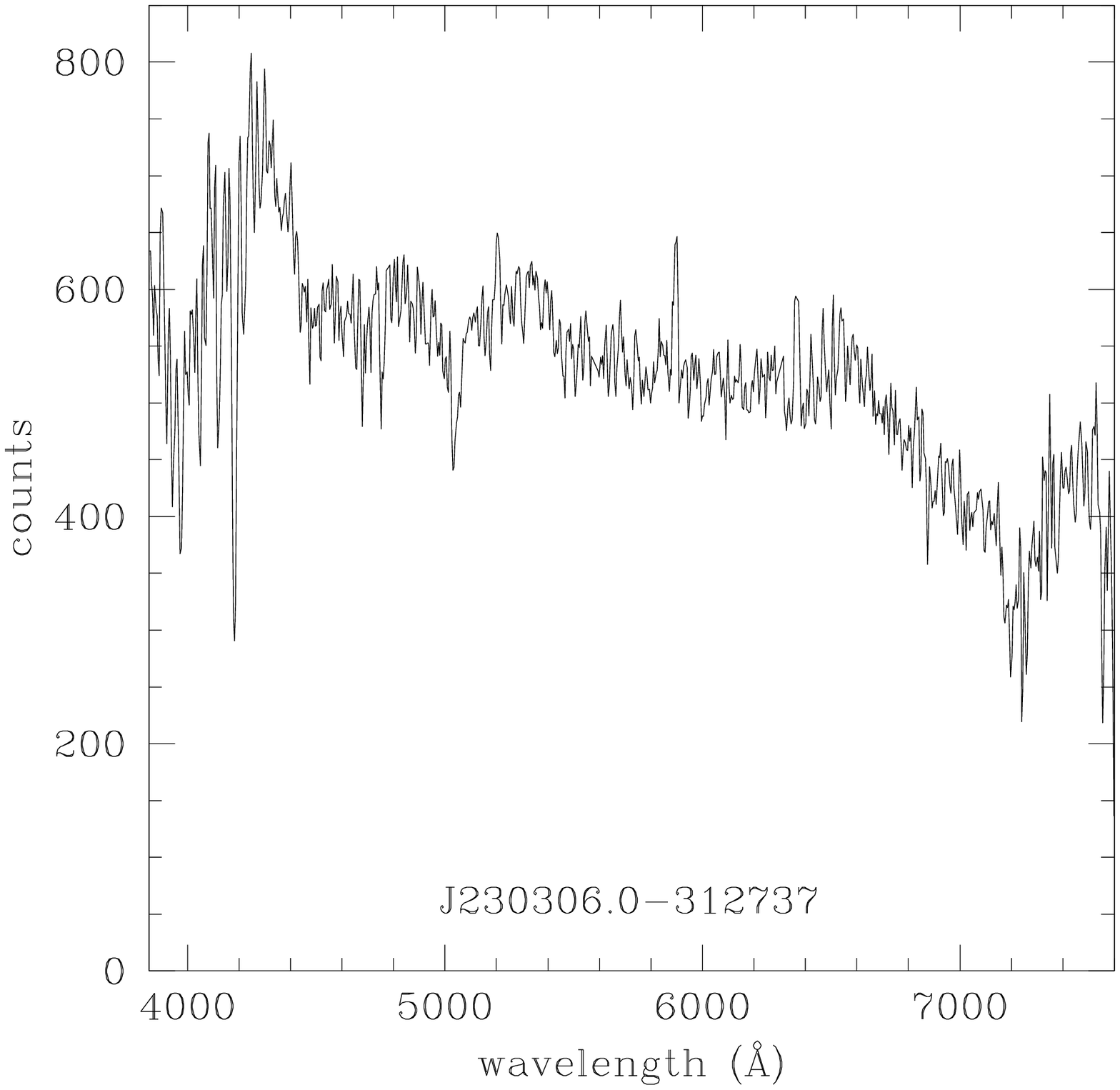,width=2.5in}\psfig{file=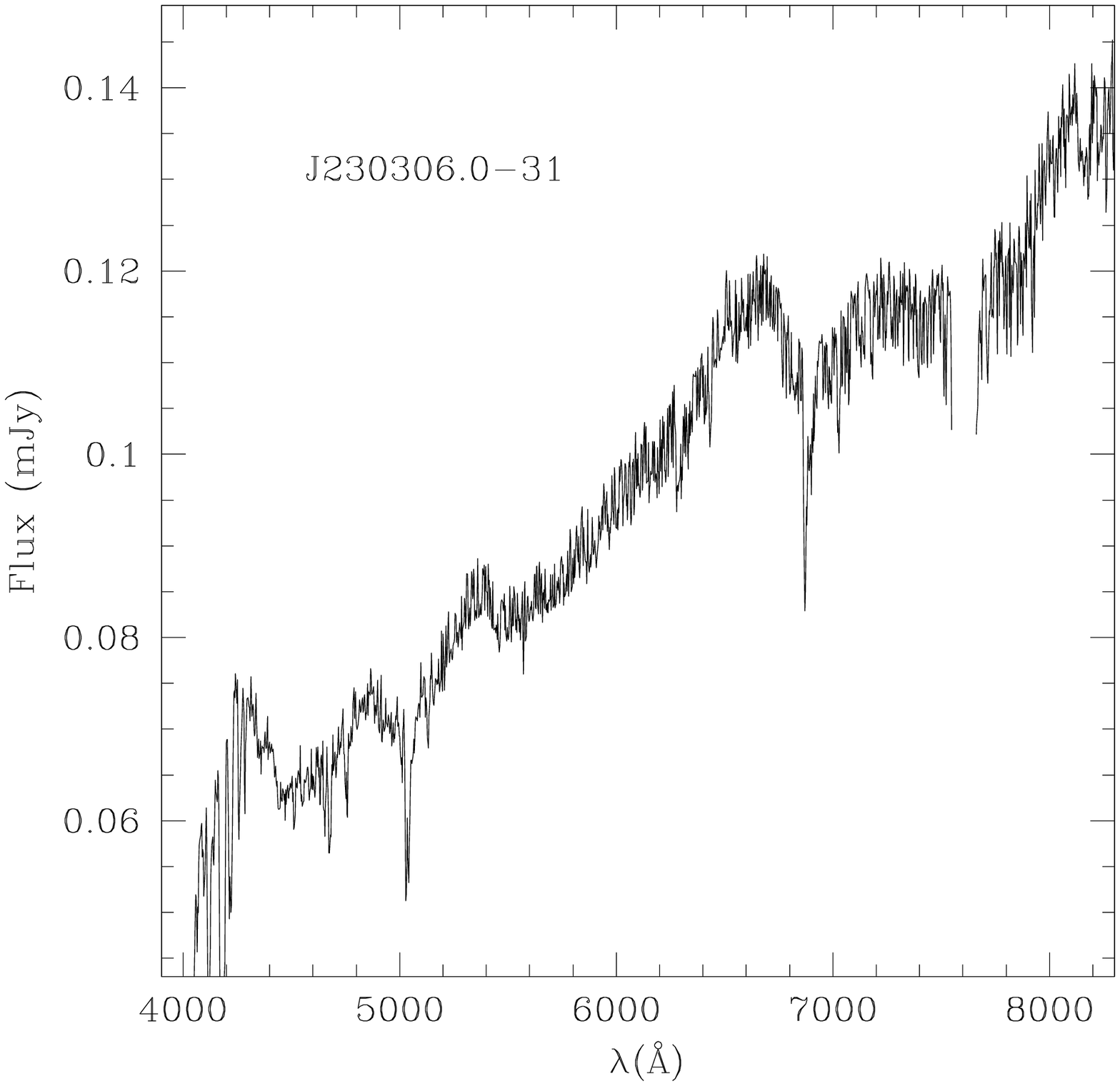,width=2.4in}}
\centerline{\psfig{file=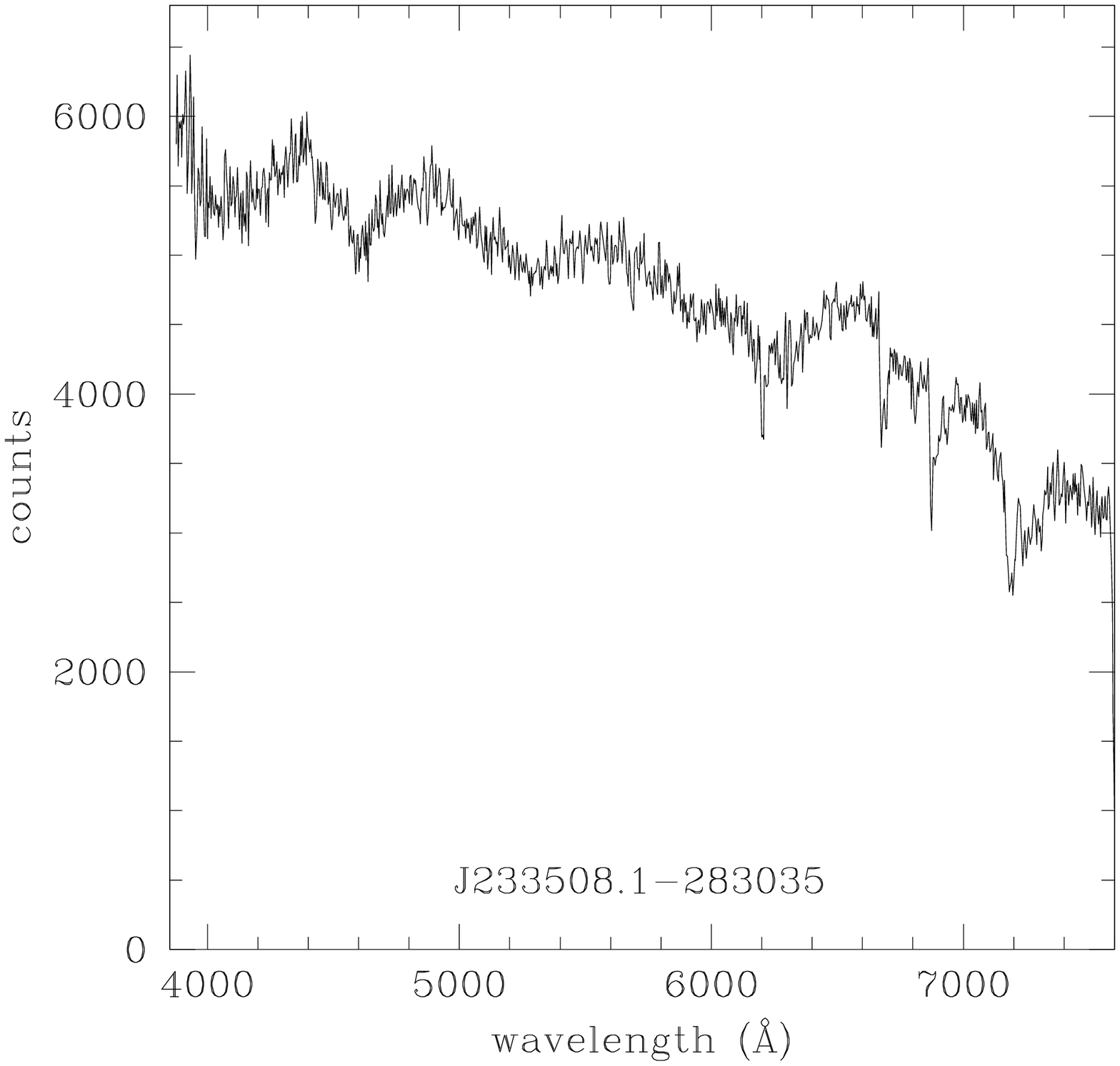,width=2.5in}\psfig{file=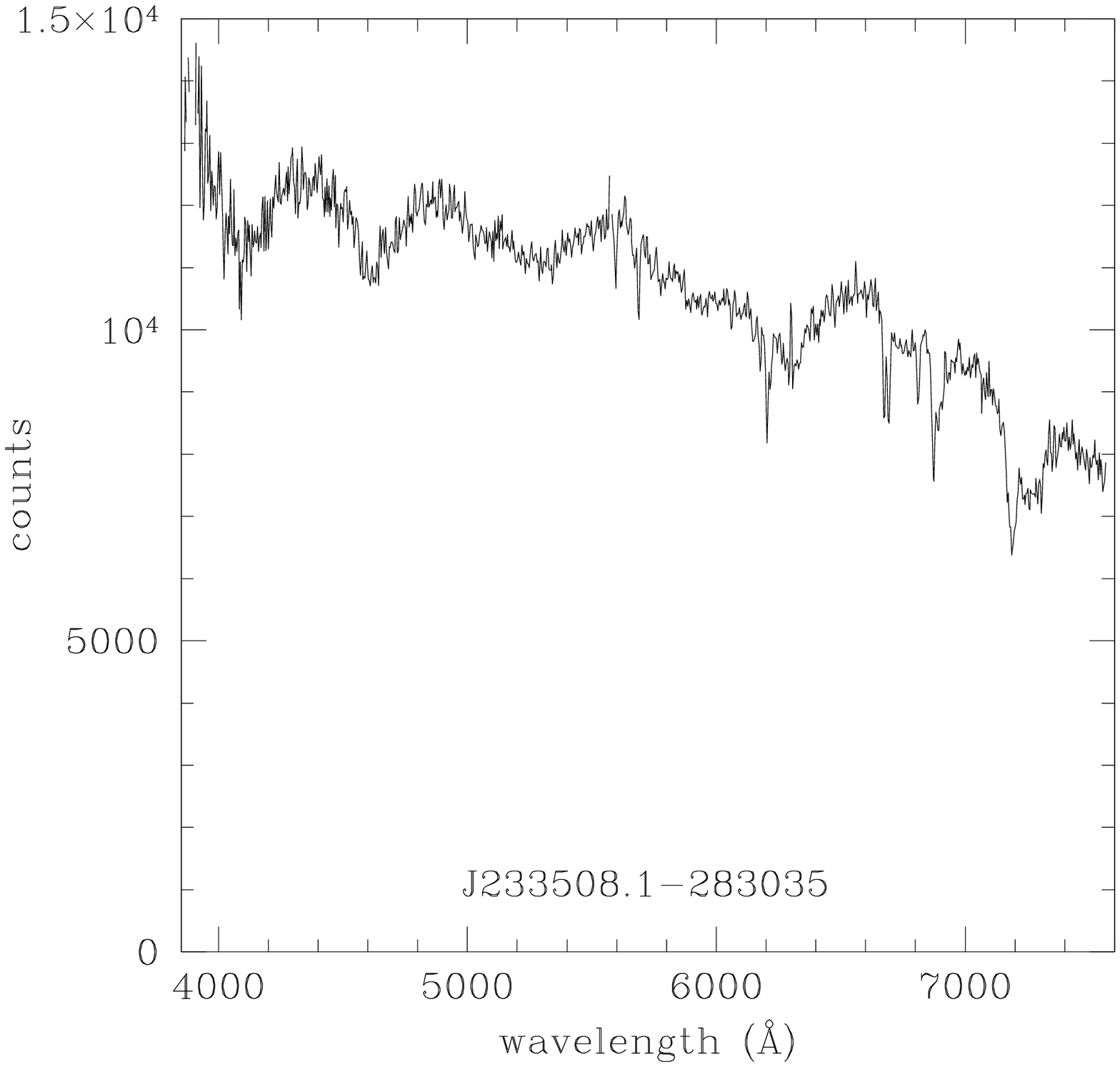,width=2.4in}}
\centerline{\psfig{file=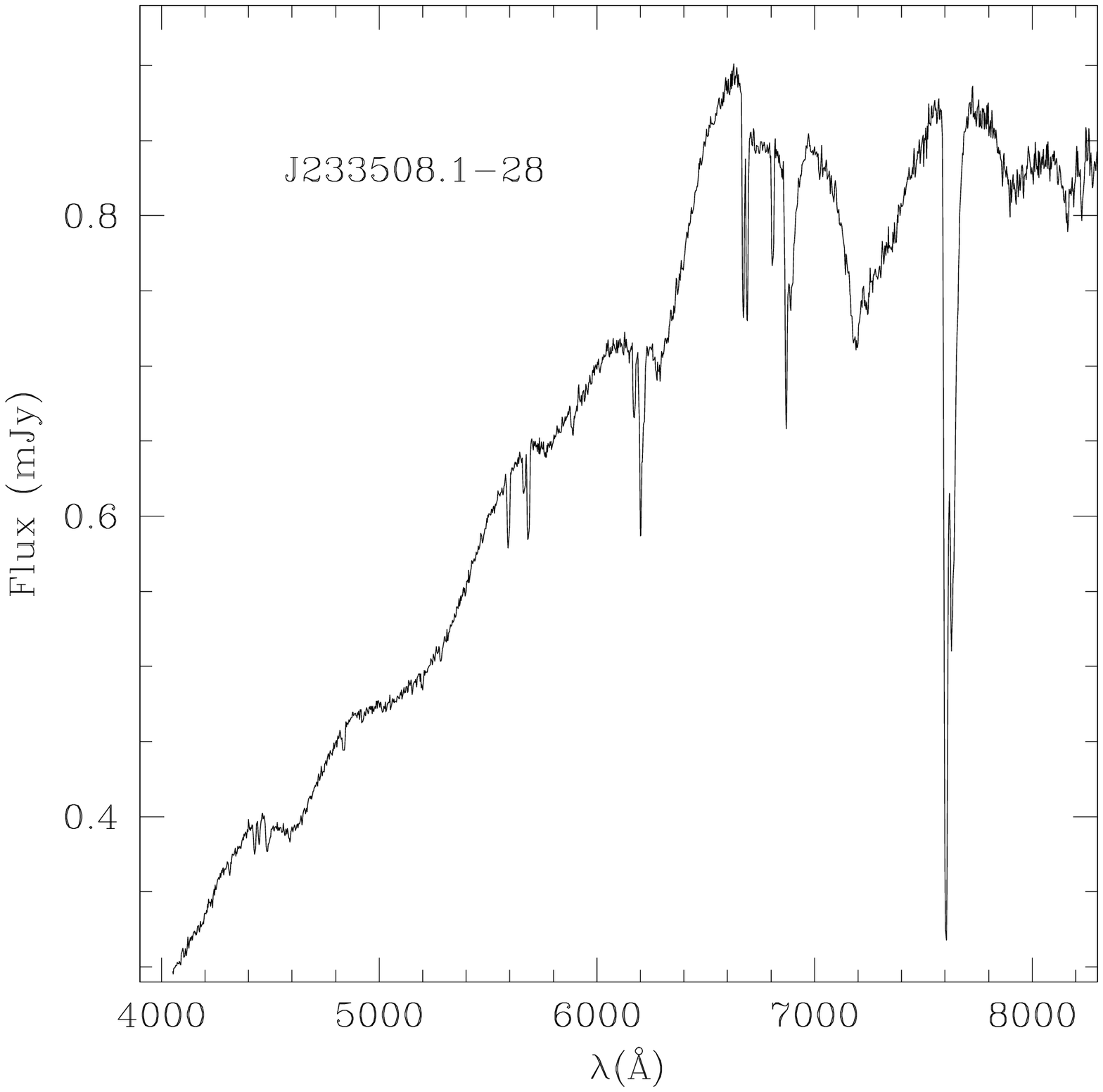,width=2.4in}}
\caption{Top: 2dF and VLT spectrum of the radio-quiet object J230306.0$-$31. Centre and bottom: 6dF spectra and VLT spectrum of the radio-quiet object J233508.1$-$28}
\end{figure*}


\end{document}